\documentclass[11pt]{article}
\usepackage{amsfonts,amssymb,amsmath,numbysec}

\usepackage{mathrsfs,fge}

\usepackage{pdfsync,color}

\usepackage[normalem]{ulem}

\usepackage{relsize,bigints}

\usepackage{hyperref}

\hsize \textwidth
\advance \hsize by -\marginparwidth
\evensidemargin \oddsidemargin
\usepackage{amssymb}
\advance\hoffset by 5mm
 \newcommand{\supp}{{\rm supp\,}}

\newcommand{\proof}{{\em Proof. }}
\newcommand{\qed}{$\Box$ }
\newcommand{\bbZ}{{\mathbb Z}}
\newcommand{\bbP}{{\mathbb P}}
\newcommand{\bbR}{{\mathbb R}}

\newcommand{\bbT}{{\mathbb T}}
\newcommand{\bW}{{\bf W}}
\newcommand{\eps}{\epsilon}

\newcommand{\Om}{\Omega}
\newcommand{\om}{\omega}

\newcommand{\ga}{\gamma}

\newcommand{\la}{\lambda}
\newcommand{\al}{\alpha}
\newcommand{\bbE}{\mathbb E}
\newcommand{\fg}{\fgeeszett}

\newtheorem{df}{Definition}[section]
\newtheorem{remark}[df]{Remark}
\newtheorem{lm}[df]{Lemma}
\newtheorem{lemma}[df]{Lemma}
\newtheorem{prop}[df]{Proposition}
\newtheorem{thm}[df]{Theorem}
\newtheorem{cor}[df]{Corollary}

\newcommand{\frakg}{\fgeeszett}

\newcommand{\farc}{\frac}

\newcommand{\commentout}[1]{{}}

\makeatother

\begin{document}

\numberbysection

\title{Kinetic limit  for a chain of harmonic oscillators with a point Langevin thermostat}

\author{Tomasz Komorowski\thanks{Institute of Mathematics, Polish Academy Of Sciences,
ul. \'{S}niadeckich 8,   00-956 Warsaw, Poland, 
e-mail: {\tt komorow@hektor.umcs.lublin.pl}}
\and
Stefano Olla\thanks{CEREMADE,  
  Universit\'e de Paris Dauphine, PSL Research University,
{Institut Universitaire de France}, GSSI, L'Aquila,
e-mail:{ \tt olla@ceremade.dauphine.fr}} 
}

\maketitle

\begin{abstract}
  We consider an infinite chain of coupled harmonic oscillators
 whose  {Hamiltonian dynamics is perturbed by a random exchange of momentum
    between particles such that total energy and momentum are conserved,
    modelling collision between atoms. This random exchange is rarefied in the limit,
    that corresponds to the hypothesis that in the macroscopic unit time only a finite number
    of collisions takes place (the Boltzmann-Grad
    limit). Furthermore,}  {the system is in contact    with  a Langevin thermostat at temperature $T$ through a single particle.} We prove that,
after the hyperbolic space-time rescaling, the Wigner distribution,
describing the energy density of phonons in space-frequency domain,
converges to a positive energy density function $W(t, y, k)$ that
evolves according to a 
linear kinetic equation,
with the interface condition at $y=0$ that corresponds to reflection,
transmission and absorption of phonons caused by the presence of the thermostat.
The  paper extends the results of
\cite{kors}, where a harmonic chain (with no inter-particle
scattering)  {  in contact with a Langevin thermostat} has been considered.
 \end{abstract}

\section{Introduction}
\label{intro}

The mathematical analysis of macroscopic energy transport in anharmonic chain of
oscillators constitutes  a very hard mathematical problem, see \cite{spohn2006}.
One approach to it is to replace the non-linearity by a
stochastic exchange of momentum between nearest neighbor particles
{in such a way that the total kinetic energy and momentum  are conserved}.
This stochastic exchange can be modeled in various ways: e.g. for each couple of
nearest neighbor particles the exchange of their momenta can occur independently at an exponential time (which models their elastic collision).
 {Otherwise, for each triple of consecutive particles, exchange of momenta can be performed
in a continuous, diffusive fashion, so  that its energy and  momentum are preserved.
In the present article we adopt  the latter choice, see Section
\ref{sec2.2.1} below for a detailed description of the dynamics. In fact, with
no significant changes, all our results can be extended to some
other stochastic noises that are both local in the spatial variable and of
independent increments in time, such as e.g. the Poisson random
exchange mentioned above
(see \cite[Section 5.1.3]{lepri})}.

A small parameter $\eps>0$, describing the  ratio between macroscopic and microscopic
space-time units, is introduced. The intensity
of the noise is adjusted so that in a (macroscopic) finite interval of time,
there is only  a finite amount of momentum exchanged by the stochastic mechanism.
In terms of the random exchanges, it means that, on average, each particle undergoes
only a finite number of stochastic collisions in a finite
time. 
 {Letting $\eps\to 0$ corresponds therefore to taking the kinetic limit
for the system, since in the case of a  chain  with no microscopic
boundary present the energy density evolution is described by a linear kinetic
equation, see \cite{BOS}}.

The Wigner distribution is a useful tool to localize in space the energy per frequency mode.
In the absence of the thermostat,  it is proven in \cite{BOS}, that as $\eps\to 0$, the Wigner distribution
converges to the solution of the kinetic transport equation
\begin{equation}
  \label{eq:bos}
  \partial_t W(t,y,k) + \bar\omega'(k) \partial_y W(t,y,k) =
  2\gamma_0 \int_{\bbT} R(k,k') \left(W(t,y,k') - W(t,y,k)\right)  dk,\, (t,y,k)\in [0,+\infty)\times\bbT\times \mathbb R,
\end{equation}
with an explicitly given  scattering kernel $ R(k,k') \ge 0$. It  is symmetric and the total scattering kernel
behaves as 
\begin{equation}
\label{Rk}
R(k):=\mathlarger{\int}_{\bbT}R(k,k')dk' \sim |k|^2\quad\mbox{ for }|k| \ll 1.
\end{equation} Here $\bbT$ is the unit torus, which is the interval $[-1/2,1/2]$, with identified
endpoints. Furthermore $\gamma_0>0$ is the scattering rate for the microscopic chain (see \eqref{eq:bas1} below) and $\bar\omega(k) = \omega(k)/2\pi$, where
$\omega(k)$ is the dispersion relation of the chain (see definition \eqref{mar2602}).

In the present paper we are interested in the macroscopic effects of a heat bath at temperature $T$,
modeled by a Langevin dynamics, applied to one particle, say the one labelled $0$,
with a coupling strength $\gamma_1>0$ (see \eqref{eq:bas1} below for a detailed description).
 {Unlike the conservative stochastic dynamics
acting on the bulk, the action of the heat bath is not rescaled with
$\eps$. In the
macroscopic coordinates it is represented by  large terms (a
deterministic and stochastic ones of orders $\eps^{-1}$ and
$\eps^{-1/2}$ respectively), see \eqref{basic:sde:2a} below. 
Therefore the presence of a thermostat can be considered as a singular
perturbation of the dynamics of the closed system.}
The effect in the limit, as $\eps\to 0$, is to introduce the following interface conditions
at $y=0$ on \eqref{eq:bos}:
\begin{equation}\label{bc-int}
  \begin{split}
W(t,0^+,k) &=p_-(k) W(t,0^+, -k) + p_+(k)W(t, 0^-,k)+{\frakg}(k)T, \quad\hbox{ for $0< k\le 1/2$},\\
W(t,0^-, k) &=p_-(k)W(t,0^-,-k) + p_+(k) W(t,0^+, k) +{\frakg}(k)T,\quad \hbox{ for $-1/2< k< 0$.}
\end{split}
\end{equation}
Interpreting $W(t,y,k)$ as the density of the energy of the phonons of mode $k$ at time $t$ and
position $y$, then $ p_+(k)$, $ p_-(k)$ and ${\frakg}(k)$ are respectively the probabilities for
transmission, reflection and absorption of a phonon of mode $k$ when it crosses $y=0$,
while ${\frakg}(k)T$ is the rate of creation of a phonon of that mode. These probabilities are
functions of $k$ and depend only on the dispersion relation $\omega(\cdot)$ and  intensity
{ $\gamma_1$} of the thermostat (cf. \eqref{033110}).
They are properly normalized, i.e. $ p_+(k) + p_-(k) + {\frakg}(k) = 1$,
so that $W(t,y,k) = T$ is a stationary solution (thermal equilibrium).

This result was recently proven in the absence of the conservative noise in the bulk
(i.e. $\gamma_0= 0$ in \eqref{eq:bos}), see  \cite{kors}. 
Then, the resulting dynamics outside the interface, given by \eqref{eq:bos},
reduces itself to pure transport as $\ga_0=0$.
Obviously, the coefficients appearing in the
interface conditions \eqref{bc-int} do not depend on the presence of the bulk noise.

The goal of the present paper is  to extend the result of \cite{kors}
to the case when the inter-particle noise is present, i.e. $\ga_0>0$,
see Theorem \ref{main-thm} below for the precise formulation of our main result.
We emphasize that in the situation when $\gamma_0=0$,
both the equations for the microscopic and macroscopic dynamics,
given below by \eqref{basic:sde:2a} and  \eqref{eq:bos} respectively,
can be solved  explicitly, in terms of the initial condition,
and this fact has been  extensively used in the proof
in \cite{kors}.
The argument can be extended
to the dynamics where only the damping terms of the noise are present,
i.e. with no noise input both from the inter-particle scattering and the thermostat,
see \eqref{012503-19}.
The equation for the macroscopic limit of the respective
Wigner distribution $W^{\rm un}(t,y,k)$ reads (cf \eqref{Rk2.34} below),
see Theorem \ref{main:thm-un} below,
\begin{equation}
  \label{eq:damp}
   \partial_t W^{\rm un}(t,y,k) + \bar\omega'(k) \partial_y W^{\rm un}(t,y,k) =
  -2 \gamma_0 R(k) W^{\rm un}(t,y,k), \quad y\not=0,
\end{equation}
with the boundary conditions as in \eqref{bc-int}. In the next step we
add the stochastic part corresponding to the inter-particle scattering,
which corresponds to $T=0$ for the thermostat,
see equation \eqref{basic:sde:2av1} formulated for the respective wave function.
Next, we use the previously described dynamics
to represent the solution of the equation with the help of the
Duhamel formula, see \eqref{eq:sol1}.
The corresponding representation for the Wigner distribution
is given in \eqref{020805-19}. Having already established the macroscopic limit
for the dynamics with no stochastic noise, we can use the Duhamel representation
to identify the kinetic limit of the noisy microscopic dynamics
when the thermostat temperature $T=0$, see Theorem \ref{cor020805-19}.
The extension to the case when the temperature $T>0$ is possible by
another application of the Duhamel formula, see Section
\ref{sec10}.   {We end this brief description of our argument with the
remark that our method relies quite
substantially on the fact that the noise in the bulk has independent increments in time,
so we can treat the respective stochastic terms as a perturbation of
the microscopic dynamics (involving the It\^o correction terms) whose
limit can be found by explicit calculations}.

 {Concerning the derivation of
 the kinetic limit from the wave equation in the presence of
 boundaries, or sources, as far as we know, there exist only a few
 results
 dealing with the subject. We mention here the papers \cite{miller,miller97,miller00},
 where the high-frequency asymptotics of the solutions of the Cauchy problem for either
the Schr\"{o}dinger, or  scalar
wave equation with discontinuous coefficients, corresponding to a
two component medium, has been considered and the semiclassical limit of the
probability density (in the case of the Schr\"{o}dinger)  and energy
density (for the wave equation) are identified.  
 In Benamou et al \cite{castella03} the Wigner distribution method is
 used to study the geometrical optics limit of the Helmholtz equation
with source terms. In  the aforementioned results the  coefficients
of the equation are not random. The kinetic
limit  has been derived from the random Schr\"{o}dinger equation in
the whole space only
when no sources, or boundaries are present, see
\cite{spohn,ey,chen} for time independent potential and
\cite{bpr1,bpr2,bkr,fan}, where the time
dependent potentials have been treated. The derivation from the wave
equation has been shown in \cite{BOS,ks,boutz}. The wave
equation with
no randomness in the bulk and an interface, given by a point Langevin
thermostat, has been considered in the aforementioned paper
\cite{kors}.  To the best of our
knowledge the present paper contains the first derivation of the
kinetic limit  from the wave equation in  the presence of both randomness
in the bulk and an interface.}  {The long time, large
space scale asymptotics  of the kinetic limit obtained
in the present article has been discussed in \cite{BOK,koran}.}

Concerning the organization of the paper, Section \ref{prelim} is devoted to  preliminaries
and the formulation of the main result, see Theorem \ref{main-thm}.
Among things discussed there is the rigorous definition of a solution of a kinetic equation \eqref{eq:bos}
with the interface condition \eqref{bc-int}, see Sections \ref{sec2.6.3} and \ref{sec2.6.4}. 
Section \ref{sec3} deals with  the basic properties of the microscopic dynamics obtained
by removal of stochastic noises, both between the particles of the chain and the thermostat.
This dynamics is an auxiliary tool for the mild formulation of the
microscopic dynamics corresponding to the chain with inter-particle
scattering and thermostat. 
We discuss first the case when  thermostat is set at $T=0$,
see Section \ref{sec4}. In this section we obtain also 
basic  estimates for the microscopic Wigner distributions,
see  Proposition \ref{prop012105-19}, that follow from the energy
balance equation established in \eqref{energy-balance100}.
A similar result is also formulated for the auxiliary dynamics
with no stochastic noise in Section \ref{sec5.1}.
In Section \ref{sec5.3} we formulate the result concerning the kinetic limit for this dynamics,
see Theorem \ref{main:thm-un}.
Its proof is quite analogous to the argument of \cite{kors}
and is given in Appendix \ref{appb}. Section \ref{sec5.5}
is essentially devoted to the proof of the main result  (Theorem \ref{main-thm})
for the case $T=0$ and the proof for $T>0$ is presented in Section \ref{sec10}.
Some properties of the dynamics corresponding to the macroscopic kinetic limit are proven
in Appendix \ref{appa}. Section \ref{appC} of the appendix is devoted to
the proof of some properties of the interface coefficients appearing in the limit.

\section*{Acknowledgements}

  T.K.  acknowledges the support of the National Science Centre:
  NCN grant DEC-2016/23/B/ST1/00492. S.O. acknowledges the ANR-15-CE40-0020-01 LSD 
 grant of the French National Research Agency.

\section{Preliminaries and statement of the main results}

\label{prelim}

\subsection{Basic notation}

We shall use the following notation: let
$\bbR_*:=\bbR\setminus\{0\}$, $\bbR_+:=(0,+\infty)$,
$\bbR_-:=(-\infty,0)$   and likewise
$\bbT_*:=\bbT\setminus\{0\}$, $\bbT_+:=(0,1/2)$,
$\bbT_-:=(-1/2,0)$.
Throughout the paper we use the short hand notation
\begin{equation}
\label{011909}
{\frak s}(k):=\sin(\pi k)\quad {\frak c}(k):=\cos(\pi k),\quad k\in\bbT.
\end{equation} 
Let
$
e_x(k):=\exp\{2\pi i xk\}$ for  $x$ belonging to the set of integers $\bbZ.$ 
The Fourier series corresponding to a complex valued sequence
$(f_x)_{x\in\bbZ}$ belonging to $\ell_2$ - the Hilbert space of square
integrable sequences of complex numbers - is given by
  \begin{equation}
  \label{fourier}
  \hat f(k)=\sum_{x\in\bbZ}f_xe_x^\star(k), \quad k\in\bbT.
\end{equation}
Here $z^\star$ is the complex conjugate of $z\in\mathbb C$.
By the Parseval identity $\hat f\in L^2(\bbT)$ - the space of complex
valued, square integrable functions - and $\|\hat f\|_{L^2(\bbT)}=\|f\|_{\ell_2}$. 

Given $\eps>0$ we let $\bbZ_{\eps}:=(\eps/2)\bbZ$ and 
$\bbT_\eps:=(2/\eps)\bbT$. Let $\ell_{2,\eps}$ be the space made of
all complex valued square integrable sequences $(f_y)_{y\in
  \bbZ_\eps}$ equipped with the norm
$$
\|f\|_{\ell_{2,\eps}}:=\left\{\frac{\eps}{2}\sum_{y\in\bbZ_\eps}|f_y|^2\right\}^{1/2}.
$$
Let 
\begin{equation}
  \label{fouriereps}
  \hat f(\eta)=\frac{\eps}{2}\sum_{y\in\bbZ_\eps}f_ye_y^\star(\eta), \quad \eta\in\bbT_\eps.
\end{equation}
The Parseval identity takes then the form
 $\|\hat f\|_{L^2(\bbT_\eps)}=\|f\|_{\ell_{2,\eps}}$.

For any non-negative  functions $f,g$ acting on a set $A$ the notation
$f\preceq g$ means that there exists a constant $C>0$ such that
$f(a)\le Cg(a)$ for $a\in A$.  We shall write $f\approx g$ if
$f\preceq g$ and $g\preceq f$.

Given a function $f:\bar\bbR_+\to\mathbb C$ satisfying $|f(t)|\le
Ce^{Mt}$, for fome $C,M>0$ we denote by $\tilde
f(\la)$ its Laplace transform
$$
\tilde
f(\la)=\int_0^{+\infty}e^{-\la t}f(t)dt,\quad {\rm Re}\,\la>M.
$$

\subsection{Some function spaces}

For a given  $ G\in {\cal S}(\bbR\times\bbT)$  - the class of Schwartz functions on
$\bbR\times\bbT$ -  we let
$$
\widehat G(\eta,k)=\int_{\bbR}e_y^\star(\eta)G(y,k)dy
$$
be its  Fourier transform in the first variable. Let 
\begin{equation}
\label{AC}
{\cal
  A}_c:=[G:\,\hat G\in C_c^\infty(\bbR\times\bbT)].
\end{equation}
Let ${\cal A}$ be the Banach space obtained by the completion of ${\cal A}_c$  in the norm
\begin{equation}
\label{060805-19}
\|G\|_{\cal A}:=\int_{\bbR}\sup_{k\in\bbT}|\widehat G(\eta,k)|d\eta,\quad G\in{\cal
  A}_c.
\end{equation}
Space
${\cal A}'$ - the dual to ${\cal A}$ - consists of all distributions
$G\in {\cal
  S}'(\bbR\times\bbT)$ of the form
$$
\langle G, F\rangle=\int_{\bbR\times \bbT} \widehat G^\star(\eta,k)
\widehat F(\eta,k)d\eta dk,\quad F\in {\cal A}
$$
for some measurable function $\widehat G:\bbR\times\bbT\to\mathbb C$,
equipped with the norm
\begin{equation}
\label{060805-19a}
\|G\|_{{\cal A}'}=\sup_{\eta\in\bbR}\int_{\bbT}|\widehat G(\eta,k)|d k<+\infty.
\end{equation}

We shall also consider the spaces  ${\cal
  L}_{2,\eps}:=\ell_{2,\eps}\otimes L^2(\bbT)$. The respective norms of
$G:\bbZ_\eps\times\bbT\to\mathbb C$ and $\widehat
G:\bbT_\eps\times\bbT\to\mathbb C$  are 
given by
\begin{equation}
\label{011505-19}
\|G\|_{{\cal L}_{2,\eps}}:=\left\{\frac{\eps}{2}\sum_{y\in\bbZ_\eps}\|G_y\|_{L^2(\bbT)}^2\right\}^{1/2}
=\|\widehat
G\|_{L^2(\bbT_\eps\times\bbT)}.
\end{equation}

\subsection{Infinite system of interacting harmonic oscillators}

\subsubsection{ Hamiltonian dynamics
  with momentum and energy conserving noise
  in contact with a Langevin thermostat}

\label{sec2.2.1}

 {We consider a stochastically perturbed chain of harmonic oscillators
with a Langevin thermostat at a fixed temperature $T\ge 0$ at
$x=0$}. Its dynamics is
described by the system of It\^o stochastic differential equations
\begin{eqnarray}
&&d{\frak q}_{x}(t)={\frak p}_x(t)dt \nonumber
\\
&& d{\frak p}_x(t)=\left[-(\alpha\star{\frak
   q}(t))_x-\frac{\eps\ga_0}{2}(\theta\star{\frak
   p}(t))_x\right]dt+\sqrt{\eps\ga_0}\sum_{k=-1,0,1}(Y_{x+k}{\frak
   p}_x(t))dw_{x+k}(t)\label{eq:bas1} \\
&&
+\left(-\ga_1{\frak p}_0(t)dt+\sqrt{2\ga_1 T}dw(t)\right)\delta_{0,x},\quad x\in\bbZ.\nonumber
\end{eqnarray}
Here 
\begin{equation}
\label{011210}
Y_x:=({\frak p}_x-{\frak p}_{x+1})\partial_{{\frak p}_{x-1}}+({\frak p}_{x+1}-{\frak p}_{x-1})\partial_{{\frak p}_{x}}+({\frak p}_{x-1}-{\frak p}_{x})\partial_{{\frak p}_{x+1}}
\end{equation}
and
  $\left(w_x(t)\right)_{t\ge0}$, $x\in\bbZ$ with
  $\left(w(t)\right)_{t\ge0}$,  are i.i.d. one dimensional, real
  valued,  non-anticipative  standard Brownian motions, over some filtered probability space $(\Sigma,{\cal F},\left({\cal F}_t\right)_{t\ge0},\bbP)$. In addition,
 $$
\theta_x=\Delta\theta^{(0)}_x:=\theta^{(0)}_{x+1}+\theta^{(0)}_{x-1}-2\theta^{(0)}_x
$$
with 
$$
 \theta^{(0)}_x=\left\{
 \begin{array}{rl}
 -4,&x=0\\
 -1,&x=\pm 1\\
 0, &\mbox{ if otherwise.}
 \end{array}
 \right.
 $$
  A simple calculation shows that
\begin{equation}
\label{beta}
\hat \theta(k)=8{\frak s}^2(k)\left(1+2{\frak c}^2( k)\right),\quad k\in\bbT.
\end{equation}
Parameters  $\eps\gamma_0>0$, $\ga_1$ describe the strength of the
inter-particle and thermostat noises, respectively.
In what follows we shall assume that $\eps>0$ is small, that
corresponds to the  {weak noise} hypothesis  that results in  atoms suffering finitely many
''collisions'' in a macroscopic unit of time  (the Boltzmann-Grad limit). 

Since the vector field $Y_x$ is orthogonal both to a sphere ${\frak
  p}_{x-1}^2+{\frak p}_x^2+{\frak p}_{x+1}^2\equiv {\rm const}$ and 
plane ${\frak p}_{x-1}+{\frak p}_x+{\frak p}_{x+1}\equiv  {\rm const}$, the inter-particle noise conserves locally the kinetic energy and
momentum.

Concerning the Hamiltonian part of the dynamics, 
we assume  (cf  \cite{BOS}) that the coupling constants
$(\al_x)_{x\in\bbZ}$ satisfy the following:
 \begin{itemize}
 \item[a1)] they are real valued and there exists $C>0$ such that $|\alpha_x|\le Ce^{-|x|/C}$ for all $x\in \bbZ$,
  \item[a2)] $\hat\alpha(k)$ is also real valued 
and  $\hat\alpha(k)>0$ for $k\not=0$ and in case $\hat \alpha(0)=0$ we  have  $\hat\alpha''(0)>0$.
 \end{itemize}
   The above conditions imply that both functions $x\mapsto\alpha_x$
   and $k\mapsto\hat\alpha(k)$ are even. In addition, $\hat\alpha\in
   C^{\infty}(\bbT)$ and in case $\hat\alpha(0)=0$ we have
   $\hat\alpha(k)=k^2\phi(k^2)$ for some strictly positive  $\phi\in
   C^{\infty}(\bbT)$.  The  dispersion relation  $\om:\bbT\to
   \bar\bbR_+$, given by
\begin{equation}\label{mar2602}
 \om(k):=\sqrt{\hat \alpha (k)}
\end{equation}
is even.
Throughout the paper it is assumed to be unimodal, i.e. increasing on
$[0,1/2]$ and then, in consequence, decreasing on
$[-1/2,0]$. Its 
unique minimum and maximum, attained at $k=0$, $k=1/2$, respectively
are denoted by~$\om_{\rm min}\ge 0$ and $\om_{\rm max}$,
correspondingly. We denote the two branches of its inverse
by~$\om_\pm:[\om_{\rm min},\om_{\rm max}]\to\bar \bbT_\pm$.

\subsubsection{Initial data}

We assume that the initial data is random and, given $\eps>0$,
distributed according to probabilistic
measure $\mu_\eps$ and
\begin{equation}
\label{mu-eps}
{\cal E}_*:=\sup_{\eps\in(0,1]}\eps\sum_{x\in\bbZ}\langle{\frak e}_x\rangle_{\mu_\eps}<+\infty.
\end{equation}
 Here $\langle\cdot\rangle_{\mu_\eps}$ is the
expectation with respect to $\mu_\eps$ and the microscopic energy
density
$$
{\frak e}_x:=\frac12\left({\frak p}_x^2+\sum_{x'\in\bbZ}\al_{x-x'}{\frak q}_x{\frak q}_{x'}\right).
$$ The assumption \eqref{mu-eps} ensures
that the macroscopic energy density of the chain is
finite.

\subsection{Kinetic scaling of the wave-function}

To observe macroscopic effects of the inter-particle scattering
consider time of the order $t/\eps$.
It is also convenient to introduce the wave function that, adjusted to the macroscopic time, is given by
\begin{equation}
\label{011307}
\psi^{(\eps)}(t):=\tilde{\om} \star {\frak q}\left(\frac{t}{\eps}\right)+i{\frak p}\left(\frac{t}{\eps}\right),
\end{equation}
where  $({\frak p}(t),{\frak q}(t))$ satisfies \eqref{eq:bas1} and 
 $\left(\tilde \om_x\right)_{x\in\bbZ}$ are the
Fourier coefficients of the dispersion relation, see \eqref{mar2602}. We shall consider the Fourier transform of the wave
function, given by 
\begin{equation}
\label{011307a}
\hat\psi^{(\eps)}(t,k)=\om(k)\hat {\frak q}^{(\eps)}\left(t,k\right)+i\hat{\frak p}^{(\eps)}\left(t,k\right).
\end{equation}
Here $\hat {\frak q}^{(\eps)}\left(t\right)$, $\hat{\frak
  p}^{(\eps)}\left(t\right)$ are given by the Fourier series
for ${\frak q}^{(\eps)}_x(t):={\frak q}_x(t/\eps)$ and   ${\frak
  p}^{(\eps)}_x(t):={\frak p}_x(t/\eps)$, $x\in\bbZ$, respectively.
They satisfy
 \begin{eqnarray}
 \label{basic:sde:2a}
&&
 d\hat\psi^{(\eps)}(t,k)=\left\{-\frac{i}{\eps} \om(k)\hat\psi^{(\eps)}(t,k)
 -2i\ga_0R(k)\hat{\frak p}^{(\eps)}\left(t,k\right)-\frac{i\ga_1}{\eps}\int_{\bbT}\hat{\frak p}^{(\eps)}\left(t,k'\right)dk'\right\}dt
 \nonumber\\
 &&
 -2\sqrt{\ga_0}\int_{\bbT}r(k,k')\hat{\frak p}^{(\eps)}\left(t,k-k'\right)B(dt,dk')+i\sqrt{\frac{2\ga_1T}{\eps}}dw(t),\\
 &&
\hat\psi^{(\eps)}(0)= \hat\psi,\nonumber
 \end{eqnarray} 
where
\begin{align*}
&
\hat{\frak p}^{(\eps)}\left(t,k\right):=\frac{1}{2i}\left[\hat\psi^{(\eps)}(t,k)-\left(\hat\psi^{(\eps)}\right)^\star(t,-k)\right],\\
&
r(k,k'):=4 {\frak s}(k) {\frak s}(k-k') {\frak s}(2k-k')
 \quad k,k'\in \bbT,\\
&
R(k):=\int_{\bbT}r^2(k,k')dk'={\frak s}^2(2k)+2 {\frak s}^2(k)=\frac{\hat\theta(k)}{4}.
 \end{align*}
Here 
$B(t,dk)=\sum_{x\in\bbZ}w_x(t)e_x(k)dk$ is a cylindrical Wiener process on $L^2(\bbT)$, i.e.
\begin{equation}
\label{cnoise}
\bbE[B(dt,dk)B^\star(ds,dk')]=\delta(k-k')\delta(t-s)dt ds dkdk'.
\end{equation}

\subsection{Energy density - Wigner function}

One can easily check that
\begin{equation}
\label{e1}
\|\hat\psi^{(\eps)}(t)\|^2_{L^2(\bbT)}=\|\psi^{(\eps)}(t)\|^2_{\ell^2}=\sum_{x\in\bbZ}\left({\frak p}^{(\eps)}_x(t)\right)^2+\sum_{x,x'\in\bbZ}\al_{x-x'}{\frak q}^{(\eps)}_x(t){\frak q}^{(\eps)}_{x'}(t).
\end{equation}
 After straightforward calculations one can verify that
  \begin{equation}
 \label{031709}
 d\|\hat\psi^{(\eps)}(t)\|^2_{L^2(\bbT)}
 =-\frac{2\ga_1}{\eps} \left[
 \left({\frak p}^{(\eps)}_0(t)\right)^2
 - T \right]dt
  +\left(\frac{2\ga_1T}{\eps}\right)^{1/2}{\frak p}^{(\eps)}_0(t) dw(t),\quad \bbP_\eps\,\mbox{-a.s.} 
 \end{equation}
Here $\bbP_\eps:=\bbP\otimes\mu_\eps$. By $\bbE_\eps$
we denote the expectation with respect to $\bbP_\eps$.
From assumptions \eqref{mu-eps} and \eqref{031709} we obtain.
\begin{prop}
\label{prop012409}
Under the kinetic scaling we have
\begin{equation}
\label{052709-18}
{\cal E}_*(t):=\sup_{\eps\in(0,1]}\frac{\eps}{2}\bbE_\eps\|\hat\psi^{(\eps)}(t)\|^2_{L^2(\bbT)}\le
{\cal E}_*+\ga_1 T t,\quad t\ge0.
\end{equation}
\end{prop}
 We can introduce the (averaged) Wigner distribution  $
W_\eps(t)\in {\cal A}'$, corresponding to
 $\psi^{(\eps)}(t)$, by the formula
\begin{equation}
\label{wigner-bas}
\langle
W_\eps(t),G\rangle:=\frac{\eps}{2}\sum_{x,x'\in\bbZ}\bbE_\eps\left[\left(\psi^{(\eps)}_{x'}(t)\right)^\star
  \psi^{(\eps)}_{x}(t)\right]e_{x'-x}(k)G\left(\eps\frac{x+x'}{2},k\right),\quad
G\in {\cal A}.
\end{equation}
Thanks to \eqref{052709-18} it is well defined  for any $t\ge0$ and $\eps\in(0,1]$. 
Using the Fourier transform in the first variable we can rewrite the
Wigner distribution as
\begin{equation}
\label{wigner-bas1}
\langle
W_\eps(t),G\rangle=\int_{\bbT_\eps\times\bbT}\widehat
W_\eps^\star(t,\eta,k)\widehat G\left(\eta,k\right)d\eta dk,\quad
G\in {\cal A},
\end{equation}
where 
\begin{equation}
\label{W+}
\widehat
W_\eps(t,\eta,k):=\frac{\eps}{2}\bbE_\eps\left[\left(\hat\psi^{(\eps)}\right)^\star\left(t,k-\frac{\eps\eta}{2}\right) \hat\psi^{(\eps)}\left(t,k+\frac{\eps\eta}{2}\right)\right].
\end{equation}
We shall refer to $\widehat
W_\eps(t)$ as the Fourier-Wigner function corresponding to the given
wave function. 
For the sake of future reference define also $Y_\eps(t)$, by its
Fourier transform 
\begin{equation}
\label{Y+}
\widehat
Y_\eps(t,\eta,k):=\frac{\eps}{2}\bbE_\eps\left[\hat\psi^{(\eps)}\left(t,-k+\frac{\eps\eta}{2}\right) \hat\psi^{(\eps)}\left(t,k+\frac{\eps\eta}{2}\right)\right].
\end{equation}

\subsection{Kinetic equation}

An important role in our analysis will be played by the function, see
Section 2 of \cite{kors},
\begin{equation}
  \label{eq:bessel0}
  J(t) = \int_{\bbT}\cos\left(\omega(k) t\right) dk.
\end{equation}
Its Laplace transform
\begin{equation}
  \label{eq:2}
\tilde J(\la):=\int_0^{\infty}e^{-\la t}J(t)dt= \int_{\bbT}
\frac{\lambda}{\lambda^2 + \omega^2(k)} dk,\quad \la\in \mathbb C_+:=[z:\,{\rm Re}\,z>0].
\end{equation} 
One can easily see that  ${\rm Re}\,\tilde J(\la)>0$ for $\la\in
\mathbb C_+$, therefore 
we can define the function
\begin{equation}
\label{tg}
\tilde g(\lambda) := ( 1 + \gamma_1 \tilde J(\lambda))^{-1},\quad \la\in \mathbb C_+.
\end{equation} 
We have
\begin{equation}
\label{012410}
|\tilde g(\lambda)|\le 1,\quad \la\in \mathbb C_+.
\end{equation} 
The function $\tilde g(\cdot)$ is analytic on $ \mathbb C_+$ so, 
by the Fatou theorem, see e.g. p. 107 of \cite{koosis},
we know that
\begin{equation}
\label{nu}
\nu(k) :=\lim_{\eps\to0+}\tilde g(\eps-i\om(k))
\end{equation}
exists a.e. in $\bbT$ and in any $L^p(\bbT)$ for
$p\in[1,\infty)$. 

Let us introduce
\begin{equation}
\label{033110}
\wp(k):=\frac{\ga_1 \nu(k)}{2|\bar\om'(k)|},\quad \fgeeszett(k):=\frac{\ga_1|\nu(k)|^2}{|\bar\om'(k)|},\quad p_+(k):=\left|1-\wp(k)\right|^2 ,\quad 
p_-(k):=|\wp(k)|^2  ,
\end{equation}
where $
\bar\om'(k):=\om'(k)/(2\pi)$. 
We have shown  in 
\cite{kors} that
\begin{equation}
\label{feb1402}
{\rm Re}\,\nu(k)=\left(1+\frac{\ga_1}{2|\bar\om'(k)|}\right)|\nu(k)|^2.
\end{equation}
The functions $p_\pm(\cdot)$ and $\fgeeszett(\cdot)$ are even. Thanks to \eqref{feb1402} we have
\begin{equation}\label{012304}
p_+(k)+p_-(k)+\fgeeszett(k)=1.
\end{equation}

\subsubsection{Linear kinetic equation with an
interface}

Let $L$ be the operator  given by
\begin{equation}
\label{L}
LF(k):=  2\int_{\bbT}R(k,k')
\left[F\left(k'\right) - F\left(k\right)\right]dk',\quad k\in\bbT,
\end{equation}
{for $F\in L^1(\bbT)$} and
\begin{align}
\label{R}
&
R(k,k'):=\frac12\left\{r^2\left(k,k-
    k'\right)+r^2\left(k,k+k'\right)\right\}\\
&
=8{\frak s}^2(k){\frak s}^2(k')\left\{{\frak s}^2(k){\frak c}^2(k')+{\frak s}^2(k'){\frak c}^2(k)\right\},\quad k,k'\in\bbT.\nonumber
\end{align}
Note that (cf \eqref{beta}) the total scattering kernel equals
\begin{equation}
\label{Rk2.34}
 R(k):=\int_{\bbT}R(k,k')dk'=\frac{\hat\theta(k)}{4}.
\end{equation}
\begin{df}
\label{df012603-19}
Given $T\in\bbR$, let ${\cal C}_T$ be a
subclass of $C_b(\bbR_*\times\bbT_*)$ that consists of continuous functions~$F$ that can be
continuously extended to $\bar\bbR_\pm\times\bbT_*$ and satisfy the
interface conditions
\begin{equation}\label{feb1408}
F(0^+,k)=p_-(k)F(0^+, -k)+p_+(k)F(0^-,k)+{\frakg}(k)T, \quad\hbox{ for $0< k\le 1/2$},\\
\end{equation}
and
\begin{equation}\label{feb1410}
F(0^-, k)=p_-(k)F(0^-,-k) + p_+(k)F(0^+, k) +{\frakg}(k)T,\quad \hbox{ for $-1/2< k< 0$.}
\end{equation}
\end{df}
Note that  $F\in{\cal C}_T$ if and only if $F-T'\in{\cal C}_{T-T'}$
for any $T,T'\in\bbR$, because of (\ref{012304}). 

Let us fix $T\ge0$. We consider 
 the kinetic interface problem
given by equation
\begin{equation}
  \label{eq:8}
\begin{aligned} 
 &\partial_tW(t,y,k) + \bar\om'(k) \partial_y W(t,y,k) = \ga_0 L_k W(t,y,k), 
  \quad (t,y,k)\in\bbR_+\times \bbR_*\times\bbT_*,
 \\
&
W(0,y,k)=W_0(y,k),
\end{aligned}
\end{equation}
with the interface condition
\begin{equation}\label{feb1408aa}
W(t)\in {\cal C}_T,\quad t\ge0.
 \end{equation}
Here $ L_k $ denotes the operator $L$ acting on the $k$ variable. We
shall omit writing the subscript if there is no danger of confusion.

\subsubsection{Simplified case. Explicit solution}

\label{sec2.6.2}

We consider first the situation when equation \eqref{eq:8}
is replaced by 
\begin{equation}
  \label{eq:8p}
\begin{aligned} 
 &\partial_tW^{\rm un}(t,y,k) + \bar\om'(k) \partial_y W^{\rm un}(t,y,k) = -2\ga_0R(k)  W^{\rm un}(t,y,k), 
  \quad (t,y,k)\in\bbR_+\times \bbR_*\times\bbT_*,
 \\
&
W^{\rm un}(0,y,k)=W_0(y,k),
\end{aligned}
\end{equation}
 with the interface conditions
\eqref{feb1408} -- \eqref{feb1410}, with $T=0$. It can be  solved explicitly, using the
method of characteristics, and we obtain
\begin{align}
\label{010304}
&W^{\rm un}(t,y,k)
=  
e^{-2\ga_0R(k)t}\left\{\vphantom{\int_0^1}W_0\left(y-\bar{\om}'(k)t,k\right)
  1_{[0,\bar{\om}'(k)t]^c}(y)
  +p_+(k)W_0\left(y-\bar{\om}'(k)t,k\right)1_{[0,\bar{\om}'(k)t]}(y)
\right. \nonumber
\\
&
+\left. \vphantom{\int_0^1}p_-(k) W_0\left(-y+\bar\om'(k) t,-k\right)1_{[0,\bar \om'(k)t]}(y)\right\}.
\end{align}
Consider a semigroup of bounded operators on
$L^\infty(\bbR\times \bbT_*)$ defined by
\begin{equation}
\label{010304s}
{\frak W}^{\rm un}_t(W_0)\left(y,k\right)
:=  
W^{\rm un}(t,y,k),
\end{equation}
with $W_0\in L^\infty(\bbR\times
\bbT_*)$, $t\ge0$  and $(y,k)\in \bbR_*\times \bbT_*$. 
From formula \eqref{010304} one can conclude that $\left({\frak
    W}^{\rm un}_t\right)_{t\ge0}$ forms a semigroup of contractions
on both $L^1(\bbR\times\bbT)$ and $L^\infty(\bbR\times\bbT)$. Thus, by
interpolation, formula \eqref{010304}  defines a semigorup of
contractions on any $L^p(\bbR\times\bbT)$, $1\le p\le+\infty$.

Note that if $W_0$ is continuous in $\bbR_*\times \bbT_*$, then
$ W^{\rm un}(t,y,k)$ satisfies the interface conditions~\eqref{feb1408} and \eqref{feb1410}, with $T=0$ for all $t>0$.
  Therefore, $( {\frak W}^{\rm un}_t)_{t\ge0}$
is a semigroup on ${\cal C}_0$ with $W^{\rm un}(t,y,k)$ (cf \eqref{010304s})
satisfying the first equation of \eqref{eq:8p},
 the interface condition \eqref{feb1408} -
\eqref{feb1410}  and the initial condition
$$
\lim_{t\to0+} W^{\rm un}(t,y,k)=W_0(y,k),\quad (y,k)\in\bbR_*\times\bbT_*.
$$

\subsubsection{Kinetic equation  - classical solution} 

\label{sec2.6.3}

\begin{df}
\label{df013001-19}
We say that a function
$W:\bar\bbR_+\times \bbR\times \bbT_*\to \bbR$
is a classical
solution to equation \eqref{eq:8} with the interface conditions
\eqref{feb1408}, \eqref{feb1410} at $y=0$, 
 if it is bounded and continuous on
$\bbR_+\times\bbR_*\times \bbT_*$, and the
following conditions hold:
\begin{itemize}
\item[(1)] 
the restrictions of $W$ to  
$\bbR_+\times\bbR_\iota\times \bbT_{\iota'}$, $\iota,\iota'\in\{-,+\}$, can be extended
to bounded and continuous functions on the respective closures
$\bar\bbR_+\times\bar\bbR_{\iota}\times \bar\bbT_{\iota'}$,
\item[(2)] for each $(t,y,k)\in \bbR_+\times\bbR_*\times  \bbT_*$ fixed, the function
$W(t+s,y+\bar\om'(k) s,k)$ is of the~$C^1$ class in
the $s$-variable in a  neighborhood of $s=0$, and the directional derivative  
\begin{equation}
\label{Dt}
D_tW(t,y,k)=\left(\partial_t+\bar\om'(k)\partial_y\right)W(t,y,k):=\frac{d}{ds}_{|s=0} W(t+s,y+\bar\om'(k) s,k)
\end{equation}
is bounded in $\bbR_+\times\bbR_*\times \bbT_*$
and satisfies
\begin{equation}
  \label{eq:8a}
 \begin{array}{ll}
 D_tW(t,y,k) = \ga_0 L_k W(t,y,k), &
\quad (t,y,k)\in\bbR_+\times \bbR_*\times\bbT_*,
\end{array}
\end{equation}
\item[3)] $W(t)$ satisfies  \eqref{feb1408},  \eqref{feb1410}  and
\begin{equation}
\label{010102-19}
\lim_{t\to0+}W(t,y,k)=W_0(y,k),\quad (y,k)\in\bbR_*\times\bbT_*.
\end{equation}
\end{itemize}
\end{df}

The following result has been shown in \cite{koran}, see Proposition 2.2.
\begin{prop}
\label{prop013001-19}
Suppose that $W_0\in{\cal C}_T$. Then, under the above
hypotheses on the scattering kernel $R(k,k')$ and the
dispersion relation $\om(k)$, there exists a
unique classical
solution to  equation \eqref{eq:8} with the interface conditions
\eqref{feb1408} and \eqref{feb1410} in the sense of Definition $\ref{df013001-19}$.   
\end{prop}

\subsubsection{$L^2$ solution}

\label{sec2.6.4}

We assume that $T=0$.
Define ${\frak W}_t(W_0):=W(t)$. Thanks to Proposition
\ref{prop013001-19} the family $\left({\frak W}_t\right)_{t\ge0}$
forms a semigroup on the linear space ${\cal C}_0$. 
Furthermore, we let
\begin{equation}
\label{cR}
{\cal R}F(k):=  \int_{\bbT}R(k,k')
F\left(k'\right) dk',\quad k\in\bbT,\quad F\in L^1(\bbT).
\end{equation} 
Let 
$
 {\cal C}_0':=  {\cal C}_0\cap L^2(\bbR\times \bbT).
$
The following result holds.
\begin{prop}
\label{prop011406-19}
We have ${\frak W}_t\left( {\cal C}_0'\right)\subset  {\cal C}_0'$ for
all $t\ge0$. The semigroup $\left({\frak W}_t\right)_{t\ge0}$  extends
by the $L^2$ closure from ${\cal C}_0'$ to a $C_0$-continuous semigroup of
contractions on $L^2(\bbR\times\bbT)$. Moreover, it is the unique
solution in $L^2(\bbR\times\bbT)$ of the integral equation
\begin{equation}
\label{integral}
{\frak W}_t={\frak W}_t^{\rm un}+2\ga_0\int_0^t {\frak W}_{t-s}^{\rm
  un}{\cal R} {\frak W}_sds,\quad t\ge0.
\end{equation}
\end{prop}
The proof of this result is contained in Appendix \ref{appa}. We shall
refer to the semigroup solution
described in Proposition \ref{prop011406-19} as the $L^2$-solution of quation \eqref{eq:8} with the interface conditions
\eqref{feb1408} and \eqref{feb1410} for $T=0$. To extend the definition of such
a solution to the case of an arbitrary $T\ge0$ we proceed as
follows. Suppose that $W_0\in L^2(\bbR\times\bbT)$. Let $\chi\in C^\infty_c(\bbR)$ be an arbitrary real valued, even function that
satisfies 
\begin{equation}
\label{011406-19}
\chi(y)=\left\{\begin{array}{ll}
1,&\mbox{ for }|y|\le 1/2, \\
0,&\mbox{ for }|y|\ge 1,\\
\mbox{belongs to }[0,1],& \mbox{ if otherwise.}
\end{array}\right.
\end{equation}
\begin{df}
\label{df011406-19}
We say that $W(t,y,k)$ is the $L^2$-solution of quation \eqref{eq:8} with the interface conditions
\eqref{feb1408} and \eqref{feb1410} for a given $T\ge0$ and an initial
condition $W_0\in L^2(\bbR\times\bbT)$, if it is of
the form
\begin{equation}
\label{021406-19}
W(t,y,k):={\frak W}_t(\widetilde W_0)(y,k)+\int_0^t {\frak
  W}_s(F)(y,k)ds+T\chi(y),\quad (t,y,k)\in \bar\bbR_+\times\bbR\times\bbT.
\end{equation}
Here
\begin{equation}
\label{eqF}
F(y,k):=-T\bar\om'(k)\chi'(y),\quad \widetilde W_0(y,k):= W_0(y,k)-T\chi(y).
\end{equation}
\end{df}

\begin{remark}
\label{rm011406-19} {\em Note that the definition of the solution
 does not depend on the choice of function $\chi$
satisfying \eqref{011406-19}. 
 {Indeed, suppose that
$W_j$, $j=1,2$ correspond to two choices of  $\chi_j(y)$,
$j=1,2$, with the respective  $\widetilde
W_0^j$, $j=1,2$. Let $\delta W:=W_2-W_1$. Hence, $\delta W(0)=0$ and
$\delta\widetilde W_0:=\widetilde W_0^2- \widetilde W_0^1=
T(\chi_1-\chi_2)
$  belongs to ${\cal C}_0$.
Moreover, $\frac{d}{dt}{\frak W}_t(\delta\widetilde W_0)=-{\frak
  W}_t(\delta F)$, where $\delta F:=
T(\chi_1'-\chi_2')\bar\om'$. Therefore, using \eqref{021406-19}, we
conclude that $\frac{d}{dt}\delta W(t)\equiv0$, which in turn
implies that $W_j$, $j=1,2$ coincide.}}
\end{remark}


\begin{remark}
\label{rm021406-19} {\em Suppose that $W_0\in {\cal C}_T$. Then, $W(t,y,k)$
given by\eqref{021406-19} is the classical solution of  \eqref{eq:8} with the interface conditions
\eqref{feb1408} and \eqref{feb1410}, in the sense of Definition
$\ref{df013001-19}$.}
\end{remark}

\medskip

\subsection{Asymptotics of the Wigner functions - the statement of the main result}

Thanks to \eqref{052709-18}
we conclude that
\begin{equation}
\label{030406-19}
\sup_{\eps\in(0,1]}\|  W_{\eps}\|_{L^\infty([0,\tau];{\cal
    A}')}<+\infty,\quad \mbox{for any }\tau>0.
\end{equation}
Therefore $\left(W_{\eps}(\cdot)\right)$ is sequentially
$\star$-weakly compact in  
$L^\infty_{\rm loc}([0,+\infty),{\cal A}')$, i.e. from any sequence
$\eps_n\to0$ we can choose a subsequence, that we still denote by the
same symbol, for which $\left( W_{\eps_n}(\cdot)\right)$
$\star$-weakly converges in $\left(L^1([0,t],{\cal
    A})\right)'$ for any $t>0$. In our main result we identify the limit as the
$L^2$ solution of the kinetic equation \eqref{eq:8} with the interface conditions
\eqref{feb1408} and \eqref{feb1410}, in the sense of Definition
$\ref{df011406-19}$.
\begin{thm}
\label{main-thm}
Suppose that 
there exist $C,\kappa>0$ such that 
\begin{equation}
\label{011812aa}
|\widehat W_\eps(0,\eta,k)|+|\widehat Y_\eps(0,\eta,k)|\le 
C\varphi(\eta),\quad (\eta,k)\in\bbT_{\eps}\times \bbT, \,\eps\in(0,1],
\end{equation} 
where
\begin{equation}
\label{011812c}
\varphi(\eta):=\frac{1}{(1+\eta^2)^{3/2+\kappa}},
\end{equation} 
and 
$$
W_{\eps}(0)\mathop{\stackrel{{\rm \tiny w}^\star}{\longrightarrow}}\limits_{\tiny{\eps\to0+}} W_0\quad\mbox{ in
}{\cal A}'.
$$ Then, $W_0\in L^2(\bbR\times\bbT)$ and for any $G\in
L^1_{\rm loc}\left([0,+\infty);{\cal A}\right)$ we have
\begin{equation}
\label{061406-19}
\lim_{\eps\to0+}\int_0^{\tau}\langle W_{\eps}(t), G(t)\rangle dt=\int_0^{\tau}\langle W(t), G(t)\rangle dt,\quad\tau>0.
\end{equation}
Here $W(t)$ is the $L^2$ solution of  the kinetic equation \eqref{eq:8} with the interface conditions
\eqref{feb1408} and \eqref{feb1410} satisfying $W(0)=W_0$.
\end{thm}
In the case $T=0$ the theorem is a direct consequence of Theorem \ref{cor020805-19}
proved below. The more general case $T\ge0$ is treated in Section \ref{sec10}.

\section{Deterministic wave equation corresponding to (\ref{basic:sde:2a})}

\label{sec3}

Here we consider a deterministic part of  the dynamics described in
(\ref{basic:sde:2a}). Its corresponding energy density function will
converge to the solution of \eqref{eq:8p}.
In microscopic time the evolution of the wave function is given by
\begin{eqnarray}
 \label{012503-19}
&&
\frac{d}{dt}\hat\phi(t,k)=-i\om(k)\hat\phi(t,k)
 -2i\eps\ga_0R(k)\hat p\left(t,k\right)-i\ga_1\int_{\bbT}\hat{ p}\left(t,k'\right)dk',
 \\
 &&
\hat\phi(0,k)= \hat\psi(k),\nonumber
 \end{eqnarray} 
where
$$
\hat p(t,k):=\frac{1}{2i}\left[\hat\phi(t,k)-\left(\hat\phi(t,-k)\right)^\star\right].
$$
 In
fact it is convenient to deal with the vector formulation of the
equation for
\begin{equation}
\label{Ut}
\hat\Phi(t,k)=\left[
\begin{array}{c}
\hat\phi_+(t,k)\\
\\
\hat\phi_-(t,k),
\end{array}\right],\quad \hat\Psi(k)=\left[
\begin{array}{c}
\hat\psi_+(k)\\
\\
\hat\psi_-(k)
\end{array}\right].
\end{equation}
Here, we use the convention  $\hat\phi_+(t,k)=\hat\phi(t,k)$ and
$\hat\phi_-(t,k):=\left(\hat\phi(t,-k)\right)^\star$
and similarly for $\hat\psi_\pm(k)$.
The equation then takes the form
\begin{eqnarray}
 \label{basic:sde:2av}
&&
 \frac{d}{dt}\hat\Phi(t,k)=\Om_\eps(k) \hat\Phi(t,k) - i\gamma_1  {\frak f}{p}_0(t) 
 ,\nonumber\\
 &&
\hat\Phi(0,k)= \hat\Psi(k).
 \end{eqnarray} 
Here
\begin{equation}
\label{Omk}
\Om_\eps(k):=\left[
\begin{array}{cc}
-\ga_0\eps R(k)-i\om(k)&\ga_0\eps R(k)\\
\ga_0\eps R(k)&-\ga_0\eps R(k)+i\om(k)
\end{array}\right]=\Om_0(k)-\ga_0\eps R(k){\bf D}.
\end{equation}
and 
$$
{\frak f}:=\left[
\begin{array}{r}
1\\
-1
\end{array}\right],\quad {\bf D}:={\frak f}^T\otimes {\frak f}=\left[
\begin{array}{rr}
1&-1\\
-1&1
\end{array}\right].
$$
The momentum at $x=0$ equals
\begin{equation}
\label{p0}
{p}_0(t)
:=\frac{1}{2i}\int_{\bbT}\hat\Phi(t,k)\cdot {\frak f}dk=\frac{1}{2i}\int_{\bbT}\left[\hat\phi (t,k)-\left(\hat\phi (t,-k)\right)^\star\right]dk.
\end{equation}
The eigenvalues of the matrix $\Om_\eps(k)$ equal
$
\la_\pm(k)=-\ga_0\eps R(k)\pm i\om_\eps(k),
$
where
\begin{equation}
\label{030904-19}
\beta(k)=\frac{\ga_0 R(k)}{\om(k)},\qquad\om_\eps(k):=\om(k) \sqrt{1-(\eps\beta(k))^2}.
\end{equation}
Note that
$
\la_+^\star=\la_-.
$

\subsection*{Solution of (\ref{basic:sde:2av})}

By the Duhamel formula, from \eqref{basic:sde:2av} we get
 \begin{equation}
 \label{022503-19}
 \hat\Phi(t,k)=e_{\Om_\eps}(k,t) \hat\Psi(k) - i\gamma_1  \int_0^t e_{\Om_\eps}(k,t-s){\frak f}{p}_0(s)ds .
 \end{equation} 
Here
\begin{align}
\label{011902-19ac}
&e_{\Om_\eps}(k,t):=\exp\left\{\Om_\eps(k)t\right\}=\left[
\begin{array}{cc}
e_{\Om_\eps}^{1,1}(k,t)&e_{\Om_\eps}^{1,2}(k,t)\\
e_{\Om_\eps}^{1,2}(k,t)&[e_{\Om_\eps}^{1,1}]^\star(k,t)
\end{array}\right]
\end{align}
and
\begin{align}
\label{011902-19a}
&
e_{\Om_\eps}^{1,1}(k,t):=\dfrac14\left\{\left(1+\sqrt{1-(\eps\beta(k))^2}\right)^2e_-(k,t)-(\eps\beta(k))^2e_+(k,t)\right\}, \nonumber\\
&e_{\Om_\eps}^{1,2}(k,t):=\dfrac{i\eps\beta(k)}{4} \left(1+\sqrt{1-(\eps\beta(k))^2}\right)(e_-(k,t)-e_+(k,t))
,\\
&
e_\pm(k,t):=e^{\la_\pm(k)
  t}.\nonumber
\end{align}
Note that $e_\pm^\star(k,t)=e_\mp(k,t)$.

Multiplying scalarly both sides of \eqref{022503-19} by ${\frak f}$
and integrating over $k$ we conclude that
\begin{equation}
 \label{042503-19}
p_0(t)+\ga_1J_\eps\star p_0(t)=p_0^0(t)
 \end{equation} 
 Here
\begin{equation}
\label{null3.11}
p_0^0(t):=\frac{1}{2i}\int_{\bbT}e_{\Om_\eps}(k,t)\hat\Psi(k) \cdot {\frak f}dk
\end{equation}
and
\begin{equation}
\label{J-eps}
{J}_\eps(t):=\frac12\int_{\bbT} \exp\left\{\Om_\eps(k)t\right\}{\frak
  f}\cdot {\frak f} dk=\int_{\bbT}j_\eps(t,k)dk,
\end{equation}
where
\begin{equation}
\label{j}
j_\eps(t,k):=\frac12\exp\left\{\Om_\eps(k)t\right\}{\frak f}\cdot
   {\frak f}
=e^{-\eps\ga_0 R(k) t}\left\{\sqrt{1-(\eps\beta(k))^2}\cos\left(\om_\eps(k)t\right)-\eps\beta(k)\, \sin\left(\om_\eps(k)t\right)\right\}.
\end{equation}
Taking the Laplace transforms of the both sides of \eqref{042503-19}
we obtain
\begin{equation}
 \label{062503-19}
\tilde p_0(\la)(1+\ga_1\tilde J_\eps(\la))=\tilde p_0^0(\la).
 \end{equation} 
By a direct calculation one concludes that
\begin{equation}
\label{052503-19}
{\rm Re}\, \tilde {J}_\eps(\la)>0,\quad\mbox{for any }\la\in \mathbb
C_+.
\end{equation}
Since $J_\eps(\cdot)$ is real valued, we have 
$$
\tilde {J}_\eps^\star(\la)=\tilde {J}_\eps(\la^\star)\quad\mbox{for any }\la\in \mathbb
C_+.
$$
From \eqref{062503-19} we get
\begin{equation}
 \label{072503-19}
\tilde p_0(\la)=\tilde g_\eps(\la)\tilde p_0^0(\la),
 \end{equation} 
with $\tilde g_\eps(\la)$ defined by
\begin{equation}
\label{012302-19}
\tilde {g}_\eps(\la):=\left(1+\ga_1\tilde {J}_\eps(\la)\right)^{-1},\quad {\rm Re}\,\la>0.
\end{equation}
Thanks to \eqref{052503-19} we obtain
\begin{equation}
\label{G}
|\tilde {g}_\eps(\la)|\le 1,\quad {\rm Re}\,\la>0
\end{equation}
and, as a result,
\begin{equation}
\label{JG}
\ga_1|\tilde {J}_\eps(\la)\tilde {g}_\eps(\la)|\le 2,\quad {\rm Re}\,\la>0.
\end{equation}
The following result shows in particular that $\tilde
{g}_\eps(\eps-i\om(k))$ approximates in some sense
$\nu(k)$, as $\eps\to0+$ (see \eqref{nu}).
\begin{prop}
\label{cor010304-19}
Suppose that $K:\bbT\to\bbR_+$ is a uniformly continuous and bounded function  satisfying
\begin{equation}
\label{lim-is3.20}
\inf_{k\in\bbT}K(k)>0.
\end{equation}
Then,
\begin{equation}
\label{010304-19}
\tilde g_\eps(\la)=\tilde g(\la)+\eps\tilde r_\eps(\la),\quad
\la\in\mathbb C_+,
\end{equation}
where
$$
\lim_{\eps\to0+}\eps^p\int_{\bbT}|\tilde r_\eps(\eps K(k)-i\om(k))|^pdk=0,
$$
for any $p\in[1,+\infty)$.
\end{prop}
The proof of Proposition \ref{cor010304-19} is shown in Appendix \ref{appC}.

Define by $g_\eps(ds)$ the distribution such that
\begin{equation}
\label{g-eps}
\tilde {g}_\eps(\la)=\int_0^{+\infty}e^{-\la t}g_\eps(ds).
\end{equation}
From \eqref{012302-19}
it satisfies
\begin{equation}
\label{g-eps1}
g_\eps(ds)=\delta(ds)-\ga_1 J_\eps\star g_\eps(s)ds.
\end{equation}
The Volterra equation \eqref{g-eps1} has a unique real-valued
solution and $\ga_1 J_\eps\star g_\eps(s)$ is a $C^\infty$ smooth function, see e.g. the argument made in Section 3
of \cite{kors}.
The solution of \eqref{022503-19} can be then written as follows
\begin{align}
\label{eq:10}
&
\hat\Phi(t,k) =U(t) \hat\Psi(k):=e_{\Om_\eps}(k,t)\hat\Psi(k) - i\gamma_1  \int_0^t
 e_{\Om_\eps}(k,t-s){\frak f}p_0^0\star g_\eps(s) ds \\
&
=e_{\Om_\eps}(k,t) \hat\Psi(k) - \frac{\gamma_1}{2}  \int_0^tds\int_0^sg_\eps(ds_1)\int_{\bbT}
  e_{\Om_\eps}(k,t-s){\bf D}e_{\Om_\eps}(k,s-s_1) \hat\Psi(\ell) d\ell .\nonumber
 \end{align}

\section{Dynamics of the energy density when $T=0$}

\label{sec4}

Starting with the present section untill Section \ref{sec10} we shall assume that the thermostat temperature
$T=0$, see \eqref{basic:sde:2a}. We maintain this assumption untill
Section \ref{sec10}.
Let
\begin{equation}
\label{Psit}
\hat\Psi^{(\eps)}(t,k)=\left[
\begin{array}{c}
\hat\psi^{(\eps)}_+(t,k)\\
\hat\psi^{(\eps)}_-(t,k)
\end{array}\right],
\end{equation}
where 
$\hat\psi^{(\eps)}_+(t,k):=\hat \psi^{(\eps)}(t,k)$ and
$\hat\psi^{(\eps)}_-(t,k):=\left(\hat
  \psi^{(\eps)}\right)^\star(t,-k)$ (cf \eqref{011307a}).
From \eqref{basic:sde:2a} we get
\begin{eqnarray}
 \label{basic:sde:2av1}
&&
 d \hat\Psi^{(\eps)}(t,k)=\frac{1}{\eps}\Om_\eps(k) \hat\Psi^{(\eps)}(t,k) dt
 +i\sqrt{\ga_0}\int_{\bbT}r(k,k'){\bf
   D} \hat\Psi^{(\eps)}\left(t,k-k'\right)B(dt,dk')-\frac{i\ga _1}{\eps}{\frak p}_0(t){\frak
   g}dt,\nonumber\\
 &&
\Psi^{(\eps)}(0,k)= \hat\Psi(k).
 \end{eqnarray}

With some abuse of notation we denote by ${\cal A}$ the Banach space of all
matrix valued functions obtained by the completion of functions of the form
\begin{align}
\label{bFS}
{\bf F}(y,k)=\left[\begin{array}{ll}
G(y,k)&H(y,k)\\
H^\star(y,k)&G(y,-k)
\end{array}\right],\quad (y,k)\in\bbR\times \bbT,
\end{align}
with $C^\infty$ smooth entries satisfying $G$ is real valued and $H$ is
even in $k$. The completion is taken in the norm given by the maximum
of the ${\cal A}$ norms of the entries, see \eqref{060805-19}.

The Wigner distribution, corresponding to the wave function
$\psi^{(\eps)}(t)$, is a $2\times 2$-matrix tensor ${\bf W}_\eps(t)$, whose
entries are distributions,  given by their respective  Fourier transforms
\begin{align}
\label{hbw}
&
\widehat \bW_\eps(t,\eta,k):=\frac{\eps}{2}\bbE\left[\hat\Psi^{(\eps)}\left(t,k+\frac{\eps\eta}{2}\right)\otimes \left(\hat\Psi^{(\eps)}\right)^\star\left(t,k-\frac{\eps\eta}{2}\right)\right]
\\
&
=\left[\begin{array}{ll}
\widehat W_{\eps,+}(t,\eta,k)&\widehat Y_{\eps,+}(t,\eta,k)\\
\widehat Y_{\eps,-}(t,\eta,k)&\widehat W_{\eps,-}(t,\eta,k)
\end{array}\right], \quad (\eta,k)\in\bbT_{\eps}\times \bbT,
\end{align}
with
\begin{align*}
&
\widehat W_{\eps,+}(t,\eta,k):=\widehat W_{\eps}(t,\eta,k)=\frac{\eps}{2}\bbE_\eps\left[\hat\psi^{(\eps)}\left(t,k+\frac{\eps\eta}{2}\right) \left(\hat\psi^{(\eps)}\right)^\star\left(t,k-\frac{\eps\eta}{2}\right)\right],
\\
&
\widehat Y_{\eps,+}(t,\eta,k):=\frac{\eps}{2}\bbE_\eps\left[\hat\psi^{(\eps)}\left(t,k+\frac{\eps\eta}{2}\right) \hat\psi^{(\eps)}\left(t,-k+\frac{\eps\eta}{2}\right)\right],
\\
&
\widehat Y_{\eps,-}(t,\eta,k):=\widehat
  Y_{\eps,+}^\star(t,-\eta,k),\quad 
\widehat W_{\eps,-}(t,\eta,k):=\widehat W_{\eps,+}(t,\eta,-k).
\end{align*}
Then ${\bf W}_\eps(t)$ belongs to $ {\cal A}'$ - the dual to ${\cal A}$ that
is made of all
distributions ${\bf W}$, whose Fourier transform in the first variable equals  
\begin{align}
\label{bW}
\widehat{\bf W}(\eta,k)=\left[\begin{array}{ll}
\widehat W_{+}(\eta,k)&\widehat Y_{+}(\eta,k)\\
\widehat Y_{-}(\eta,k)&\widehat W_{-}(\eta,k)
\end{array}\right],\quad (\eta,k)\in\bbT_{\eps}\times \bbT,
\end{align}
whose entries  belong to ${\cal A}'$ and satisfy
\begin{align*}
&\widehat W_{+}^\star(\eta,k)=\widehat W_{+}(-\eta,k),\quad \widehat
  Y_{+}(\eta,k)=\widehat Y_{+}(\eta,-k),\\
&
\widehat W_{-}(\eta,k)=\widehat W_{+}(\eta,-k),\quad \widehat Y_{-}(\eta,k)=\widehat
  Y_{+}^\star(-\eta,k).
\end{align*}
The duality pairing
between ${\cal A}'$ and ${\cal A}$ is determined by the relation
\begin{align}
\label{pairing}
&
\left\langle {\bf F},{\bf
    W}\right\rangle:=\int_{\bbT_\eps\times\bbT}\widehat {\bf
  F}(\eta,k)\cdot  {\bf
    W}(\eta,k)d\eta dk\\
&
=2\int_{\bbT_\eps\times\bbT}\left\{\widehat F_{+}(\eta,k)\widehat
  W_{+}^\star(\eta,k)+{\rm Re}\left(\widehat H_{+}(\eta,k)\widehat
  Y_{+}^\star(\eta,k)\right)\right\},\nonumber
\end{align}
wher  the scalar product of two matrices is given by
\begin{equation}
\label{scalar}
\widehat {\bf
  F}\cdot \widehat  {\bf
    W}=\sum_{\iota=\pm}\left(\widehat {
  F}_\iota \widehat {
  W}^\star_\iota+\widehat {
  H}_\iota \widehat {
  Y}^\star_\iota\right).
\end{equation}
The norm $\|{\bf
  W}\|_{{\cal A}'}$ is therefore the sum of the norm of its entries.

Thanks to \eqref{mu-eps} and \eqref{052709-18}
we conclude that
\begin{equation}
\label{030406-194.9}
\sup_{\eps\in(0,1]}\sup_{t\ge0}\|{\bf  W}_{\eps}(t)\|_{{\cal A}'}=:A_*'<+\infty.
\end{equation}
Therefore $\left({\bf  W}_{\eps}(\cdot)\right)$ is bounded in
$L^\infty([0,+\infty),{\cal A}')$. In consequence, from any sequence
$\eps_n\to0$ we can choose a subsequence, that we still denote by the
same symbol, such that $\left({\bf  W}_{\eps_n}(\cdot)\right)$
is $\vphantom{1}^\star$-weakly convergent in $\left(L^1([0,+\infty),{\cal A})\right)'$.

In what follows we shall also consider the Hilbert spaces ${\cal
  L}_{2,\eps}$ with the scalar product
$\langle\cdot,\cdot\rangle_{{\cal L}_{2,\eps}}$ given by the
formula \eqref{pairing}. The  respective Hilbert space norms are
$$
\|{\bf W}\|_{{\cal L}_{2,\eps}}:=\left\{2\left(\| W_{+}\|_{{\cal
        L}_{2,\eps}}^2+\| Y_{+}\|_{{\cal
        L}_{2,\eps}}^2\right)\right\}^{1/2}. 
$$
We introduce the following notation,
given a function $f:\bbT\to\mathbb C$, we let
\begin{equation}
\label{bar-f}
\bar f(k,\eta):=\frac{1}{2}\left[f\left(k+\frac{\eta}{2}\right)+f\left(k-\frac{ \eta}{2}\right)\right]
\end{equation}
and the difference quotient for the dispersion relation
\begin{equation}
\label{d-om}
\delta_{\eps}\om(k,\eta):=\frac{1}{\eps}\left[\om\left(k+\frac{\eps \eta}{2}\right)-\om\left(k-\frac{\eps \eta}{2}\right)\right].
\end{equation}
Equipped with this notation we introduce
\begin{align}
\label{bH}
\widehat{\bf H}_\eps(\eta,k)=\left[\begin{array}{cc}
-i\delta_{\eps}\om(k;\eta)&-\dfrac{2i}{\eps} \bar\om(k,\eps \eta)\\
&\\
\dfrac{2i}{\eps} \bar\om(k,\eps \eta)&i\delta_{\eps}\om(k;\eta)
\end{array}\right],\quad (\eta,k)\in\bbT_{\eps}\times \bbT
\end{align}
 and
\begin{eqnarray}
\label{021605-19}
&&
{L}_\eta f(k):=2{\cal R}_{ \eta} f(k) - 2\bar R (k,\eta) f(k),\qquad
{L}_{\eta}^\pm f(k):=2{\cal R}_{\eta} f(k) -2R\left(k\pm\frac
   \eta2\right) f(k),\nonumber\\
&&
 {\cal R}_\eta f(k):=\int_{\bbT} R(k,k',\eta)f(k')dk',
\\
&&
R(k,k',\ell):=\frac12\sum_{\iota=\pm1}r\left(k-\frac{ \ell}{2},k-\iota
   k'\right)r\left(k+\frac{ \ell}{2},k-\iota k'\right),\quad
   k,k'\in\bbT,\,\ell\in 2\bbT. \nonumber
\end{eqnarray}
We denote by ${\frak
    L}_{\eps\eta}$, ${\frak H}_\eps$, ${\frak T}_\eps$ the operators,
  acting on ${\cal L}_{2,\eps}$, defined by 
\begin{equation}
\label{011906-19}
\widehat{{\frak
    L}_{\eps\eta}{\bf W}}=\widehat{\frak
    L}_{\eps\eta}\widehat{\bf W},\quad \widehat{{\frak
    H}_{\eps}{\bf W}}=\widehat{\frak
    H}_{\eps}\widehat{\bf W},\quad \widehat{{\frak
    T}_{\eps}{\bf W}}=\widehat{\frak
    T}_{\eps}\widehat{\bf W}.
\end{equation}
Here $\widehat{\bf W}$ is the Fourier transform of ${\bf W}\in {\cal L}_{2,\eps}$, given
by \eqref{bW}. Operator $\hat{\frak H}_\eps$, acting on
$L^2(\bbT_\eps\times\bbT)$, is given by 
\begin{align}
\label{fH}
\widehat{\frak H}_\eps \widehat \bW(\eta,k):=\widehat{\bf H}_\eps(\eta,k) \circ\widehat \bW(\eta,k)=\left[\begin{array}{cc}
-i\delta_{\eps}\om(k;\eta)\widehat W_{+}(\eta,k)&-\dfrac{2i}{\eps} \bar\om(k,\eps \eta)\widehat Y_{+}(\eta,k)\\
&\\
\dfrac{2i}{\eps} \bar\om(k,\eps \eta)\widehat Y_{-}(\eta,k)&i\delta_{\eps}\om(k;\eta)\widehat W_{-}(\eta,k)
\end{array}\right],
\end{align}
with $\circ$ denoting the Hadamard's product of $2\times 2$ matrices.
Moreover, $\widehat{\frak
    L}_{\eps\eta}$ and  $\widehat{\frak
    T}_{\eps}$ act on $L^2(\bbT_\eps\times\bbT)$ via the formulas
\begin{align}
\label{LHT}
\widehat{\frak
    L}_{\eps\eta}\widehat \bW(\eta,k):=\left[\begin{array}{cc}
\widehat W'_{+}(\eta,k)&\widehat Y'_{+}(\eta,k)\\
\widehat Y'_{-}(\eta,k)&\widehat W'_{-}(\eta,k)
\end{array}\right],\qquad \widehat{\frak
    T}_{\eps}\widehat \bW(\eta,k):=\left[\begin{array}{cc}
\widehat W''_{+}(\eta,k)&\widehat Y''_{+}(\eta,k)\\
\widehat Y''_{-}(\eta,k)&\widehat W''_{-}(\eta,k)
\end{array}\right],
\end{align}
with
\begin{eqnarray}
\label{011605-19z4.17}
&&\widehat W_{\pm}'(\eta,k)={L}_{\eps \eta}\widehat W_{\pm} (\eta,k)-\frac{1}{2}\sum_{\iota=\pm}{L}^{\pm}_{\iota\eps \eta}\widehat Y_{-\iota}(\eta,k),
\\
&&
\widehat Y_{\pm}'(\eta,k)={L}_{\eps \eta}\widehat Y_{\pm}(\eta,k)
+ {\cal R}_{\eps \eta}(\widehat
Y_{\mp}-\widehat Y_{\pm} )(\eta,k) 
-\frac{1}{2}\sum_{\iota=\pm} {L}^\pm_{\iota\eps \eta}
\widehat W_{-\iota}(\eta,k)
\nonumber 
\end{eqnarray}
and
\begin{eqnarray}
\label{011605-19z}
&&\widehat W_{\pm}''(\eta,k)=
\frac{1}{2\eps}\int_{\bbT}\left[\widehat Y_{\pm}\left(\eta-\frac{2k'}{\eps},k+k'\right)
+\widehat Y_{\mp}\left(\eta+\frac{2k'}{\eps},k+k'\right)
\right. \nonumber\\
&&
\left.-\widehat W_{\pm}\left(\eta-\frac{2k'}{\eps},k+k'\right)-\widehat W_{\pm}\left(\eta+\frac{2k'}{\eps},k+k'\right)\right]dk',
\\
&&
\widehat Y_{\pm}''(\eta,k)=-\frac{1}{2\eps}\int_{\bbT}\left[\widehat Y_{\pm}\left(\eta+\frac{2k'}{\eps},k+k'\right)
+\widehat Y_{\pm}\left(\eta-\frac{2k'}{\eps},k+k'\right)
\right.\nonumber \\
&&
\left.-\widehat W_{\mp}\left(\eta+\frac{2k'}{\eps},k+k'\right)-\widehat W_{\pm}\left(\eta-\frac{2k'}{\eps},k+k'\right)\right]dk'.
\nonumber 
\end{eqnarray}
Using \eqref{basic:sde:2av1} we obtain the following system of equations
for the evolution of the tensor $\widehat \bW_\eps(t,\eta,k)$:
\begin{equation}
\label{011806-19}
\frac{d}{dt}\widehat \bW_\eps(t,\eta,k)=\left(\ga_0\widehat{\frak
    L}_{\eps\eta}+\widehat{\frak H}_\eps+\ga_1\widehat{\frak
    T}_\eps\right)\widehat \bW_\eps(t,\eta,k).
\end{equation}
The respective semigroups on ${\cal
    L}_{2,\eps}$ and $L^2(\bbT_\eps\times\bbT)$  shall be denoted by
\begin{equation}
\label{frakW}
{\frak W}_\eps(t):=\exp\left\{\left(\ga_0{\frak
    L}_{\eps\eta}+{\frak H}_\eps+\ga_1{\frak T}_\eps\right)t\right\},\quad \widehat{\frak W}_\eps(t):=\exp\left\{\left(\ga_0\hat{\frak
    L}_{\eps\eta}+\hat{\frak H}_\eps+\ga_1\hat{\frak
    T}_\eps\right)t\right\},\quad  t\ge0.
\end{equation}

Let us introduce the Hilbert space norms
\begin{equation}
\label{H0eps}
\|{\bf W}\|_{{\cal
  H}_{0,\eps}}:=\left\{\int_{\bbT_{\eps}\times\bbT}R\left(k-\frac{\eps
   \eta}{2}\right)\left|
 \widehat  W_{+} (\eta,k)
-\widehat
  Y_{+} (\eta,k)\right|^2d\eta dk\right\}^{1/2}
\end{equation}
and
\begin{equation}
\label{H2eps}
\|{\bf W}\|_{{\cal
  H}_{1,\eps}}:=\left\{\int_{\bbT}dk\left|\int_{\bbT_{\eps}}\left[\widehat
  W_{+}\left(t,\eta, k-\frac{\eps\eta}{2}\right)-\widehat
  Y_{+}\left(t, \eta, k-\frac{\eps\eta}{2}\right)\right]d\eta\right|^2\right\}^{1/2}.
\end{equation}
By a direct calculation one can verify the following identity
\begin{align}
&\|{ \bW}_\eps(t)\|^2_{{\cal
  L}_{2,\eps}}+8\ga_0\int_0^t\|{ \bW}_\eps(s)\|^2_{{\cal
  H}_{0,\eps}}ds+4\ga_1\int_0^t\|{ \bW}_\eps(s)\|^2_{{\cal
  H}_{1,\eps}}ds
=\|{ \bW}_\eps(0)\|^2_{{\cal
  L}_{2,\eps}}\nonumber
\\
\label{energy-balance100}
&
+4\ga_0\int_0^tds\int_{\bbT_{\eps}\times\bbT^2}R(k,k',\eps\eta)
{\rm Re}\,\left\{2\widehat
   W_{\eps,+} (s,\eta,k) \left(\widehat
   W_{\eps,+}\right)^\star (s,\eta,k') \right.
\\
&
\left.+Y_{\eps,+} (s,\eta,k) \left(\widehat
   Y_{\eps,+}\right)^\star (s,\eta,k') +Y_{\eps,-} (s,\eta,k) \left(\widehat
   Y_{\eps,+}\right)^\star (s,\eta,k')
  \right\}d\eta dk dk'\nonumber
\\
&
-8\ga_0\int_0^tds\int_{\bbT_{\eps}\times\bbT^2}R(k,k',\eps\eta)
{\rm Re}\,\left\{\widehat
   W_{\eps,+} (s,\eta,k) \left(\widehat
   Y_{\eps,+}\right)^\star (s,\eta,k')
\right. \nonumber
\\
&
\left.+\widehat Y_{\eps,+} (s,\eta,k) \left(\widehat
   W_{\eps,+}\right)^\star (s,\eta,k') \right\}d\eta dk dk',\quad t\ge0,\,\eps\in(0,1]. \nonumber
\end{align}
In particular we have
\begin{equation}
\label{012105-19}
\frac{d}{dt}\|{ \bW}_\eps(t)\|^2_{{\cal
  L}_{2,\eps}}\le 16 R_*\|{ \bW}_\eps(t)\|^2_{{\cal
  L}_{2,\eps}}
\end{equation}
and
\begin{equation}
\label{012105-19a} 
2\ga_0\|{ \bW}_\eps(t)\|^2_{{\cal
  H}_{0,\eps}}+\ga_1\|{ \bW}_\eps(t)\|^2_{{\cal
  H}_{1,\eps}}\le  4 R_*\|{ \bW}_\eps(t)\|^2_{{\cal
  L}_{2,\eps}},
\end{equation}
where
$R_*:=\sup_{k,k'\in \bbT,\,\ell\in 2\bbT}|R(k,k',\ell)|$. By
the Gronwall inequality we conclude the following.
\begin{prop}
\label{prop012105-19}
We have
\begin{equation}
\label{022105-19}
\| {\bf W}_{\eps}(t)\|_{{\cal
      L}_{2,\eps}}\le \|{\bf W}_{\eps}(0)\|_{{\cal
      L}_{2,\eps}}e^{8 R_* t}
\end{equation}
and
\begin{equation}
\label{022105-19b}
2\ga_0\|{ \bW}_\eps(t)\|^2_{{\cal
  H}_{0,\eps}}+\ga_1\|{ \bW}_\eps(t)\|^2_{{\cal
  H}_{1,\eps}}\le  4 R_*\|{\bf W}_{\eps}(0)\|_{{\cal
      L}_{2,\eps}}^2e^{16 R_* t},\quad t\ge0.
\end{equation}
\end{prop}
\bigskip
A direct consequence of Proposition \ref{prop012105-19} is the following.
\begin{cor}
\label{cor011706-19}
The semigroup $\left({\frak W}_\eps(t)\right)_{t\ge0}$ is
uniformly continuous on ${\cal L}_{2,\eps}$. Moreover, its
norm satisfies
 \begin{equation}
\label{022105-19a}
\| {\frak W}_{\eps}(t)\|_{{\cal L}_{2,\eps}}\le e^{8 R_* t},\quad t\ge0,\quad \eps\in(0,1].
\end{equation}
\end{cor}
\bigskip
\begin{cor}
\label{cor010406-19}
Suppose that ${\bf  W}(\cdot)$ is a  $\vphantom{1}^\star$-weak limit of
$\left({\bf  W}_{\eps_n}(\cdot)\right)$ in
$\left(L^1([0,+\infty),{\cal A})\right)'$ and assume that
\begin{equation}
\label{Wstar}
W_*:=\limsup_{\eps\to0+}\|{\bf W}_{\eps}(0)\|_{{\cal
    L}_{2,\eps}}<+\infty. 
\end{equation}
Then, ${\bf
  W}(\cdot)\in L^\infty_{\rm loc}([0,+\infty);L^2(\bbR\times \bbT))$. In fact
we have
\begin{equation}
\label{040406-19}
\left\|{\bf  W}(\cdot) \right\|_{L^\infty([0,\tau];L^2(\bbR\times
  \bbT))}\le e^{8R_*\tau}W_*,\quad \tau\ge0.
\end{equation}
\end{cor}
\proof 
 Fix some $\tau>0$. Suppose that ${\bf G}\in{\cal A}_c$.
Suppose that  $A\subset [0,\tau]$ is a Borel measurable set. We know that
$$
\lim_{n\to+\infty}\int_A du\int_{\bbR\times\bbT}\widehat{\bf G}(\eta,k)\cdot\widehat{\bf W}_{\eps_n}(u,\eta,k) d\eta dk=\int_ Adu\int_{\bbR\times\bbT}\widehat{\bf G}(\eta,k)\cdot\widehat{\bf  W}(u,\eta,k) \hat
d\eta dk.
$$
Suppose that $n_0$ is such that 
$
{\supp} \widehat {\bf G}\subset
[-\eps_n^{-1},\eps_n^{-1}]\times\bbT$, $n\ge n_0.
$
Then for these $n$ we can write
\begin{align*}
&
\left|\int_A du\int_{\bbR\times\bbT}\widehat{\bf
  G}(\eta,k)\cdot\widehat{\bf W}_{\eps_n}(u,\eta,k) 
d\eta dk\right|
=\left|\int_A du\int_{\bbT_{\eps_n}\times\bbT}\widehat{\bf G}(\eta,k)\cdot\widehat{\bf W}_{\eps_n}(u,\eta,k) d\eta dk\right|\\
&
\le \|{\bf  G}\|_{L^2(\bbR\times\bbT)}\int_A \|{\bf W}_{\eps_n}(u) \|_{{\cal L}_{2,\eps_n}}du.
\end{align*}
Therefore, by \eqref{022105-19},
$$
\left|\int_A du\int_{\bbR\times\bbT}\widehat{\bf G}(\eta,k)\cdot\widehat{\bf  W}(u,\eta,k)  d\eta dk\right|\le e^{8R_*\tau}m_1(A)\| {\bf G}\|_{L^2(\bbR\times\bbT)}\limsup_{\eps\to0+}\|{\bf W}_{\eps}(0)\|_{{\cal L}_{2,\eps}}.
$$
By the density argument,
the above inequality holds for all ${\bf G}\in L^2(\bbR\times\bbT)$. 
We can further  extend the above estimate by taking a simple function 
of the form ${\bf F}(t):=\sum_{i=1}^{n}{\bf G}_i1_{A_i}(t)$, where
${\bf G}_i\in {\cal A}_2$ and $A_1,\ldots, A_n$ are disjoint, Borel measurable
subsets of $[0,\tau]$. The above argument generalizes easily and we
conclude that 
$$
\left|\int_ 0^{\tau}\int_{\bbR\times\bbT}\widehat
{\bf F}(u,\eta,k) \cdot\widehat{\bf  W}(u,\eta,k) d\eta dk\right|\le e^{8R_*\tau}\|
{\bf F}\|_{L^1([0,\tau]:L^2(\bbR\times\bbT))}\limsup_{\eps\to0+}\|{\bf W}_{\eps}(0)\|_{{\cal L}_{2,\eps}}.
$$
Since the functions ${\bf F}(\cdot)$ are dense in $L^1([0,\tau]; L^2(\bbR\times\bbT))$ we
conclude the proof of the corollary. Estimate \eqref{040406-19} is a
consequence of the results of Section IV.1 of \cite{diestel}.
\qed

\section{Fourier-Wigner functions for the
  deterministic dynamics}

Throughout the present section we shall assume that $T=0$.

\subsection{Dynamics of the Wigner distributions for the solution of (\ref{012503-19})} 

\label{sec5.1}

We  consider the Wigner tensor $\bW^{\rm un}_{\eps}(t)$,
corresponding to the wave function
$\hat\Phi^{(\eps)}(t,k)=\hat\Phi(t/\eps,k)$, where $\hat\Phi(t,k)$ is given by \eqref{Ut}. Its Fourier
transform (in the first variable) is given by
(see \eqref{eq:10}) 
$$
\widehat\bW^{\rm un}_{\eps}(t,\eta,k):=\frac{\eps}{2}\bbE\left[\hat\Phi^{(\eps)}\left(t,k+\frac{\eps\eta}{2}\right)\otimes \left(\hat\Phi^{(\eps)} \right)^\star\left(t,k-\frac{\eps\eta}{2}\right)\right]
=\left[\begin{array}{ll}
\widehat W^{\rm un}_{\eps,+}(t,\eta,k)&\widehat Y^{\rm un}_{\eps,+}(t,\eta,k)\\
\widehat Y^{\rm un}_{\eps,-}(t,\eta,k)&\widehat W^{\rm un}_{\eps,-}(t,\eta,k)
\end{array}\right],
$$
with
\begin{align*}
&
\widehat W^{\rm un}_{\eps,+}(t,\eta,k):=\frac{\eps}{2}\bbE\left[\hat\phi^{(\eps)}\left(t,k+\frac{\eps\eta}{2}\right) \left(\hat\phi^{(\eps)}\right)^\star\left(t,k-\frac{\eps\eta}{2}\right)\right],
\\
&
\widehat Y^{\rm un}_{\eps,+}(t,\eta,k):=\frac{\eps}{2}\bbE\left[\hat\phi^{(\eps)}\left(t,k+\frac{\eps\eta}{2}\right) \hat\phi^{(\eps)}\left(t,-k+\frac{\eps\eta}{2}\right)\right],
\\
&
\widehat Y^{\rm un}_{\eps,-}(t,\eta,k):=\left(\widehat Y^{\rm un}_{\eps,+}(t,-\eta,k)\right)^\star,\quad
\widehat W^{\rm un}_{\eps,-}(t,\eta,k):=\widehat W^{\rm un}_{\eps,+}(t,\eta,-k).
\end{align*}

Using \eqref{basic:sde:2av} we conclude that
the dynamics of the tensor  $\bW^{\rm un}_{\eps}(t)$ is described by
the ${\cal L}_{2,\eps}$ strongly continuous semigroup 
\begin{equation}
\label{frakWun}
{\frak W}_\eps^{\rm un}(t):=\exp\left\{\left(\ga_0{\frak
    L}_{\eps\eta}'+{\frak H}_\eps+\ga_1{\frak T}_\eps\right)t\right\},\quad t\ge0,
\end{equation}
where ${\frak H}_\eps$, ${\frak T}_\eps$ are defined in  \eqref{fH},
\eqref{LHT} and \eqref{011605-19z}. On the other hand ${\frak
    L}_{\eps\eta}'$ is given by the respective Fourier transform
\begin{align}
\label{LHT1}
\widehat{\frak
    L}_{\eps\eta}'\widehat \bW(\eta,k):=\left[\begin{array}{cc}
\widehat W'_{+}(\eta,k)&\widehat Y'_{+}(\eta,k)\\
\widehat Y'_{-}(\eta,k)&\widehat W'_{-}(\eta,k)
\end{array}\right],
\end{align}
with
\begin{eqnarray}
\label{exp-wigner-eqt-1a}
&&\widehat W'_{\pm}(\eta,k):=-2\bar R(k,\eps \eta)\widehat W^{\rm
   un}_{\pm} (\eta,k)+\left\{R\left(k\mp\frac{\eps
   \eta}{2}\right)\widehat  Y^{\rm un}_{+}
   (\eta,k)+R\left(k\pm\frac{\eps \eta}{2}\right)\widehat Y^{\rm
   un}_{-}(\eta,k)\right\}, \nonumber\\
&&
\widehat Y'_{\pm} (\eta,k)=-2\bar R(k,\eps \eta)\widehat Y _{\pm} (\eta,k)
+ R\left(k\mp\frac{\eps \eta}{2}\right)\widehat W_{+}(\eta,k)+
   R\left(k\pm\frac{\eps \eta}{2}\right)\widehat
   W_{-}(\eta,k).
\end{eqnarray}
By a direct calculation we can verify the following.
\begin{prop}
\label{prop011505-19}
The following identity holds
\begin{align}
\label{energy-balance10}
&\|{ \bW}_\eps^{\rm un}(t)\|^2_{{\cal
  L}_{2,\eps}}+8\ga_0\int_0^t\|{ \bW}_\eps^{\rm un}(s)\|^2_{{\cal
  H}_{0,\eps}}ds+4\ga_1\int_0^t\|{ \bW}_\eps^{\rm un}(s)\|^2_{{\cal
  H}_{1,\eps}}ds
=\|{ \bW}^{\rm un}_\eps(0)\|^2_{{\cal
  L}_{2,\eps}},\quad t\ge0,\,\eps\in(0,1].
\end{align}
\end{prop}

A direct consequence of the proposition is the following.
\begin{cor}
\label{cor011906-19}
$\left({\frak W}_\eps^{\rm un}(t)\right)_{t\ge0}$ forms a
uniformly continuous semigroup of contractions on ${\cal L}_{2,\eps}$ for any $\eps\in(0,1]$.
\end{cor}

Let ${  p}_0^{(\eps)}\left(t\right):={  p}_0\left(t/\eps\right)$ (see \eqref{p0})
and
\begin{align}
\label{091505-19}
&
d_\eps(t,k):=i\bbE_\eps\left[
  \left(\hat\phi^{(\eps)}\right)^\star\left(t,k\right){
  p}_0^{(\eps)}\left(t\right)\right] .
\end{align}
By a direct calculation we obtain
\begin{equation}
\label{021906-19}
\left\|{\bf  W}^{\rm un}_\eps(t)\right\|_{{\cal
    H}_{1,\eps}}=2\|d_\eps(t)\|_{L^2(\bbT)},\quad t\ge0,\,\eps\in(0,1].
\end{equation}
From \eqref{energy-balance10} it follows directly.
\begin{cor}
\label{cor011505-19}
We have
\begin{equation}
\label{1015-5-19}
\int_0^{+\infty}\|d_\eps(t)\|_{L^2(\bbT)}^2dt\le \frac{\|{\bW}_\eps(0)\|^2_{{\frak
  A}_{2,\eps}}}{16\ga_1},\quad\,\eps\in(0,1].
\end{equation}
\end{cor}



\subsection{Duhamel representation of the energy density dynamics}

Using the Duhamel formula we can reformulate \eqref{basic:sde:2av1} as follows
\begin{align}
  \label{eq:sol1}
  &
  \hat\Psi^{(\eps)}(t,k)= U\left(\frac{t}{\eps}\right)\hat\Psi(k) 
     +\sqrt{\ga_0}\int_0^{t} U\left(\frac{t-s}{\eps}\right)\left(i\int_{\bbT} r(\cdot,k'){\bf
   D}\hat\Psi^{(\eps)}\left(s,\cdot-k'\right)B(ds,dk')\right).
\end{align}
Here $U(t)$ is given by \eqref{Ut} and \eqref{basic:sde:2av}.
Using \eqref{eq:sol1}
to express
the Fourier-Wigner tensor $\bW_\eps(t)$ we obtain the following
equality 
\begin{equation}
\label{012904-19}
\bW_\eps(t)=\bW_\eps^{\rm un}(t)+ \bW_\eps'(t),
\end{equation}
where 
\begin{equation}
\label{012904-19-1}
\bW_\eps^{\rm un}(t)={\frak W}_\eps^{\rm
  un}(t)\left(\bW_\eps(0)\right)
\end{equation}
and 
$$
\widehat \bW_\eps'(t,\eta,k) 
=\frac{\eps\ga_0}{2}\sum_{n}\int_0^{t}\bbE\left\{\left(U\left(\frac{t-s}{\eps}\right)\big(\hat\Psi_n^{\eps}(s)\big)\right)^\star\left(k-\frac{\eps\eta}{2}\right)\otimes U\left(\frac{t-s}{\eps}\right)\big(\hat\Psi_n^{\eps}(s)\big) \left(k+\frac{\eps\eta}{2}\right)\right\}ds.
$$
Here 
$$
\hat\Psi_n^{(\eps)}\big(s,k\big)=i\int_{\bbT} r(k,k'){\bf
   D}\hat\Psi^{(\eps)}\left(s,k-k'\right)e_n(k')dk',
$$
where $(e_n)$ is an orthonormal base in $L^2(\bbT)$. Using
 \eqref{eq:10}
we conclude  that
\begin{align*}
&\widehat \bW_\eps'(t,\eta,k) =\ga_0 \int_0^{t}\widehat {\frak W}_\eps^{\rm
  un}(t-s)\big({\bf V}_\eps(s)\big)(\eta,k)ds.
\end{align*}
Here $\widehat {\frak W}_\eps^{\rm
  un}(t)$ is the Fourier transform of the semigroup \eqref{frakWun}
and  ${\bf V}_\eps(t)$ is given by its Fourier transform
$\widehat {\bf V}_\eps(t,\eta,k)
:=\widehat{\frak R}_\eps \widehat {\bf W}_\eps(t,\eta,k)
$,
where $\widehat{\frak R}_\eps:\hat{\cal L}_{2,\eps}\to \hat{\cal
  L}_{2,\eps}$ is defined by 
\begin{align}
\label{030805-19}
&
\widehat{\frak R}_\eps\widehat {\bf W}:=\widehat V_\eps(\eta,k){\bf D},\\
&
\widehat V_{\eps}(\eta,k)=\int_{\bbT}r\left(k-\frac{\eps\eta}{2},k-k'\right)
  r\left(k+\frac{\eps\eta}{2},k-k'\right)
\left[\widehat W_{+}(\eta,k')+\widehat W_{-}(\eta,k')-\widehat Y_{+}(\eta,k')-\widehat Y_{-}(\eta,k')\right]dk'.\nonumber
\end{align}
Summarizing, we have shown that
\begin{equation}
\label{020805-19}
\bW_\eps(t)={\frak W}_\eps^{\rm un}(t)(\bW_\eps(0))+ \ga_0 \int_0^{t} {\frak W}_\eps^{\rm
  un}(t-s)\big({\frak R}_\eps \bW_\eps(s)\big)(\eta,k)ds,\quad t\ge0,\,\eps\in(0,1].
\end{equation}
Operator ${\frak R}_\eps $ is given by its Fourier transform, see
\eqref{030805-19}. A direct calculation yields the following.
\begin{prop}
\label{prop030706-19}
Operators ${\frak R}_\eps$ can be defined by \eqref{030805-19}  as
bounded operators both on ${\cal L}_{2,\eps}$ and
$L^2(\bbR\times\bbT)$. They   are uniformly
bounded in $\eps\in(0,1]$. More precisely, there exists ${\frak r}_*>0$ such that
\begin{equation}
\label{050706-19}
\|{\frak R}_\eps\|_{{\cal L}_{2,\eps}}\le {\frak r}_*,\quad \|{\frak
  R}_\eps\|_{L^2(\bbR\times\bbT)}\le {\frak r}_*,\quad \eps\in(0,1].
\end{equation}
In addition, we have
\begin{equation}
\label{050702-19}
\lim_{\eps\to0+}{\frak R}_\eps{\bf W}={\frak R}{\bf W},\quad\mbox{in }L^2(\bbR\times\bbT)
\end{equation}
for any ${\bf W}\in L^2(\bbR\times\bbT)$.  Here (cf \eqref{cR})
\begin{equation}
\label{fR}
{\frak R}{\bf W}(y,k)={\bf D}{\cal R}\left(W_{+}(y,\cdot)+
  W_{-}(y,\cdot)-Y_{+}(y,\cdot)-
  Y_{-}(y,\cdot)\right)(k),\quad (y,k)\in\bbR\times\bbT.
\end{equation}
\end{prop}
Theorem \ref{main-thm} is a direct corollary from the following. 
\begin{thm}
\label{cor020805-19}
Under the assumptions of Theorem \ref{main-thm}
for any $G\in L^1\left([0,+\infty);{\cal A}\right)$ we have
\begin{equation}
\label{012404-19z}
\lim_{\eps\to0+}\int_0^{+\infty}dt\int_{\bbR\times\bbT}  \widehat{Y}^\star_{\eps,\pm}(t,\eta,k)  \widehat G(t,\eta,k)d\eta dk=0
\end{equation}
and
\begin{equation}
\label{012404-19}
\lim_{\eps\to0+}\int_0^{+\infty}dt\int_{\bbR\times\bbT}  \widehat{ W}^\star_{\eps,\pm}(t,\eta,k)  \widehat G(t,\eta,k)d\eta dk=
\int_0^{+\infty}dt\int_{\bbR\times\bbT} \widehat{ W}^\star(t,\eta,\pm k)  \widehat G(t,\eta,k)d\eta dk,
\end{equation}
where 
$
W(t,y,k)
$ 
is the unique solution of the equation \eqref{integral} with the
initial condition $W_0$.
\end{thm}
The proof is presented in Section \ref{sec5.4}.

\subsection{Laplace transform}

\label{sec5.3}

As we have already mentioned, see \eqref{030406-194.9}, 
$\left({\bf  W}_{\eps}(\cdot)\right)$ is  $\vphantom{1}^\star$-weakly
sequentially compact in the dual to $L^1([0,+\infty),{\cal A})$,
therefore the proof of Theorem \ref {cor020805-19} comes down
to showing uniqueness of  limiting points, as $\eps\to0+$.
For that purpose it is convenient to work with the Laplace transform
\begin{align}
\label{012906-19}
&
{\bf w}_{\eps}(\la)=\left[\begin{array}{ll}
w_{\eps,+}(\la)&y_{\eps,+}(\la)\\
y_{\eps,-}(\la)&w_{\eps,-}(\la)
\end{array}\right]:=\int_0^{+\infty}e^{-\la t}{\bf  W}_{\eps}(t)dt.
\end{align}
Sometimes, when we wish to highlight the dependence on the initial
data, we shall also write 
$$
{\bf w}_{\eps}(\la; {\bf  W}_{\eps}(0))=:\widetilde{\frak W}_\eps(\la)( {\bf  W}_{\eps}(0)).
$$

Thanks to \eqref{052709-18} the Laplace transform
is well defined as an element in ${\cal A}'$ for any ${\rm Re}\,\la>0$.
From Proposition \ref{prop012105-19} we conclude the following
estimate
\begin{equation}
\label{032105-19}
\| \widetilde{\frak W}_\eps(\la)\|_{{\cal
      L}_{2,\eps}}\le {[({\rm Re}\,\la-8R_*){\rm Re}\,\la]^{-1/2}},\quad {\rm Re}\,\la>8R_*.
\end{equation}
The argument used in the proof of Corollary \ref{cor010406-19} 
shows that  
\begin{align*}
&
{\bf w}(\la)
=\left[\begin{array}{cc}
{w}_{+}(\la)&{y}_{+}(\la)\\
{y}_{-}(\la)&{w}_{-}(\la)
\end{array}\right],
\end{align*} 
- any $\vphantom{1}^\star$-weak limiting point in ${\cal A}'$  of  ${\bf
  w}_{\eps}(\la)$, as $\eps\to0+$ - belongs to $L^2(\bbR\times\bbT)$
and satisfies the estimate
\begin{equation}
\label{032105-19a}
\| {\bf w}(\la)\|_{L^2(\bbR\times\bbT)}\le {[({\rm Re}\,\la-8R_*){\rm Re}\,\la]^{-1/2}}\limsup_{\eps\to0+}\| {\bf W}_\eps(0)\|_{L^2(\bbR\times\bbT)},\quad {\rm Re}\,\la>8R_*.
\end{equation}

Equation  \eqref{020805-19} leads to the following equation for the
Laplace transform
\begin{equation}
\label{020805-19a}
{\bf w}_\eps(\la;  {\bf  W}_{\eps}(0))={\bf w}_\eps^{{\rm
    un}}(\la ; {\bf  W}_{\eps}(0))+\ga_0 {\bf w}_\eps^{\rm
  un}\left(\la; {\bf v}_\eps\left(\la\right)\right),\quad {\rm Re}\, \la>0.
\end{equation}
Here,
\begin{equation}
\label{010701-19}
{\bf w}_\eps^{{\rm
    un}}(\la ;{\bf W}):= \widetilde{\frak W}_\eps^{\rm un}(\la) {\bf
  W}:=\int_0^{+\infty}e^{-\la t}{\frak W}_\eps^{\rm un}(t) {\bf
  W}dt,\quad {\bf W}\in {\cal L}_{2,\eps}
\end{equation}
and 
\begin{equation}
\label{020701-19}
{\bf v}_\eps\left(\la\right):={\frak R}_\eps {\bf w}_\eps(\la;  {\bf  W}_{\eps}(0)).
\end{equation}
Let $W_0\in L^2(\bbR\times\bbT)$. We define
$$
w^{\rm un}_{+}(\la, W_0)=\widetilde {\frak W}^{\rm un}(\la)W_0:=\int_0^{+\infty}e^{-\la t}{{\frak
    W}^{\rm un}_tW_0}(\eta,k)dt.
$$
It follows directly from \eqref{010304} that
its Fourier transform satisfies the equality
\begin{align}
\label{020911absz}
&  \left(\vphantom{\int_0^1}\la +2\ga_0 R(k)+
  i\om'(k)\eta\right)\widehat w^{\rm un}_{+}(\la,\eta,k; W_0)=\widehat W_0(\eta,k)
\\
&
  -\ga_1\left\{ (1-p_+(k))\int_{\bbR}\frac{ \widehat W_0(\eta',k)d\eta' dk}{\la+2\ga_0 R(k)+i\om'(k)\eta'}-p_-(k) \int_{\bbR}\frac{ 
\widehat W_0(\eta',-k)d\eta' }{ \la+2\ga_0 R(k)-i\om'(k)\eta'}\right\}.\nonumber
\end{align}
It is clear from the above formula that $w^{\rm un}_{+}(\la, W_0) \in {\cal A}'$, provided that
$W_0\in {\cal A}'$. Thanks to formulas \eqref{033110} and
\eqref{feb1402} it can also be seen that 
$w^{\rm un}_{+}(\la, W_0) \in   L^2(\bbR\times\bbT)$, if $W_0 \in  L^2(\bbR\times\bbT)$.

The  first step
towards the limit identification of $\left({\bf
    W}_{\eps}(\cdot)\right)$ consists in showing the following result. 
\begin{thm}
\label{main:thm-un}
Under the assumptions on the initial data made in Theorem \ref{main-thm}, the family $\left({\bf w}^{\rm
  un}_{\eps}(\la; {\bf W}_{\eps}(0))\right)$  converges
$\vphantom{1}^\star$-weakly to ${\bf w}^{\rm un}(\la;W_0)$ of the form
\begin{align*}
&
{\bf w}^{\rm un}(\la,y,k;W_0)
=\left[\begin{array}{cc}
w^{\rm un}_{+}(\la,y,k;W_0)&0\\
0&w^{\rm un}_{+}(\la,y,-k;W_0)
\end{array}\right],\quad (y,k)\in \bbR\times\bbT,
\end{align*}
where the Fourier transform of $w^{\rm un}_{+}(\la,y,k;W_0)$ is given
by \eqref{020911absz}.
\end{thm}
The proof of this result follows closely the argument contained in
\cite{kors}. We present its outline in Appendix \ref{appb}.


The identification of the limit of $\left({\bf W}_\eps(t)\right)$ is possible thanks to the following result.
\begin{thm}
\label{main:thm2}
Assume the hypotheses of Theorem $\ref{main-thm}$.
Furthermore, suppose that 
 ${\bf w}(\la)$ is the $\vphantom{1}^\star$-weak limit of  $\left({\bf
    w}_{\eps_n}(\la)\right)$ in ${\cal A}'$ for some sequence
$\eps_n\to0+$. Then, there exists $\la_0>0$ such that 
${\bf w}(\la)  \in L^2(\bbR\times\bbT)$ for ${\rm Re}\,\la>\la_0$ and 
\begin{equation}
\label{010702-19}
{\bf w}(\la,y,k)
=\left[\begin{array}{cc}
{w}_{+}(\la,y,k)&0\\
0&{w}_{+}(\la,y,-k)
\end{array}\right],\quad (y,k)\in L^2(\bbR\times\bbT),
\end{equation}
where ${w}_{+}(\la,y,k)$ satisfies the equation
\begin{equation}
\label{integral-bis}
w_+(\la)=w^{\rm un}_{+}(\la;W_0)+2\ga_0\widetilde{\frak W}^{\rm
  un}(\la){\cal R} { w}_+(\la),
\end{equation}
and ${\cal R}:L^2(\bbR\times\bbT)\to L^2(\bbR\times\bbT)$ (cf
\eqref{cR}) is given by 
\begin{equation}
\label{cRW}
{\cal R}F(y,k):=  \int_{\bbT}R(k,k')
F\left(y,k'\right) dk',\quad (y,k)\in\bbR\times \bbT,\quad F\in L^2(\bbR\times \bbT).
\end{equation} 
\end{thm}
We present the proof of Theorem \ref{main:thm2} in Section \ref{sec5.5} below.

\subsection{Proof of Theorem \ref{cor020805-19}} 

\label{sec5.4}
 Since $\left({\frak W}^{\rm
  un}_t\right)_{t\ge0}$ is a semigroup of contractions,  see Proposition
\ref{prop011406-19}, we have
\begin{equation}
\label{032105-19b}
\|\widetilde{\frak W}^{\rm
  un}(\la)\|_{L^2(\bbR\times\bbT)}\le \frac{1}{{\rm
    Re}\,\la},\quad {\rm
  Re}\,\la>0.
\end{equation}
Thanks to the fact that ${\cal R}$ is bounded, it is straightforward
to see that equation \eqref{integral-bis} has a unique
$L^2(\bbR\times\bbT)$ solution for  $\la$ with a  sufficiently large
real part. Therefore, the Laplace transform
${w}_{+}(\la)$  of any
limiting point of $\left(W_\eps(t)\right)$ is uniquely determined by   \eqref{integral-bis}. This in turn implies the
conclusion of the theorem.

\subsection{Proof of Theorem \ref{main:thm2}}
\label{sec5.5}

To avoid using double subscript
notation we assume that  ${\bf w}(\la)=\lim_{\eps\to0+}{\bf
    w}_{\eps}(\la)$. We wish to show that the limit is of the
form \eqref{010702-19} with ${w}_{+}(\la)$ satisfying \eqref{integral-bis}.

\subsubsection{Proof of  (\ref{010702-19})}

\label{sec5.5.1}

We prove that 
\begin{equation}
\label{031405-19z}
\lim_{\eps\to0+}\int_{\bbR\times\bbT}\widehat
y_{\eps,\iota}^\star(\la,\eta,k)\widehat G(\eta,k)d\eta dk=0,\quad \iota=\pm
\end{equation}
for any  $G\in {\cal A}_c$ (see \eqref{AC}) and ${\rm Re}\,\la>0$.
Consider only  the case of $\iota=+$, the other one is analogous. 
Assume that $K>0$ is fixed and
$\eps$ is so small that 
\begin{equation}
\label{GK}
{\rm supp}\,\hat
G\subset[-K,K]\times\bbT\subset
[\eps^{-1},\eps^{-1}]\times\bbT.
\end{equation}
Let 
\begin{equation}
\label{chik}
\chi_K(\eta,k):=1_{[-K,K]}(\eta)1_{\bbT}(k).
\end{equation}
By virtue of estimate \eqref{032105-19} we conclude that
$\left(\widehat y_{\eps,+} (\la)\chi_K\right)$ is bounded, thus
weakly compact in $L^2(\bbR\times\bbT)$ for a fixed $K>0$.
It converges weakly  in
$L^2(\bbR\times\bbT)$, due to its $\vphantom{1}^\star$- weak
convergence in ${\cal A}'$. Denote the limit, belonging to
$L^2(\bbR\times\bbT)$,  by $\widehat y_{+} (\la)\chi_K$.

Let
\begin{align}
\label{fd-eps}
&
{\frak d}_\eps(\la,k):=i\int_0^{+\infty}e^{-\la t}\bbE_\eps\left[
  \left(\hat\psi^{(\eps)}\right)^\star\left(t,k\right){
  \frak p}_0^{(\eps)}\left(t\right)\right]dt \\
&
=\frac{1}{2}\int_{\bbT_{\eps}}\left[\widehat
  y_{\eps,+}\left(\la, \eta,
  k-\frac{\eps\eta}{2}\right)-\widehat
  w_{\eps,+}\left(\la,\eta, k-\frac{\eps\eta}{2}\right)\right]d\eta .\nonumber
\end{align}
Taking the Laplace transforms of both sides of \eqref{011806-19} and
multiplying by $\chi_K$ we
obtain in particular the equation 
\begin{align}
\label{020911ab1z}&
-\eps\ga_0 R\left(k-\frac{\eps \eta}{2}\right)\widehat w_{\eps,+}(\la)\chi_K +\left(\vphantom{\int_0^1}\eps\la +2\ga_0\eps \bar R(k,\eps \eta)+ 2i\bar\om(k,
\eps \eta)\right) \widehat y_{\eps,+}(\la) \chi_K-\ga_0\eps
   R\left(k+\frac{\eps \eta}{2}\right) \widehat w_{\eps,-}(\la) \chi_K
  \nonumber\\
&
=\eps\widehat Y_{\eps,+}(\eta,k) \chi_K +\frac{\eps\ga_1}{2}\left\{{\frak
   d}_\eps^\star\left(\la,-k+\frac{\eps \eta}{2}\right)+{\frak
   d}_\eps^\star\left(\la,k+\frac{\eps \eta}{2}\right)\right\}\chi_K\\
&
+\ga_0{\cal R}_{\eps \eta}\left\{\vphantom{\int_0^1}\widehat y_{\eps,+}(\la)
+\widehat
y_{\eps,-}(\la)-
\widehat w_{\eps,+}(\la) -
\widehat w_{\eps,-}(\la)\right\}\chi_K.\nonumber
\end{align}
Thanks to estimate \eqref{022105-19b}, see also \eqref{H2eps}, we conclude that
\begin{equation}
\label{121505-19a}
\|{\frak d}_\eps(\la)\|_{L^2(\bbT)}\le \left(\frac{R_*}{ \ga_1{\rm Re}\,\la({\rm Re}\,\la-8R_*)}\right)^{1/2}\|{ \bW}_\eps(0)\|_{{\cal
  L}_{2,\eps}},\quad
\eps\in(0,1],\,{\rm Re}\,\la>8R_*.
\end{equation}
From the strong convergence in
$L^2(\bbR\times\bbT)$ of the coefficients of \eqref{020911ab1z} and estimate \eqref{121505-19a} we conclude that
$2i\om(k) \widehat  y_{+} (\la)\chi_K=0$ for any $\la$ such that ${\rm
  Re}\,\la>8R_*$. This  in turn implies that $\widehat
y_{+}(\la) =0$ for such $\la$-s, which by analytic continuation, implies  \eqref{031405-19z}.


\subsubsection{Proof of  (\ref{integral-bis})}

  To avoid   writing double subscript we maintain the convention to
  denote a
 subsequence of $\left({\bf w }_\eps(\la)\right)$ by the same symbol as the entire sequence.
From the already proved part of the theorem we know that the limiting
element is of the form \eqref{010702-19}.
We let the test
matrix valued function from ${\cal A}_c$ be of the form
$$
{\bf G}(y,k):=\left[\begin{array}{cc}
G(y,k)&0\\
0&G(y,-k)
\end{array}\right],\quad (y,k)\in \bbR\times\bbT,
$$
with $G$ satisfying \eqref{GK}. 
Using the inclusion $[-K,K]\subset \bbT_\eps$ we can treat
${\bf G}$ as an element of ${\cal L}_{2,\eps}$.
Applying both sides of \eqref{020805-19a} to this test matrix we
obtain
the following equality
\begin{equation}
\label{020702-19}
\left\langle{\bf w}_{\eps}(\la),
 {\bf
  G}\right\rangle_{{\cal L}_{2,\eps}}=I_\eps +I\! I_\eps,
\end{equation}
where
$$
I_\eps:=\left\langle{\bf w}_{\eps}^{\rm un}(\la),{\bf
  G}\right\rangle_{{\cal 
L}_{2,\eps}},\quad I\! I_\eps:=\ga_0\left\langle{\bf w}_{\eps}(\la),
  {\frak R}_{\eps}^\star\left(\widetilde {\frak W}_{\eps}^{\rm
      un}(\la)\right)^\star{\bf
 G}\right\rangle_{{\cal L}_{2,\eps}}.
$$
Here $ {\frak R}_{\eps}^\star$ and $\left(\widetilde {\frak
    W}_{\eps}^{\rm un}(\la)\right)^\star$ are the adjoints of the respective operators (see
\eqref{030805-19} and \eqref{010701-19}), in ${\cal
  L}_{2,\eps}$.
Invoking Theorem  \ref{main:thm-un} we conclude that
\begin{equation}
\label{030702-19}
\lim_{\eps\to0+}I_\eps=2\int_{\bbR\times \bbT}\left({ w}^{\rm un}_+(\la,y,k;W_0)\right)^\star{
  G}(y,k)dy dk,\quad {\rm Re}\,\la>0.
\end{equation}
Concerning
 the term $ I\! I_\eps$, we are going  to show that there exists
 $\la_0>0$ such that
\begin{equation}
\label{wish}
\lim_{\eps\to0+}I\!I_\eps=4\ga_0\int_{\bbR\times\bbT} w_+(\la,k,y){\cal R} \bar{ g}_+(\la)(y,k) dydk
\end{equation} 
for ${\rm Re}\,\la>\la_0$. Here
$$
\bar{ g}_+(\la,y,k):=\left(\widetilde{\frak W}^{\rm un} (\la)\right)^\star(G)(y,k).
$$


Denote by 
$$
\widehat{\bf g}_\eps(\la,\eta,k):=\left[\begin{array}{cc}
\widehat{ g}_{\eps,+}(\la,\eta,k)& \widehat{ h}_{\eps,+} (\la,\eta,k)\\
\widehat{ h}_{\eps,-}(\la,\eta,k)& \widehat{ g}_{\eps,-} (\la,\eta,k)
\end{array}\right],
$$ 
the Fourier transform of
the distribution
$
{\bf g}_\eps(\la)=\left(\widetilde{\frak W}^{\rm
      un}_\eps(\la)\right)^\star {\bf G}
$ and let
\begin{equation}
\label{whF}
\widehat{\bf f}_\eps(\la,\eta,k):=
\widehat{\frak R}_\eps^\star\widehat{\bf g}_\eps(\la)(\eta,k),\quad \eps\in(0,1].
\end{equation}
Note that
\begin{align}
\label{030805-195.39}
&
\widehat{\bf f}_\eps(\la,\eta,k)=\widehat f_\eps(\la,\eta,k){\bf D},\quad\mbox{with}\nonumber\\
&
\widehat f_{\eps}(\la,\eta,k):=\int_{\bbT}r\left(k'-\frac{\eps\eta}{2},k'-k\right)
  r\left(k'+\frac{\eps\eta}{2},k'-k\right)\\
&
\times\left[\vphantom{\int_0^1} g_{\eps,+}(\la,\eta,k')+ g_{\eps,-}(\la,\eta,k')- h_{\eps,+}(\la,\eta,k')- h_{\eps,-}(\la,\eta,k')\right]dk'.\nonumber
\end{align}

The argument used in the proof of Theorem
 \ref{main:thm-un} (it suffices only to replace the terms containing
 the dispersion relation 
  by their conjugates)
 shows that in fact 
 \begin{equation}
 \label{060702-19a}
 \lim_{\eps\to0+}\left\langle {\bf g}_\eps(\la),{\bf
   F}\right\rangle= \left\langle \bar{\bf g}(\la), {\bf
   F}\right\rangle
 \end{equation}
 for any ${\bf F}\in {\cal A}_c$
 and ${\rm Re}\,\la>0$.
 Here
 $$
 \bar {\bf g}(\la,y,k):=\left[\begin{array}{cc}
 \bar { g}_+(\la,y,k)&0\\
 0&\bar { g}_+(\la,y,-k)
 \end{array}\right].
 $$



 Suppose now that $\varphi$ is an arbitrary, bounded and compactly
 supported measurable function. Then, using \eqref{060702-19a}, we
 conclude that
\begin{equation}
\label{050703-19}
\lim_{\eps\to0+}\varphi(\eta)\widehat{\bf f}_\eps(\la,\eta,k)=2
\varphi(\eta){\bf D}{\cal R}\widehat{\bar { g}}_+(\la,\eta,k),\quad\mbox{weakly in $L^2(\bbR\times\bbT)$},
\end{equation}
where $\widehat{\bar { g}}_+(\la)$ is the Fourier transform
of ${\bar { g}}_+(\la)$ in the first variable.


In order to show \eqref{wish}, we prove 
the following two results.
\begin{lm}
\label{lm010606-19}
There exists $C>0$ such that
\begin{equation}
\label{111505-19a1}
\left\| \nabla\widehat{\bf f}_\eps(\la)\right\|_{L^2(\bbT_\eps\times\bbT)}\le \frac{C}{ {\rm Re}\,\la},\quad
\eps\in(0,1],\,{\rm Re}\,\la>0.
\end{equation}
The gradient operator in \eqref{111505-19a1} is in the $\eta$ and $k$ variables.
\end{lm}

\bigskip

\begin{lm}
\label{lm011006-19}
For any $\rho>0$ there exists $M>0$ such that 
\begin{equation}
\label{011006-19}
\limsup_{\eps\to0+}\int_{[(\eta,k)\in\bbT_\eps\times\bbT,\,|\eta|>M]}\left|\widehat{\bf w}_{\eps}(\la,\eta,k)\cdot
 \widehat {\bf
  f}_\eps(\la,\eta,k)\right| d\eta dk<\rho
\end{equation}
\end{lm}

Having the above results  we can write that
$
I\!I_\eps=I\!I_{\eps,1}+I\!I_{\eps,2},
$
where
\begin{align*}
&
I\!I_{\eps,1}:=\int_{\bbT_{\eps}\times\bbT}\varphi^2(\eta)\widehat{\bf w}_{\eps}(\la,\eta,k)
  \cdot
 \widehat {\bf
  f}_\eps(\la,\eta,k) d\eta dk,\\
&
I\!I_{\eps,2}:=\int_{\bbT_{\eps}\times \bbT}[1-\varphi^2(\eta)]\widehat{\bf w}_{\eps}(\la,\eta,k)
  \cdot
 \widehat {\bf
  f}_\eps(\la,\eta,k) d\eta dk.
\end{align*}
Here $\varphi:\bbR\to[0,1]$ is a $C^\infty$ smooth function
satisfying $\varphi(\eta)\equiv 1$, $|\eta|\le M$ and
$\varphi(\eta)\equiv 0$, $|\eta|\ge 2M$.

Choose an arbitrary $\rho>0$.
Using Lemma \ref{lm011006-19} we conclude that
$M$ can be adjusted in such a way that
\begin{equation}
\label{031006-19}
\limsup_{\eps\to0+}|I\!I_{\eps,2}|<\rho.
\end{equation}
On the other hand, by virtue of Lemma \ref{lm010606-19} the set $\left(\varphi \widehat{\bf
  f}_\eps(\la) \right)$, $\eps\in(0,1]$ is compact in $L^2(\bbR\times\bbT)$ in the strong
topology. 
Since it converges also in the weak topology
(see \eqref{050703-19}), it
has to also converge strongly in $L^2(\bbR\times\bbT)$.
Combining this with
 the fact that $\left(\varphi\widehat{\bf w}_{\eps}(\la)\right)$ 
is weakly compact in $L^2(\bbR\times\bbT)$ we conclude that
\begin{equation}
\label{031006-19}
\limsup_{\eps\to0+}I\!I_{\eps,1}=
4\int_{\bbR\times \bbT}\varphi^2(\eta) \widehat w_{+}(\la,\eta,k)d\eta dk\left\{\int_{\bbT}R(k,k')\widehat{\bar  g} _{+}^\star(\la,\eta,k')dk'\right\}.
\end{equation}
This ends the proof of \eqref{wish}. The only items
yet to be proven are Lemmas \ref{lm010606-19} and \ref{lm011006-19}.

\subsubsection{Proof of Lemma \ref{lm010606-19}}

Let 
${ \bf G}_\eps(t):=\left({\frak W}_\eps^{\rm un}\right)^\star(t){\bf
  G}$ and $\widehat{ \bf G}_\eps(t,\eta,k)$ be its Fourier transform in the
first variable. From \eqref{frakWun} we conclude that it satisfies
\begin{equation}
\label{frakWun1}
\frac{d}{dt}\widehat{ \bf G}_\eps(t)=\left(\ga_0 \widehat{\frak
    L}_{\eps\eta}'+\widehat{\frak H}_\eps^\star+\ga_1 \widehat{\frak T}_\eps\right) \widehat{ \bf G}_\eps(t).
\end{equation}
One can formulate the respective
the energy balance equation, 
see \eqref{energy-balance10},
\begin{align}
\label{energy-balance11}
&\|\widehat{ \bf G}_\eps(t)\|^2_{L^2(\bbT_\eps\times\bbT)}+8\ga_0\int_0^t\|{ \bf G}_\eps(s)\|^2_{{\cal
  H}_{0,\eps}}ds+4\ga_1\int_0^t\|{ \bf G}_\eps(s)\|^2_{{\cal
  H}_{1,\eps}}ds
=\|\widehat{ \bf G}\|^2_{L^2(\bbT_\eps\times\bbT)},\quad t\ge0,\,
\end{align}
and $\eps$ sufficiently small that \eqref{GK} holds. 
It allows us to obtain estimates
\begin{equation}
\label{060703-19}
\|\widehat{\bf g}_\eps(\la)\|_{L^2(\bbT_\eps\times\bbT)}\le\frac{1}{{\rm Re}\,\la}\|{\bf
  G}\|_{L^2(\bbR\times\bbT)}
\end{equation}
for $ {\rm Re}\,\la>0$ and $\eps$ sufficiently small, as above.

Thanks to \eqref{030805-195.39} and  estimate \eqref{060703-19} we
conclude that
\begin{equation}
\label{111505-19a2}
\left\| \partial_k\widehat{\bf f}_\eps(\la)\right\|_{L^2(\bbT_\eps\times\bbT)}\le \frac{R_*' \|{\bf
  G}\|_{L^2(\bbR\times\bbT)}}{ {\rm Re}\,\la},\quad
\eps\in(0,1],\,{\rm Re}\,\la>0,
\end{equation}
with $R'_*:=\sup_{k,k'\in\bbT,\ell\in
  2\bbT}|\partial_{k'}R(k,k',\ell)|$ (cf \eqref{021605-19}).

To estimate the $L^2$-norm of $ \partial_\eta\widehat{\bf
  f}_\eps(\la,\eta,k)$ we differentiate in $\eta$ both sides of 
\eqref{frakWun1} and obtain
\begin{equation}
\label{frakWun1p}
\frac{d}{dt}\widehat{ \bf G}_{\eps,\eta}'(t)=\left(\ga_0 \widehat{\frak
    L}_{\eps\eta}'+\widehat{\frak H}_\eps^\star+\ga_1 \widehat{\frak T}_\eps\right) \widehat{ \bf G}_{\eps,\eta}'(t)+\left(\ga_0\partial_\eta \widehat{\frak
    L}_{\eps\eta}'+\partial_\eta \widehat{\frak H}_\eps^\star\right) \widehat{ \bf G}_{\eps}(t).
\end{equation}
Both here and below 
$$
\widehat{ \bf G}_{\eps,\eta}'(t):= \partial_\eta \widehat{ \bf
  G}_{\eps}(t) =
\left[\begin{array}{cc}
 \widehat  G_{\eps,\eta,+} (t,\eta,k)&\widehat  H_{\eps, \eta,+} (t,\eta,k)\\
 \widehat  H_{\eps, \eta,-} (t,\eta,k)&\widehat  G_{\eps, \eta,-} (t,\eta,k)
 \end{array}\right]
$$
and $\widehat{ \bf G}_{\eta}':= \partial_\eta \widehat{ \bf
  G}$. Let  ${ \bf G}_{\eps}^{(1)}(t)$ denote the inverse Fourier
transform of $\widehat{ \bf G}_{\eps,\eta}'(t)$.  Analogously to \eqref{energy-balance10} we conclude
the following identity
\begin{align}
\label{energy-balance21}
&\|\widehat{ \bf G}_{\eps,\eta}'(t)\|^2_{L^2(\bbT_\eps\times\bbT)}+8\ga_0\int_0^t\|{ \bf G}_{\eps}^{(1)}(s)\|^2_{{\cal
  H}_{0,\eps}}ds+4\ga_1\int_0^t\|{ \bf G}_{\eps}^{(1)}(s)\|^2_{{\cal
  H}_{1,\eps}}ds
=\|\widehat{ \bf G}_{\eta}\|^2_{L^2(\bbT_\eps\times\bbT)}\nonumber\\
&
+2  {\rm Im}\,\left\{\int_0^tds \int_{\bbT_{\eps}\times\bbT} \left\{\left[\om'\left(k+\frac{\eps \eta}{2}\right)+\om'\left(k-\frac{\eps
  \eta}{2}\right)\right]\widehat G^{\rm  un}_{\eps,+} (s,\eta,k) \left[\widehat G_{\eps,\eta,+}
  (s,\eta,k)\right]^\star
\right. \right.\nonumber\\
&
\left. \left.+\left[\om'\left(k+\frac{\eps \eta}{2}\right)-\om'\left(k-\frac{\eps
  \eta}{2}\right)\right]\widehat H^{\rm
  un}_{\eps,+} (s,\eta,k) \left[\widehat H_{\eps,\eta,+}
  (s,\eta,k)\right]^\star d\eta dk\vphantom{\int_0^1}\right\}\right\}\\
&
-2\eps\ga_0 \int_0^tds \int_{\bbT_{\eps}\times\bbT} 
  [R'(k+\eps \eta/2)-R'(k-\eps \eta/2)]\nonumber\\
&
\times {\rm Re}\, \left\{\widehat
  G_{\eps,+} (s,\eta,k) \left[\widehat G^{\rm
  un}_{\eps,\eta,+} (s,\eta,k)\right]^\star +\widehat H^{\rm
  un}_{\eps,+} (s,\eta,k) \left[\widehat H_{\eps,\eta,+}
  (s,\eta,k)\right]^\star\vphantom{\int_0^1}\right\} d\eta dk
 \nonumber
\end{align}
\begin{align*}
&
-4\eps\ga_0 \int_0^tds \int_{\bbT_{\eps}\times\bbT} R'\left(k-\frac{\eps
   \eta}{2}\right)  {\rm Re}\,\left\{ \widehat H_{\eps,+}(s,
   \eta,k) \left[\widehat G_{\eps,\eta,+}
  (s,\eta,k)\right]^\star+\widehat G^{\rm
  un}_{\eps,+} (s,\eta,k) \left[\widehat H_{\eps,\eta,+}
  (s,\eta,k)\right]^\star\right\}d\eta dk .\nonumber
\end{align*}

From \eqref{energy-balance11} and \eqref{energy-balance21} it follows directly
that
for any $\delta>0$ there exists a constant $C>0$ such that
\begin{align}
\label{energy-balance22}
&\|\widehat {\bf G}_{\eps, \eta}(t)\|^2_{L^2(\bbT_\eps\times\bbT)}+8\ga_0\int_0^t\|{ \bf G}^{(1)}_{\eps}(s)\|^2_{{\cal
  H}_{0,\eps}}ds+4\ga_1\int_0^t\|{ \bf G}^{(1)}_{\eps}(s)\|^2_{{\cal
  H}_{1,\eps}}ds\\
&
\le \|\widehat {\bf
  G}_{\eta}\|^2_{L^2(\bbT_\eps\times\bbT)}+\delta\int_0^t 
\|\widehat {\bf G}_{\eps, \eta}(s)\|^2_{L^2(\bbT_\eps\times\bbT)} ds+Ct,\quad t\ge0,\,\eps\in(0,1],\,t\ge0.\nonumber
\end{align}
Therefore, for any $\la_0>0$ we can find $C>0$ such that
\begin{align}
\label{energy-balance23}
&\|\partial_\eta\widehat{ \bf f}(\la)\|^2_{L^2(\bbT_\eps\times\bbT)}\le \frac{C}{{\rm Re}\,\la-\la_0},\quad {\rm Re}\,\la>\la_0,\,\eps\in(0,1] 
\end{align}
and the conclusion of Lemma \ref{lm010606-19}  follows.\qed

\subsubsection{Proof of Lemma \ref{lm011006-19}}

We show that for any $\rho>0$ there exists $M>0$ such that 
\begin{align}
\label{051006-19}
\limsup_{\eps\to0+}{\cal I}_\eps(M)<\rho,
\end{align}
where
\begin{align}
\label{051006-19a}
&
{\cal I}_\eps(M):=\int_{[\eta \in\bbT_\eps,\,|\eta|>M]}d\eta
\int_{\bbT^2}dk dk'\left|{\frak s}
   \left(k-\frac{\eps\eta}{2}\right)
{\frak s}
   \left(k+\frac{\eps\eta}{2}\right)
{\frak s}
   \left(k'-\frac{\eps\eta}{2}\right)
{\frak s}   \left(k'+\frac{\eps\eta}{2}\right)
\right.\nonumber\\
&
\times \left. {\frak   s}\left(k+k'-\eps\eta\right){\frak
   s}\left(k+k'+\eps\eta\right)\widehat{ w}_{\eps,+}(\la,\eta,k)
 \widehat{  g}_{\eps,+}^\star(\la,\eta,k')\right|.
\end{align}
The proof in the case of the remaining terms appearing in the expression
\eqref{011006-19}
carries out in a similar fashion. We shall also consider the case of
an optical dispersion relation, that is somewhat more involved than
the accoustic one, as then
the dispersion relation has two critical points.

Recalling well known trigonometric identities we can write 
$$
 {\frak   s}\left(k+k'-\eps\eta\right){\frak
   s}\left(k+k'+\eps\eta\right)=\sum_{\iota_1,\iota_2=0,1}{\frak s}_{\iota_1}
   \left(k'-\frac{\eps\eta}{2}\right) {\frak s}_{1-\iota_1}
   \left(k-\frac{\eps\eta}{2}\right) {\frak s}_{\iota_2}
   \left(k'+\frac{\eps\eta}{2}\right) {\frak s}_{1-\iota_2}
   \left(k+\frac{\eps\eta}{2}\right).
$$
Here ${\frak   s}_0(k):={\frak c}(k)$ and  ${\frak   s}_1(k):={\frak s}(k)$.
Correspondingly, expression \eqref{051006-19a}
 can be rewritten in the form $\sum_{\iota_1,\iota_2\in\{0,1\}}{\cal
  I}_{\iota_1,\iota_2}$. The analysis of each term is similar, so we
only deal with $\iota_1=\iota_2=1$.
The respective expression is of the form
\begin{align}
\label{051006-19b}
&
{\cal I}_{1,1}=\int_{[\eta \in\bbT_\eps,\,|\eta|>M]}d\eta
\int_{\bbT^2}dk dk'\left|\vphantom{\int_0^1}{\frak s}
   \left(k-\eps\eta/2\right) {\frak c}
   \left(k-\eps\eta/2\right){\frak s}
   \left(k+\eps\eta/2\right) {\frak c}
   \left(k+\eps\eta/2\right) \widehat{ w}_{\eps,+}(\la,\eta,k)
\right.\nonumber\\
&
\times \left. {\frak s}^2
   \left(k'-\frac{\eps\eta}{2}\right)
{\frak s} ^2  \left(k'+\frac{\eps\eta}{2}\right)
 \widehat{  g}_{\eps,+}^\star(\la,\eta,k')\right|.
\end{align}
We partition
the domain of  integration  in \eqref{051006-19b} into two sets
$T_{1}$ and $T_2$. To $T_1$ belong all those $(\eta,k,k')$,
for which either 
\begin{equation}
\label{010707-19}
\left|k\pm \frac{\eps\eta}{2}\right|\le
\delta, \quad \mbox{or}\quad   1/2-\delta\le \left|k\pm\frac{\eps\eta}{2}\right|\le
1/2,
\end{equation}
 while to $T_2$ belong all other $(\eta,k,k')$-s.
Parameter 
$\delta\in(0,1/2)$ is to be chosen later on. We can write then 
${\cal I}_{1,1}={\cal I}_{1,1}^{1}+{\cal I}_{1,1}^{2}$, where $
{\cal I}_{1,1}^{i}$ correspond to integration over
$T_{i}$, $i=1,2$.
We can write
\begin{equation}
\label{061106-19}
{\cal I}_{1,1}^{1}\preceq \delta \int_{\bbT_\eps}d\eta
\int_{\bbT^2}dk dk'\left|\widehat{ w}_{\eps,+}(\la,\eta,k)
 \widehat{ g}_{\eps,+}^\star(\la,\eta,k')\right|\preceq \delta,
\end{equation}
for $\eps\in(0,1]$, by virtue of \eqref{060703-19} and \eqref{032105-19}.

Taking the Laplace transforms of both sides of \eqref{011806-19} we
obtain in particular that
\begin{align}
\label{010704-19}
&  \widehat w_{\eps,+}(\la,\eta,k) =\left(\vphantom{\int_0^1}\la +2\ga_0\bar R(k,\eps \eta)+ i\delta_\eps\om(k,
\eta)\right)^{-1}D_{\eps}(\la,\eta,k),
\end{align}
where
\begin{align*}
&
D_{\eps}(\la,\eta,k) 
:=\widehat W_{\eps,+}(\eta,k) -\frac{\ga_1}{2}\left\{\vphantom{\int_0^1}{\frak d}_\eps\left(\la,k-\frac{\eps \eta}{2}\right)+{\frak
   d}_\eps^\star\left(\la,k+\frac{\eps \eta}{2}\right)\right\}
\nonumber\\
&
+\frac{\ga_0}{2}\left(\vphantom{\int_0^1}{\cal L}^{+}_{\eps
                   \eta}\widehat y_{\eps,-}(\la, \eta,k) 
+{\cal L}^{+}_{-\eps \eta}\widehat y_{\eps,+}(\la,
  \eta,k)\right) -
2\ga_0{\cal R}_{\eps \eta}\widehat w_{\eps,+}
                   (\la,\eta,k).\nonumber
\end{align*}
Hence,
\begin{align}
\label{021106-19}
& {\cal I}_{1,1}^2=\int_{T_2}d\eta
dk dk'\left|\frac{{\frak s}
   \left(k-\eps\eta/2\right) {\frak c}
   \left(k-\eps\eta/2\right){\frak s}
   \left(k+\eps\eta/2\right) {\frak c}
   \left(k+\eps\eta/2\right)  d_{\eps}(\la,\eta,k)}{\la+2\ga_0\bar R(k,\eps\eta)+i\delta_{\eps}\om(k,\eta)}
\right.\nonumber\\
&
\times \left. 
{\frak s}^2
   \left(k'-\frac{\eps\eta}{2}\right)
{\frak s}^2   \left(k'+\frac{\eps\eta}{2}\right) \widehat{ g}_{\eps,+}^\star(\la,\eta,k')\vphantom{\int_0^1}\right|.
\end{align}
Thanks to \eqref{032105-19} and \eqref{121505-19a} there exists
$\la_0$ such that 
\begin{equation}
\label{011106-19}
d_{*}:=\sup_{\eps\in
  (0,1],\,\eta\in\bbT_{\eps}}\|D_{\eps}(\la,\eta,\cdot)\|_{L^2(\bbT)}<+\infty,\quad
\mbox{for }{\rm Re}\,\la>\la_0.
\end{equation}
Since $\om'(k)\not=0$, except for $k=0,1/2$ we can find $c_*(\delta)>0$
such that
$$
|\la+2\ga_0\bar R(k,\eps\eta)+i\delta_{\eps}\om(k,\eta)|\ge
\la+c_*(\delta)|\eta|\quad \mbox{ for $(\eta,k)$ such that
  \eqref{010707-19} does not hold}.
$$
Therefore
\begin{align}
\label{021106-19a}
& {\cal I}_{1,1}^2\le \int_{[\eta \in\bbT_\eps,\,|\eta|>M]}d\eta
\int_{\bbT^2}\frac{|D_{\eps}(\la,\eta,k) \widehat{ 
  g}_{\eps,+}^\star(\la,\eta,k')|}{\la+c_*(\delta)|\eta|}dk dk'\\
&
\le \left\{\int_{[|\eta|>M]}\frac{d\eta }{(\la+c_*(\delta)|\eta|)^2}
\sup_{\eps\in(0,1],\eta'\in\bbT_\eps}\int_{\bbT}|D_{\eps}(\la,\eta',k)|^2dk\right\}^{1/2}
  \|{ 
  g}_{\eps,+}(\la)\|_{{\cal L}_{2,\eps}}.\nonumber
\end{align}
In light of \eqref{060703-19} for any $\rho>0$ we can choose a sufficiently large $M$
so that
$\limsup_{\eps\to0+}{\cal I}_{1,1}^2\le{\rho}/{2}$.
Adjusting suitably $\delta>0$, cf \eqref{061106-19}, we have also
$\limsup_{\eps\to0+}{\cal I}_{1,1}^1\le{\rho}/{2}$. Combining these
two estimates we conclude that there exists $\la_0$ such that for any
$\rho>0$ we can find $M>0$ for which
$$
\limsup_{\eps\to0+}{\cal I}_{1,1}<\rho\quad \mbox{for all }{\rm Re}\,\la>\la_0
$$ and the
conclusion of the lemma follows.\qed

\section{The case of arbitrary thermostat temperature $T$}

\label{sec10}
Setting the thermostat temperature at $T$ leads to the following
dynamics of the Wigner functions (cf \eqref{011806-19})
\begin{equation}
\label{011806-19a}
\frac{d}{dt}\widehat \bW_\eps(t,\eta,k)=\left(\ga_0\widehat{\frak
    L}_{\eps\eta}+\widehat{\frak H}_\eps+\ga_1\widehat{\frak
    T}_\eps\right)\widehat \bW_\eps(t,\eta,k) +\frac{\ga_1T}{\eps}{\bf D}.
\end{equation}

\noindent
Suppose that $\chi\in C^\infty_c(\bbR)$ is an arbitrary real valued,
even function satisfying   \eqref{011406-19}.
Then  $\widehat \chi\in {\cal S}(\bbR)$ and let 
 $\widehat \chi_\eps\in C^\infty(\bbT_{\eps})$ be given by 
$$
\widehat \chi_\eps(\eta):=\sum_{n\in\mathbb Z}\widehat
\chi\left(\eta+\frac{2n}{\eps}\right),\quad \eta\in\bbT_{\eps}.
$$
Note that
$$
\int_{\bbT_{\eps}}\widehat \chi_\eps(\eta)d\eta=\int_{\bbR}\widehat \chi(n)d\eta=\chi(0)=1.
$$
Define 
$$
\widehat {\bf V}_\eps(t,\eta,k)=\left[\begin{array}{cc}
 \widehat V_{\eps,+}(t,\eta,k)&\widehat U_{\eps,+}(t,\eta,k)\\
 \widehat U_{\eps,-}(t,\eta,k)&\widehat V_{\eps,-}(t,\eta,k)
 \end{array}\right]:=\widehat {\bf W}_\eps(t,\eta,k)-T \widehat
\chi_\eps(\eta){\bf I}_{2},
$$
where ${\bf I}_2$ is the $2\times 2$ identity matrix,
 i.e.
$$
\widehat V_{\eps,\pm}(t,\eta,k):=\widehat W_{\eps,\pm}(t,\eta,k)-T \widehat
\chi_\eps(\eta),\quad \widehat U_{\eps,\pm}(t,\eta,k):=\widehat Y_{\eps,\pm}(t,\eta,k),\quad  t\ge0,\,(\eta,k)\in\bbT_{\eps/2}\times\bbT.
$$

It satisfies 
\begin{equation}
\label{011806-19b}
\frac{d}{dt}\widehat {\bf V}_\eps(t,\eta,k)=\left(\ga_0\widehat{\frak
    L}_{\eps\eta}+\widehat{\frak H}_\eps+\ga_1\widehat{\frak
    T}_\eps\right)\widehat {\bf V}_\eps(t,\eta,k)+\widehat {\bf F}_\eps(\eta,k), 
\end{equation}
where
$$
\widehat {\bf F}_\eps(\eta,k):=-i\delta_{\eps}\om(k;\eta)T \widehat
\chi_\eps(\eta){\bf J}_2
$$
and
$$
{\bf J}_2:= 
\left[\begin{array}{cc}
 1&0\\
 0&-1
 \end{array}\right].
$$
The solution can be then written in the form, cf \eqref{frakW},
\begin{equation}
\label{111206-19}
{\bf V}_\eps(t)={\frak W}_\eps(t){\bf V}_\eps(0)+\int_0^t {\frak
  W}_\eps(s){\bf F}_\eps ds.
\end{equation}

Using the already proved part of Theorem \ref{main-thm} for $T=0$, we conclude that for any ${\bf
  G}\in L^1([0,+\infty),{\cal A})$ we have
\begin{equation}
\label{121206-19}
\lim_{\eps\to0+}\int_0^{+\infty}\langle {\bf V}_\eps(t),{\bf G}(t)\rangle dt=\int_0^{+\infty}\langle {\bf V}(t),{\bf G}(t)\rangle dt,
\end{equation}
where
$$
{\bf V}(t,y,k)= 
\left[\begin{array}{cc}
 V_+(t,y,k)&0\\
 0&V_+(t,y,-k)
 \end{array}\right]
$$
and
\begin{equation}
\label{111206-19a}
V_+(t,y,k):={\frak W}(t){ V}_{0,+}(y,k)-\int_0^t {\frak
  W}(s)F(y,k)ds.
\end{equation}
Here
$
{ F}$ is given by \eqref{eqF} and ${ V}_{0,+}(y,k):=W_{0,+}(y,k)-T\chi(y)$. 
 This ends the proof of Theorem \ref{main-thm} for an arbitrary $T\ge0$.
\qed

\appendix

\section{Proof of Proposition \ref{prop011406-19}}

\label{appa}

Using formula \eqref{010304} we can see that ${\frak W}^{\rm un}_t({\cal
  C}_0')\subset {\cal C}_0'$, $t\ge0$.  In addition (see Section
\ref{sec2.6.2}) $\left({\frak W}^{\rm un}_t\right)_{t\ge0}$, given by \eqref{010304},
is a $C_0$-semigroup of contractions on $L^2(\bbR\times\bbT)$.
According to Section A, of the Appendix of \cite{koran} the semigroup
$\left({\frak W}_t\right)_{t\ge0}$ is defined by the Duhamel series that
corresponds to the equation \eqref{integral}. Since ${\cal R}$ is a
bounded operator on $L^2(\bbR\times\bbT)$ and $\left({\frak W}^{\rm
    un}_t\right) _{t\ge0}$ is a semigroups of contractions, the semigroup
defined by the series is a $C_0$-semigroup of bounded
operators on $L^2(\bbR\times\bbT)$. From here we conclude also that ${\cal
  C}_0'$ has to be invariant under $\left({\frak W}_t\right)_{t\ge0}$. For
$W_0\in {\cal C}_0'$ we conclude by a direct calculation that
$W(t,y,k):={\frak W}_t(W_0)$ satisfies the following identity
\begin{align}
  \label{eq:14a}
&  \frac12 \frac{d}{dt} \|W(t)\|_{L^2(\bbR\times\bbT)}^2 = -\gamma_0\int_{\bbR\times\bbT^2} R(k,k')\left[ W(t,y,k) -W(t,y,k')\right]^2  dydkdk'\\
  &
   - \frac{1}{2} \int_{\bbT} \;
    \bar\omega'(k)  \left[W(t,0^-, k)^2-W(t,0^+, k)^2
    \right] dk ,\quad t\ge0.\nonumber
\end{align}
 Taking into account \eqref{feb1408} and \eqref{feb1410} we obtain
\begin{align}
\label{010709-19}
&\int_{\bbT} \; \bar\omega'(k)  \left\{ [W(t,0^-, k)]^2-[W(t,0^+, k)]^2 \right\} dk
\\
&
=\int_{\bbT_+} \; \bar\omega'(k)  \left\{ [W(t,0^-, k)]^2-\left[p_-(k) W(t,0^+, -k)+p_+(k) W(t,0^-,k)\right]^2 \right\} dk\nonumber\\
&
+\int_{\bbT_-} \; \bar\omega'(k)  \left\{ \left[p_-(k) W(t,0^-,-k) + p_+(k) W(t,0^+, k)\right]^2-[W(t,0^+, k)]^2 \right\} dk.\nonumber
\end{align}
After straightforward calculations (recall that coefficients $p_\pm(k)
 $ are even, while $\bar\omega'(k)$ is odd) we conclude that the right hand
 side equals
\begin{align*}
                 \int_{\bbT_+} \; \bar\omega'(k) & \left\{ \left(W(t,0^-, k)^2+ W(t,0^+, -k)^2\right)
                 \left(1-p_+^2(k)-p_-^2(k)\right)\right. \\
&
\left.-
4p_-(k)p_+(k)W(t,0^+, -k)W(t,0^-,k)\right\} dk.
\end{align*}
 Since
 $p_+(k)+p_-(k)\le 1$
 we have $1-p_+^2(k)-p_-^2(k)\ge 0$. In addition,
\begin{align*}
 &
 {\rm det}
 \left[
 \begin{array}{ll}
 1-p_+^2(k)-p_-^2(k)&-2p_-(k)p_+(k)\\
 &\\
 -2p_-(k)p_+(k)&1-p_+^2(k)-p_-^2(k)
 \end{array}
 \right]
 \\
 &
 =\left[1-(p_+(k)+p_-(k))^2\right]\left[1-(p_+(k)-p_-(k))^2\right]\ge0.
 \end{align*}
 Using \eqref{012304} we conclude that the  quadratic form
 \begin{equation}
\label{020709-19}
 (x,y)\mapsto \left(1-p_+^2(k)-p_-^2(k)\right)(x^2+y^2)-4p_-(k)p_+(k)xy
 \end{equation}
 is non-negative definite (since $p_+(k)+p_-(k)\le1$). 
Hence, in particular 
$$\frac{d}{dt}
\|W(t)\|_{L^2(\bbR\times\bbT)}^2\le0,\quad t\ge0,
$$ which in turn proves that
$\left({\frak W}_t\right)_{t\ge0}$ is a semigroup of contractions.\qed

\section{Outline of the proof of Theorem  \ref{main:thm-un}}

\label{appb}

The proof of Theorem \ref{main:thm-un}, for the most part,  follows
closely the argument contained in \cite{kors}. We shall present here its
outline, invoking the relevant parts of \cite{kors} and focus on
the necessary modifications. We start with the following.
\begin{prop}
\label{thm021405-19}
Suppose that the initial data satisfies the assumptions of Theorem $\ref{main-thm}$. Then,
\begin{equation}
\label{031405-19}
\lim_{\eps\to0+}\int_{\bbR\times\bbT}\widehat y^{\rm
  un}_{\eps,\iota}(\la,\eta,k)\widehat G^\star(\eta,k)d\eta dk=0,\quad \iota=\pm
\end{equation}
for any ${\rm Re}\,\la>0$ and $G\in {\cal A}$.
\end{prop}
The proof of the proposition follows the argument presented in Section
\ref{sec5.5.1}, with the simplification consisting in the fact that
the equation of $\widehat y^{\rm
  un}_{\eps,+}(\la,\eta,k)$ corresponding to \eqref{020911ab1z} does
not contain the scattering terms involving the operator ${\cal R}_{\eps\eta}$.

\bigskip

Next, we show that
\begin{equation}
\label{011605-19}
\lim_{\eps\to0+}\int_{\bbR\times\bbT}\widehat w^{\rm
  un}_{\eps,+}(\la,\eta,k)\widehat G^\star(\eta,k)d\eta dk=\int_{\bbR\times\bbT}\widehat w^{\rm
  un}_{+}(\la,\eta,k)\widehat G^\star(\eta,k)d\eta dk,
\end{equation}
where $w^{\rm
  un}_{+}(\la)$ is given by \eqref{020911absz} and $G\in
{\cal A}_c$.

We let ${\rm supp}\,\hat
G\subset[-K,K]\times\bbT$.
Taking the Laplace transforms of both sides of \eqref{frakWun} 
we
obtain in particular the equation 
\begin{align}
\label{010712-19}
&  \left(\vphantom{\int_0^1}\la +2\ga_0\bar R(k,\eps \eta)+ i\delta_\eps\om(k,
\eta)\right)\widehat w^{\rm un}_{\eps,+}(\la,\eta,k)
  +\ga_0R\left(k-\frac{\eps \eta}{2}\right)  \widehat y^{\rm
  un}_{\eps,+} (\la, \eta,k)+\ga_0R\left(k+\frac{\eps \eta}{2}\right)
  \widehat  y^{\rm un}_{\eps,-}(\la, \eta,k)
\\
&
=\widehat W_{\eps,+}(0,\eta,k) -\frac{\ga_1}{2}\left\{\vphantom{\int_0^1}{\frak d}_\eps ^{\rm un}\left(\la,k-\frac{\eps \eta}{2}\right)+\left({\frak
   d}_\eps ^{\rm un}\right)^\star\left(\la,k+\frac{\eps \eta}{2}\right)\right\}.\nonumber
\end{align}
Here (cf \eqref{091505-19})
\begin{align}
\label{fdeps}
&{\frak
   d}_{\eps}^{\rm un}(\la,k):=\int_0^{+\infty}e^{-\la
  t}d_\eps(t,k)ds=\frac{1}{2}\int_{\bbT_{\eps}}\left[\widehat
  y_{\eps,+}^{\rm un}\left(\la, \eta,
  k-\frac{\eps\eta}{2}\right)-\widehat
  w_{\eps,+}^{\rm un}\left(\la,\eta, k-\frac{\eps\eta}{2}\right)\right]d\eta .
\end{align}
Thanks to \eqref{1015-5-19} we conclude
\begin{equation}
\label{121505-19}
\|{\frak d}_\eps^{\rm un}(\la)\|_{L^2(\bbT)}\le \frac{\|{ \bW}_\eps(0)\|_{{\cal
  L}_{2,\eps}}}{(2^5 \ga_1{\rm Re}\,\la)^{1/2}},\quad
\eps\in(0,1],\,{\rm Re}\,\la>0.
\end{equation}

  Suppose that $\eps_n\to0$ as a sequence that corresponds to a
  $\vphantom{1}^\star$-weakly convergent  in
${\cal A}'$  subsequence  $\left(w^{\rm un}_{\eps_n,+}(\la,\cdot)\right)$ and $[-K,K]\subset [-\eps_n^{-1},\eps_n^{-1}]$, $n\ge1$.
By virtue of Proposition \ref{prop011505-19} families $\left(\widehat w^{\rm
    un}_{\eps_n,+}(\la)\chi_K\right)$ (see \eqref{chik}) and $\left({\frak
   d}_{\eps_n}\chi_K\right)$ are $\vphantom{1}^\star$-weakly compact in
$L^2(\bbR\times\bbT)$.
Let  $\widehat w^{\rm un}_{+}(\la,\cdot)$ and ${\frak
   d}(\la,\cdot)$ be their respective limits.
Multiplying equation \eqref{010712-19} by $\chi_K$ and letting
$\eps\to0+$ we conclude that
\begin{align}
\label{020911abs}
&  \left(\vphantom{\int_0^1}\la +2\ga_0 R(k)+
  i\om'(k)\eta\right)\widehat w^{\rm un}_{+}=\widehat W_0(\eta,k)
  -\ga_1{\rm Re}\,{\frak d}^{\rm un}\left(\la,k\right),
\end{align}
where
$$
\int_{\bbT}{\rm Re}\,{\frak
  d}^{\rm un}\left(\la,k\right) \widehat G^\star(k)dk=\lim_{\eps\to0+}\int_{\bbT}{\rm
  Re}\,{\frak d}^{\rm un}_\eps\left(\la,k\right) \widehat G^\star(k)dk,\quad \widehat G\in C^\infty(\bbT).
$$
The
conclusion of Theorem \ref{main:thm-un} follows from.
\begin{thm}
\label{thm010504-19}
For any $\la>0$ we have
\begin{equation}
\label{013110}
{\rm Re}\,{\frak d}(\la,k)=(1-p_+(k))\int_{\bbR}\frac{ \widehat W_0(\eta,k)d\eta dk}{\la+2\ga_0 R(k)+i\om'(k)\eta}-p_-(k) \int_{\bbR}\frac{ 
\widehat W_0(\eta,-k)d\eta }{ \la+2\ga_0 R(k)-i\om'(k)\eta},\quad k\in\bbT.
\end{equation}
\end{thm}

\subsection{Proof of Theorem \ref{thm010504-19}}

To simplify somewhat  our presentation we shall assume   that
\begin{equation}
\label{null}
\langle\hat\psi(k)\hat\psi(\ell) \rangle_{\mu_\eps}=0,\quad k,\ell\in\bbT.
\end{equation}
The result remains
valid without this hypothesis, although the calculations become more
extensive.
Assumption \eqref{null} results in the condition
\begin{equation}
\label{nullY}
\widehat Y_{\eps,\pm}(0,\eta,k)=0,\quad (\eta,k)\in\bbR\times \bbT.
\end{equation}

Using \eqref{eq:10}
we may write
\begin{equation}
\label{d-12}
{\frak d}_\eps^{\rm un}\left(\la,k\right)=  {\frak  d}_\eps^1\left(\la,k\right)+{\frak  d}_\eps^2\left(\la,k\right).
\end{equation}
Here, ${\frak d}_\eps^j\left(\la,k\right)$, $j=1,2$ are the  respective
Laplace transforms of  
\begin{equation}
\label{012703}
I_\eps(t,k):= i \int_0^t\left\langle { p}^0_0(t-s) {\frak
    e}_1\cdot e_{\Om_\eps}(k,t)\hat\Psi(k) \right\rangle_{\mu_\eps}g_\eps(ds),
\end{equation}
$$
 I\!I_\eps(t,k):=  
    - \gamma_1  \int_0^{t}  g_\eps\left(ds'\right) \int_0^{t} \Theta\left(t-s,k\right)\langle{ p}_0^0(s){ p}_0^0(t-s')\rangle_{\mu_\eps}ds,
$$
Here $e_{\Om_\eps}(k,t)$ is given by \eqref{011902-19a} and 
\[
p_0^0(t)=\frac{1}{2i}\int_{\bbT} e_{\Om_\eps}(k,t)\hat\Psi(k) \cdot {\frak f}dk,\quad
   \Theta(t,k) := \int_0^{t}{\frak
    e}_1\cdot  e_{\Om_\eps}(k,t-\tau){\frak f} g_\eps(d\tau) .
\]
We  introduce 
\begin{equation}\label{feb1416}
{\frak L}^\eps_{\rm scat}(\la):=-\ga_1\int_{\bbT}\hat G^*(k) {\rm Re}\,{\frak
d}_\eps\left(\la,k\right)dk =\sum_{j=1}^2{\frak L}_{{\rm scat},j}^\eps(\la).
\end{equation}  
with
\begin{equation}
{\frak L}_{{\rm scat},j}^{\eps}(\la):=-\ga_1
\int_{\bbT}\hat G^*(k) {\rm Re}\,{\frak d}_\eps^j\left(\la,k\right)dk,
~~j=1,2.
\end{equation}

\subsubsection{The limit of ${\frak L}_{{\rm scat},1}^{\eps}(\la)$}

We will show the following.
\begin{lm} \label{lem-feb1504}
For any test function $\widehat G\in C^\infty(\bbT)$ and
  $\la>0$ we have
\begin{equation}
\label{020811}
\lim_{\eps\to0+}{\frak L}_{{\rm scat},1}^\eps(\la)=
 -\ga_1 \int_{\bbT}{\rm Re} [\nu(k)]\widehat G^*(k)dk 
 \int_{\bbR}\frac{\widehat W_0(\eta,k)d\eta}{\la+2\ga_0R(k)+i\om'(k)\eta} 
d\eta'.
\end{equation}
\end{lm}
\proof
From \eqref{012703} and \eqref{null} we get
\begin{align}
\label{Ibis}
&I_\eps(t,k)= \frac{1}{2}\left\{ e_{\Om_\eps}^{1,1}(k,t)\int_0^{t} g_\eps\left(ds\right) \int_{\bbT}
  e_{\Om_\eps}(\ell,t-s){\frak e}_1\cdot {\frak f}\langle\hat\psi^\star(k)\hat\psi(\ell) \rangle_{\mu_\eps}
\right.\nonumber\\
&
\left.+e_{\Om_\eps}^{1,2}(k,t)\int_0^{t} g_\eps\left(ds\right) \int_{\bbT}
  e_{\Om_\eps}(\ell,t-s){\frak e}_1\cdot {\frak f}\langle\hat\psi(-k)\hat\psi^\star(-\ell) \rangle_{\mu_\eps}\right\} d\ell.
\end{align}
Using formula \eqref{011902-19a}
we can write
 \begin{align}
\label{010705-19}
&
 e_{\Om_\eps}^{1,1}(k,t)=e^{\la_+(k)
  t}+\eps\sum_{\iota=\pm} r^{1}_{\iota,\eps}(k)e^{\la_\iota
  t},\quad 
e_{\Om_\eps}^{1,2}(k,t)=\eps\sum_{\iota=\pm} r^{2}_{\iota,\eps}(k)e^{\la_\iota
  t},\\
&
 e_{\Om_\eps}(\ell,t-s){\frak e}_1\cdot {\frak f}=e^{\la_-(\ell)
  (t-s)}+\eps\sum_{\iota=\pm} r^{3}_{\iota,\eps}(\ell)e^{\la_\iota(\ell)
  (t-s)},\nonumber
\end{align}
where 
\begin{equation}
\label{020304-19}
 \sup_{k\in\bbT,\eps\in(0,1]}|r^{j}_{\iota,\eps}(k)|=r^{j}_{\iota,*}<+\infty,\quad,
 \iota\in\{-,+\},\,j=1,2,3.
\end{equation}
The following result allows us to replace the entries of
$e_{\Om_\eps}(k,t)$ by the leading terms appearing in \eqref{010705-19}.
Define
\begin{align}
\label{Ibis1}
&\tilde I^{\iota_1,\iota_2}_\eps(t,k):= \eps r_{1,\eps}(k)
  e^{\la_{\iota_1}(k) t}\int_0^{t} g_\eps\left(ds\right)
  \int_{\bbT}r_{2,\eps}(\ell) e^{\la_{\iota_2}(\ell)
  (t-s)}\langle\hat\psi^\star(k)\hat\psi(\ell)
  \rangle_{\mu_\eps}d\ell
\end{align}
and
$$
\tilde I^{\iota_1,\iota_2}_\eps(\la,k):=
\int_0^{+\infty}e^{-\la t}\tilde
 I^{\iota_1,\iota_2}_\eps\left(\frac{t}{\eps},k\right)dt .
$$
\begin{lemma}
\label{lm010304-19}
Suppose that \eqref{mu-eps} holds, 
\begin{equation}
\label{020304-19a}
 \sup_{k\in\bbT,\eps\in(0,1]}|r_{i,\eps}(k)|=r_{i,*}<+\infty,\quad i=1,2.
\end{equation}
Then, for any $\iota_1,\iota_2\in\{-,+\}$ and $\la>0$ we have
\begin{equation}
\label{020304-19a}
\Big\|\tilde I^{\iota_1,\iota_2}_\eps(\la)\Big\|_{L^1(\bbT)}\preceq
\eps^{1/2}\log\eps^{-1},\quad \eps\in(0,1].
\end{equation}
\end{lemma}
The proof of this lemma is presented in Section \ref{C3}

Using Lemma \ref{lm010304-19} we conclude that
\begin{eqnarray*}
&&\int_{\bbT}{\frak d}_\eps^1(\la,k)
 \widehat G^\star(k)dk =\frac\eps2\int_{\bbT^2}\eps \langle\hat \psi^\star (k)\hat \psi(\ell)
   \rangle_{\mu_\eps}\widehat G^\star(k)d\ell dk
\\
&&
\times\int_0^{+\infty}\exp\left\{-\la_-(\ell)s\right\}g_\eps(ds)
\int_s^{+\infty}e^{-\la\eps t}
\exp\left\{[\la_+(k)+\la_-(\ell)]t\right\}dt+O(\eps)
\\
&&
=\frac12\int_{\bbT^2}\eps \langle\hat \psi^\star (k)\hat \psi(\ell)
   \rangle_{\mu_\eps} \widehat G^\star(k)\tilde g_\eps\Big(\eps(\la+\ga_0R(k))-i\om_\eps(k) \Big)d\ell
  dk
\\
&&
\times\left\{\eps(\la+\ga_0R(k)+\ga_0R(\ell))+i[\om_\eps(\ell) -\om_\eps(k) ]\right\}^{-1}
+O(\eps)
\end{eqnarray*}
Changing variables
$
k:=k'-\eps \eta'/2,$ $\ell:=k'+\eps \eta'/2$
we can write (cf  \eqref{012302-19} and \eqref{030904-19})
\begin{equation}
\label{050404-19}
\int_{\bbT}{\frak d}_\eps^1(\la,k)
 \widehat G^\star(k)dk 
={\cal I}_\eps
+O(\eps),
\end{equation}
where
$$
{\cal I}_\eps:=\int_{T_\eps}\widehat
   W_{\eps,+}(0,\eta,k)\tilde g_\eps\Big(\eps(\la+\ga_0R(k-\eps\eta/2))-i\om_\eps(k-\eps\eta/2) \Big) \Delta_\eps(k,\eta) d\eta
  dk.
$$
Here  (cf \eqref{bar-f} and \eqref{d-om})
\begin{align*}
&\Delta_\eps(k,\eta):=
\left\{\la+2\ga_0\bar R(k,\eps\eta)+i\delta_\eps\om_\eps(k;\eta) \right\}^{-1}\widehat G^\star
 (k-\eps\eta/2)
\end{align*}
and $T_\eps$ is the image of $\bbT^2$ under the inverse map
$
k':=(\ell+k)/2,$ $\eta':=(\ell-k)/\eps$.

We claim that in fact
\begin{equation}
\label{020705-19}
\lim_{\eps\to0+}({\cal I}_\eps-{\cal I}_\eps')=0,
\end{equation}
where the definition of the expression ${\cal I}_\eps'$ differs from ${\cal I}_\eps$
only by replacing $\tilde g_\eps$ by $\tilde g$.
Changing variable $k\mapsto k-\eps\eta/2$ we can write
$$
{\cal I}_\eps-{\cal I}_\eps'=\int_{T_\eps'}\widehat
   W_{\eps,+}(0,\eta, k+\eps\eta/2)(\tilde g_\eps-\tilde g)\Big(\eps(\la+\ga_0R(k))-i\om_\eps(k) \Big) \Delta_\eps\left(k+\eps\eta/2,\eta\right) d\eta
  dk.
$$
Here $T_\eps'$ is the impage of $T_\eps$ under the change of
variable.
Equality \eqref{020705-19} follows 
from Proposition \ref{cor010304-19}  and the fact that the expression
under the integral in the right hand side is bounded by an integrable
function, see \eqref{011812aa}. 

Next, thanks to Lemma 7.3 of \cite{kors}, we have
\begin{align*}
&\lim_{\eps\to0}\tilde g\Big(\eps(\la+\ga_0R(k-\eps\eta/2))-i\om_\eps(k-\eps\eta/2)  \Big) \Delta_\eps(k,\eta) 
  =\frac{\nu(k)\widehat
   G^\star(k)}{\la+2\ga_0R(k)+i\om'(k)\eta}
\end{align*}
a.e. in $(\eta,k)$. Using bounds \eqref{011812aa} and
\eqref{G} we can argue, via the dominated convergence theorem, as in
the proof of Lemma 5.1 of \cite{kors}, 
that 
\begin{equation}
\label{020706b}
\lim_{\eps\to0+}{\cal I}_\eps'=\int_{\bbR\times \bbT} \frac{\widehat
   W_0(\eta,k) \nu(k)
   \widehat
   G^*(k) d\eta dk}{\la+2\ga_0R(k)+i\om'(k)\eta}
\end{equation}
and conclusion of  Lemma \ref{lem-feb1504} follows.\qed


Concerning the second term in the utmost right hand side of
\eqref{feb1416}
we have the following.
\begin{lemma}\label{lem-feb1502}
For any $\la>0$ and $G\in C^\infty(\bbT)$ we have
\begin{equation}
\label{021511}
\lim_{\eps\to 0}{\frak L}_{scat,2}^\eps(\la)=
\frac{\gamma_1}{4}\sum_{\iota=\pm}\int_{\bbT}\frac{\hat G^\star(k)\fgeeszett(k)\widehat
  W_0(\eta,\iota k)d\eta dk}{\la+2\ga_0R(k)+\iota i\om'(k)\eta} 
 .
\end{equation}
\end{lemma}
The proof of the lemma follows very closely the proof of Lemma 6.4 of 
\cite{kors}. We present its outline in Section \ref{sec:6} below.

\subsection{The limit of ${\frak L}_{{\rm scat}}^{\eps}(\la)$}

\label{sec:4.3}


Putting together the results of Lemmas~\ref{lem-feb1504} and \ref{lem-feb1502},
we see that
\begin{equation} 
\lim_{\eps\to 0}{\frak L}_{scat}^{\eps}(\la)=\int_{\bbT} \Big({\cal W}_{tr}(k)+{\cal W}_{ref}(k)\Big)\hat G^*(k)
dk,
\label{feb1506}
\end{equation}
with the transmission term
\begin{eqnarray}\label{feb1508}
&&{\cal W}_{tr}(k)=\farc{\gamma_1}{|\bar\omega'(k)|}
\Big\{- {\rm Re}[\nu(k)]+\frac{\fg(k) }{4}\Big\}
\int_{\bbR}\frac{ \widehat W_0(\eta,k)d\eta }{\la+2\ga_0
  R(k)+i\om'(k)\eta}\\
&&
=
(p_+(k)-1)\int_{\bbR}\frac{ \widehat W_0(\eta,k)d\eta }{\la+2\ga_0 R(k)+i\om'(k)\eta}.\nonumber
\end{eqnarray}
We have used (\ref{feb1402}) in the last step. The reflection term ${\cal
  W}_{ref}(k)$  equals (cf \eqref{033110})
\begin{eqnarray}\label{feb1510}
{\cal W}_{ref}(k)=\frac{\gamma_1\fg(k)}{4|\bar\omega'(k)| }
\int_{\bbR}\frac{ 
\widehat W_0(\eta,-k)d\eta }{ \la+2\ga_0 R(k)-i\om'(k)\eta} =
p_-(k) \int_{\bbR}\frac{ 
\widehat W_0(\eta,-k)d\eta }{ \la+2\ga_0 R(k)-i\om'(k)\eta}.
\end{eqnarray}
Combining the scattering terms in
(\ref{feb1506})-(\ref{feb1510})
we obtain (\ref{013110}). Thus, the proof of Theorem~\ref{thm010504-19}
is reduced to showing Lemma~\ref{lem-feb1502}.

\subsection{Outline of the proof of Lemma~\ref{lem-feb1502}: 
the limit of ${\frak L}_{scat,2}^{\eps}(\la)$}

\label{sec:6}

Let
\begin{align}
\label{Ibis1a}
&\widetilde{I\!I}^{\iota_1,\iota_2,\iota_3}_\eps(t,k):= \eps r_{1,\eps}(k)
 \int_0^{t} ds \int_0^{t} g_\eps\left(ds'\right)\int_{0}^{s}g_\eps\left(ds_1\right) 
 \\
&
\times \int_{\bbT^2}r_{2,\eps}(\ell) r_{3,\eps}(\ell')  e^{\la_{\iota_2}(\ell) (s-s_1)} e^{\la_{\iota_3}(\ell')
  (t-s')} e^{\la_{\iota_1}(k)
  (t-s)}\langle\hat\psi^\star(\ell)\hat\psi(\ell')
  \rangle_{\mu_\eps}d\ell d\ell'\nonumber
\end{align}
and
\begin{equation}
\label{020504-19a}
 \widetilde{
 I\!I}^{\iota_1,\iota_2,\iota_3}_\eps (\la,k):=\int_0^{+\infty}e^{-\la t}\widetilde{
 I\!I}^{\iota_1,\iota_2,\iota_3}_\eps\left(\frac{t}{\eps},k\right)dt.
\end{equation}
We start with the following.
\begin{lemma}
\label{lm010504-19}
Suppose that condition \eqref{011812aa} holds,
\begin{equation}
\label{020504-19a}
 \sup_{k\in\bbT,\eps\in(0,1]}|r_{i,\eps}(k)|=r_{i,*}<+\infty,\quad i=1,2,3.
\end{equation}
Then, for any $\iota_1,\iota_2,\iota_3\in\{-,+\}$ and ${\rm Re}\,\la>0$ we have
\begin{equation}
\label{020304-19aa}
\lim_{\eps\to0+} \Big\|\widetilde{
 I\!I}^{\iota_1,\iota_2,\iota_3}_\eps (\la)\Big\|_{L^1(\bbT)} =0.
\end{equation}
\end{lemma}
The proof of the lemma is presented in Section \ref{C3.1} below.

Using Lemma \ref{lm010504-19} we can write
\begin{equation}
\label{020904-19}
{\frak L}_{scat,2}^\eps(\la)=\bar{\frak L}_{scat,2}^\eps(\la)+o(1),
\end{equation}
as $\eps\to0+$, where
\begin{equation}
\label{010904-19}
\bar{\frak L}_{scat,2}^\eps(\la):=-\ga_1\int_{ \bbT}{\rm Re}\,\bar{\frak
   d}_\eps^2\left(\la,k\right)\hat
   G^\star(k) dk
\end{equation}
and
\begin{eqnarray*}
 \bar{\frak
   d}_\eps^2\left(\la,k\right):=  - \gamma_1\eps
    \int_0^{+\infty}e^{-\eps\la t}dt\left\langle \bar p_0^0\star
    g_\eps\left(t\right) \int_0^{t}
    \bar p^0_0 \star g_\eps(s)ds\right\rangle_{\mu_\eps}e^{\la_+(k)
  (t-s)}.
\end{eqnarray*}
Here
\begin{align}
 &
\label{eq:1}
\bar p^0_0(t):=\int_{\bbT} e^{-\eps\ga_0 R(k)t}{\rm Im}\left(\hat\psi(k) e^{-i\omega(k) t}\right) dk.
\end{align}

Since $ \bar p_0^0\star
    g_\eps$ is real valued we have
\begin{equation}
\label{021207-19}
{\rm Re}\,\bar{\frak
   d}_\eps^2\left(\la,k\right):=  - \gamma_1\eps
    \int_0^{+\infty}e^{-\eps\la t}dt\left\langle \bar p_0^0\star
    g_\eps\left(t\right) \int_0^{t}
    \bar p^0_0 \star g_\eps(s)ds\right\rangle_{\mu_\eps}e^{-\ga_0\eps R(k) (t-s)}\cos(\om_\eps(k)(t-s)).
\end{equation}
After rather lengthy, cut starightforward calculation, see Section
\ref{d-eps2} below for details, we get
\begin{equation}
\label{010104}
2{\rm Re}\,\bar{\frak d}_\eps^2(\la,k)
=R_\eps(\la,k)+\rho_\eps(\la,k),
\end{equation}
with
\begin{eqnarray}
\label{R-eps}
&&
R_\eps(\la,k):=-\frac{\gamma_1 (\la+2\ga_0\eps R(k))}{2^3\cdot\pi\eps^2}\int_{\bbR}\frac{d\xi}{(\la/2+\ga_0R(k))^2+\xi^2} \int_{\bbT^2}d\ell d\ell' \eps\langle \hat\psi(\ell)\hat\psi^*(\ell')\rangle_{\mu_\eps}\\
&&
\times \frac{|\tilde g_\eps\left(\la\eps/2-i[\eps\xi+\om_\eps(k)]\right)|^2 }{\la/2-i\{\xi+\eps^{-1}[\om_\eps(k)-\om_\eps(\ell)]\}}
\times \frac{1
   }{\la/2+i\{\xi+\eps^{-1}[\om_\eps(k)-\om_\eps(\ell')]\}}\nonumber
\end{eqnarray}
and
\begin{eqnarray}
\label{rho-e}
&&
\rho_\eps(\la,k):=-\frac{\eps\gamma_1 (\la+2\ga_0\eps R(k))}{2^4\cdot\pi}\int_{\bbR}\frac{d\xi}{(\la/2+\ga_0R(k))^2+\xi^2} 
\int_{\bbT^2}d\ell d\ell' \langle \hat\psi(\ell)
\hat\psi^*(\ell')\rangle_{\mu_\eps}\nonumber
\\
&&
\times
\Bigg\{ \frac{\tilde g\left(\la\eps/2-i[\eps\xi+\om_\eps(k)]\right) }{\eps\la/2-i[\eps\xi+\om_\eps(k)-\om_\eps(\ell)]}\nonumber\\
&&\times
 \Big\{\frac{\tilde g\left(\la\eps/2+i[\eps\xi+\om_\eps(k)]\right)
   }{\eps\la/2+i[\eps\xi+\om_\eps(k)+\om_\eps(\ell')]}+\frac{\tilde
   g\left(\la\eps/2+i[\eps\xi-\om_\eps(k)])\right)
   }{\eps\la/2+i\{\eps\xi-[\om_\eps(k)+\om_\eps(\ell')]\}}\Big\} \nonumber\\
&&
+\frac{\tilde
   g\left(\la\eps/2-i[\eps\xi-\om_\eps(k)]\right)}{\eps\la/2-i[\eps\xi+\om_\eps(k)+\om_\eps(\ell)]}\\
   &&\times
 \Big\{\frac{\tilde g\left(\la\eps/2+i[\eps
      \xi+\om_\eps(k)]\right)}{\eps\la/2+i[\eps\xi+\om_\eps(k)+\om_\eps(\ell')]}+\frac{\tilde
      g\left(\la\eps/2+i[\eps\xi-\om_\eps(k)])\right)
      }{\eps\la/2+i\{\eps\xi-[\om_\eps(k)+\om_\eps(\ell')]\}}\Big\}\nonumber\\
&&{-\frac{\tilde g\left(\la\eps/2-i[\eps\xi+\om_\eps(k)]\right)
   }{\eps\la/2-i[\eps\xi+\om_\eps(k)-\om_\eps(\ell)]}\cdot\frac{\tilde
   g\left(\la\eps/2+i[\eps\xi-\om_\eps(k)])\right)
   }{\eps\la/2+i\{\eps\xi-[\om_\eps(k)+\om_\eps(\ell')]\}}}\nonumber \\
&&
{-\frac{\tilde
   g\left(\la\eps/2-i[\eps\xi-\om_\eps(k)]\right)}{\eps\la/2-i[\eps\xi+\om_\eps(k)+\om_\eps(\ell)]}\cdot\frac{\tilde
   g\left(\la\eps/2+i[\eps\xi -\om_\eps(k)]\right)}{\eps\la/2+i[\eps\xi+\om_\eps(k)-\om_\eps(\ell')]}}
\Bigg\}.\nonumber
\end{eqnarray}
Substituting for ${\rm Re}\,\bar{\frak d}_\eps^2(\la,k)$ from \eqref{010104} into \eqref{010904-19} we obtain
$\bar{\frak L}_{scat,2}^\eps(\la)=\bar{\frak
  L}_{scat,21}^\eps(\la)+\bar{\frak L}_{scat,22}^\eps(\la)$, where
the terms in the right hand side correspond to $R_\eps(\la,k)$, $\rho_\eps(\la,k)$
respectively.
As for $\rho_\eps(\la,k)$  we expect its
contribution to be small in the limit and $\lim_{\eps\to0+}\bar{\frak
  L}_{scat,22}^\eps(\la)=0$. 
In fact, we have.
\begin{lemma}
\label{lm011311}
For each $\la>0$ we have
\begin{equation}
\label{011311}
\lim_{\eps\to0+}\int_{\bbT}|\rho_\eps(\la,k)|dk=0.
\end{equation}
\end{lemma}
The proof of the lemma follows closely the argument presented in the
proof Lemma 6.1 in \cite{kors}, so we will not present it here.

Concerning $\bar{\frak
  L}_{scat,21}^\eps(\la)$, we note 
first that, by
the same type of estimate as in \eqref{011804-19}, 
\begin{equation}
\label{031804-19}
\lim_{\eps\to0+}\Big[\bar{\frak
  L}_{scat,21}^\eps(\la)-{\frak
  L}_{scat,21}^\eps(\la)\Big]=0,
\end{equation}
where
\begin{equation}\label{Hpm0}
{\frak
  L}_{scat,21}^\eps(\la):=-\frac{\ga_1}{2}\int_{
  \bbT}R_\eps^0\left(\la,k\right) \hat
   G^\star(k)  dk
\end{equation}
and 
\begin{eqnarray}
\label{R-eps0}
&&
R_\eps^0(\la,k):=-\frac{\gamma_1 (\la+2\ga_0\eps R(k))}{2^3\cdot\pi\eps^2}\int_{\bbR}\frac{d\xi}{(\la/2+\ga_0R(k))^2+\xi^2} \int_{\bbT^2}d\ell d\ell' \eps\langle \hat\psi(\ell)\hat\psi^*(\ell')\rangle_{\mu_\eps}\\
&&
\times \frac{|\tilde g_\eps\left(\la\eps/2-i[\eps\xi+\om(k)]\right)|^2 }{\la/2-i\{\xi+\eps^{-1}[\om(k)-\om(\ell)]\}}
\times \frac{1
   }{\la/2+i\{\xi+\eps^{-1}[\om(k)-\om(\ell')]\}}.\nonumber
\end{eqnarray}

Using \eqref{R-eps}  and the change of variables 
\begin{equation}
\label{ell-k}
\ell=:k'+\frac{\eps \eta'}{2},\quad \ell'=:k'-\frac{\eps \eta'}{2}
\end{equation}
we can write  
\begin{eqnarray}
\label{011811}
&&
{\frak
  L}_{scat,21}^\eps(\la):
=\frac{\gamma_1^2 (\la+2\ga_0\eps R(k))}{2^3\pi\eps}\int_{\bbR}d\xi 
\int_{\bbT\times T_\eps}  \frac{\widehat W_{\eps,+}^{\rm un}(0,\eta',k') G^*(k) dk d\eta' dk'
}{\la/2-i\{\xi+\eps^{-1}[\om(k)-\om(k'+\eps \eta'/2)]\}}\nonumber\\
&&
\times \frac{|\tilde g_\eps\left(\la\eps/2-i[\eps\xi+\om(k)]\right)|^2
   }{\la/2+i\{\xi+\eps^{-1}[\om(k)-\om(k'-\eps \eta'/2)]\}}\times \frac{1}{(\la/2+\ga_0R(k))^2+\xi^2}
  . 
\end{eqnarray}
Here $T_\eps$ is the image of $\bbT_\eps\times\bbT$ under the change
of variables.
In fact, we may discard the contribution due to large $\eta'$, thanks
  to assumption \eqref{011812c}. 
The main contribution to the limit comes therefore from the regions where $\omega(k)\approx\omega(k')$, that is, where 
either~$k\approx k'$ -- this generates the transmission term, or $k\approx -k'$ -- this is responsible for the reflection term in the limit.
The conclusion of Lemma \ref{lem-feb1502} follows directly from the
following result.
\begin{lemma}\label{lem-feb1508}
We have
\begin{equation}
\label{021511c}
\lim_{\eps\to0+}{\frak
  L}_{scat,21}^\eps(\la)=
\frac{\gamma_1}{4}\sum_{\iota=\pm}\int_{\bbT}\frac{\hat G^\star(k)\fgeeszett(k)\widehat
  W_0(\eta,\iota k)d\eta dk}{\la+2\ga_0R(k)+\iota i\om'(k)\eta} .
\end{equation} 
\end{lemma}
The proof of the lemma follows the argument presented in the
proof of Lemma 6.2 in \cite{kors}, so we omit it here.

\subsection{Proofs of auxiliary results}

\subsubsection{Proof of Lemma \ref{lm010304-19}}

\label{C3}


We can write
\begin{align*}
&
\int_{\bbT}\Big|\tilde I^{\iota_1,\iota_2}_\eps(\la,k)\Big|dk
=
\eps^2  \int_{\bbT}dk \Big|
\int_{\bbT}d\ell \int_0^{+\infty} \exp\left\{-\{\eps(\la+\ga_0 R(k))- i\iota_1 \om_\eps(k)\} s\right\}
  g_\eps\left(ds\right)\\
&\times
\frac{ r_{1,\eps}(k) r_{2,\eps}(\ell) }{\eps(\la
  +\ga_0R(k)+\ga_0 R(\ell))-i[\iota_1 \om_\eps(k)+\iota_2 \om_\eps(\ell)]}
  \langle\hat\psi^\star(k)\hat\psi(\ell)
  \rangle_{\mu_\eps}\Big|
\end{align*}
\begin{align*}
&
\le 
\eps^2 
\int_{\bbT^2}dkd\ell \Big|\tilde g_\eps\left(\vphantom{\int_0^1}\eps(\la+\ga_0 R(k))+ i\iota_1 \om_\eps(k)\right)\Big|
 \\
&\times
\frac{| r_{1,\eps}(k) r_{2,\eps}(\ell)| }{\eps(\la
  +\ga_0R(k)+\ga_0 R(\ell))-i[\iota_1 \om_\eps(k)+\iota_2 \om_\eps(\ell)]|}
  \Big|\langle\hat\psi^\star(k)\hat\psi(\ell)
  \rangle_{\mu_\eps}\Big|.
\end{align*}
The expression in the right hand side  can be estimated by
$I_\eps^{\iota_1\iota_2}$, where
\begin{align*}
&
I_\eps^{\iota}:=\eps^2  r_{1,*}r_{2,*}\|\tilde g_\eps\|_\infty
\int_{\bbT^2}
\frac{\Big|\langle\hat\psi^\star(k)\hat\psi(\ell)
  \rangle_{\mu_\eps}\Big| dkd\ell  }{\eps(\la
  +\ga_0R(k)+\ga_0 R(\ell))+|\om_\eps(k)+\iota \om_\eps(\ell)|}.
\end{align*}
We need to show that
\begin{equation}
\label{050905-19}
I_\eps^{\iota}\preceq
\eps^{1/2}\log\eps^{-1},\quad \eps\in(0,1]
\end{equation}
for $\iota=\pm$.
Note that
\begin{align}
\label{011804-19}
&
{\frak r}_\eps:=\sup_{A\in\bbR,k\in\bbT}\left|\frac{1}{|\eps\la/2-i\left(\om_\eps(k)-A\right)|}-\frac{1}{|\eps\la/2-i\left(\om(k)-A\right)|}\right|\\
&
=
\sup_{A\in\bbR,k\in\bbT} \frac{\eps^2dk}{|\eps\la/2-i\left(\om_\eps(k)-A\right)||\eps\la/2-i\left(\om(k)-A\right)|}
\preceq 1, \quad \eps\in(0,1].\nonumber
\end{align}
Let
\begin{equation}
\label{050905-19a}
\tilde I_\eps^{\iota}:=\eps^2  r_{1,*}r_{2,*}\|\tilde g_\eps\|_\infty
\int_{\bbT^2}
\frac{\Big|\langle\hat\psi^\star(k)\hat\psi(\ell)
  \rangle_{\mu_\eps}\Big| dkd\ell  }{\eps(\la
  +\ga_0R(k)+\ga_0 R(\ell))+|\om(k)+\iota \om(\ell)|},\quad \eps\in(0,1].
\end{equation}
We have
\begin{eqnarray*}
&&
 |\tilde I^\iota_{\eps}-I^\iota_{\eps}|\le \eps^2  r_{1,*}r_{2,*}{\frak r}_\eps\|\tilde g_\eps\|_\infty
\int_{\bbT^2}
\Big|\langle\hat\psi^\star(k)\hat\psi(\ell)
  \rangle_{\mu_\eps}\Big| dkd\ell \preceq \eps,\quad \eps\in(0,1],
\end{eqnarray*}
as $\eps\to0$.

In the case $\iota=+$ we can write
\begin{align*}
&\tilde I_\eps^+
\le 2  r_{1,*}r_{2,*}\|\tilde g_\eps\|_\infty
\Big\{\eps\int_{\bbT}\Big|\langle\hat\psi^\star(k) \rangle_{\mu_\eps}\Big|^2dk\Big\}
\int_{\bbT}\frac{\eps d\ell  }{\eps\la
  + \om(\ell)}\preceq \eps\log\eps^{-1},\quad \eps\in(),1].
\end{align*}
In the case $\iota=-$ we can write
\begin{align*}
&\tilde I_\eps^-
\le 2  r_{1,*}r_{2,*}\Gamma_\eps\|\tilde g_\eps\|_\infty
\Big\{\eps\int_{\bbT}\Big|\langle\hat\psi^\star(k) \rangle_{\mu_\eps}\Big|^2dk\Big\}
,
\end{align*}
with 
$$
 \Gamma_\eps :=\sup_{A\in\bbR}\int_{\bbT}\frac{\eps dk}{\eps\la+|\om(k)-A|}.
$$
Note that
$
 \Gamma_\eps =\Gamma_{\eps}^++\Gamma_{\eps}^-,
$
with
$$
  \Gamma_{\eps}^\pm:=\sup_{A\in\bbR}\int_{\om_{\rm min}}^{\om_{\rm max}}\frac{\eps du}{(\eps\la+|u-A|)|\om'(\om_\pm(u))|}.
 $$
Recall that $\om_-$, $\om_+$ are the decreasing and increasing
branches of the inverse function of the dispersion relation
$\om(\cdot)$. 
Our assumptions on the dispersion relation imply that  
\begin{equation}
\label{020707-19}
|\om'(\om_\pm(u))|\approx (\om_{\rm max}-u)^{1/2},
\hbox{ for $\om_{\rm max}-u\ll1$.}
\end{equation}
The consideration near the minimum of $\omega$ is 
identical unless $\omega_{\rm min}=0$,
in which case $|\omega'(k)|$ stays uniformly positive near the minimum.
Therefore, we have
$$
\Gamma_{\eps}^\pm\preceq \sup_{A\in [0,1]}\int_0^1\frac{\eps du}{(\eps+|u-A|)\sqrt{u}}
\preceq \eps^{1/2}\log\eps^{-1}
 $$
and 
we conclude that
\eqref{050905-19} holds. 
\qed

\subsubsection{Proof of Lemma \ref{lm010504-19}}

\label{C3.1}

 We have
\begin{align*}
&\Big\|\widetilde{
 I\!I}^{\iota_1,\iota_2,\iota_3}_\eps (\la)\Big\|_{L^1(\bbT)} 
=\eps^2 \int_{\bbT}dk\Big|\int_0^{+\infty}\int_0^{+\infty}e^{-\eps \la
  (t+t')/2}\delta(t-t') dt dt'\int_0^{t} ds \int_0^{t'}
  g_\eps\left(ds'\right)\int_{0}^{s}g_\eps\left(ds_1\right)
  \int_{\bbT^2}
d\ell d\ell'\\
&
\times  r_{1,\eps}(k)r_{2,\eps}(\ell) r_{3,\eps}(\ell')  e^{\la_{\iota_2}(\ell) (s-s_1)} e^{\la_{\iota_3}(\ell')
  (t'-s')} e^{\la_{\iota_1}(k)
  (t-s)}\langle\hat\psi^\star(\ell)\hat\psi(\ell')
  \rangle_{\mu_\eps}\Big|
\end{align*}
\begin{align*}
&
=\frac{\eps^2}{2\pi}\int_{\bbT}dk\Big|\int_{\bbR}d\xi\int_0^{+\infty}\int_0^{+\infty}e^{-\eps \la
  (t+t')/2}e^{i\xi(t-t')} dt dt'\int_0^{t} ds \int_0^{t'}
  g_\eps\left(ds'\right)\int_{0}^{s}g_\eps\left(ds_1\right)
  \int_{\bbT^2}
d\ell d\ell'\\
&
\times  r_{1,\eps}(k)r_{2,\eps}(\ell) r_{3,\eps}(\ell')  e^{\la_{\iota_2}(\ell) (s-s_1)} e^{\la_{\iota_3}(\ell')
  (t'-s')} e^{\la_{\iota_1}(k)
  (t-s)}\langle\hat\psi^\star(\ell)\hat\psi(\ell')
  \rangle_{\mu_\eps}\Big|.
\end{align*}
Note that
\begin{align*}
&
\int_0^{+\infty}e^{-(\eps \la
  /2-i\xi)t} dt\int_0^{t} ds \int_{0}^{s}g_\eps\left(ds_1\right)
e^{\la_{\iota_2}(\ell) (s-s_1)} e^{\la_{\iota_1}(k)
  (t-s)}\\
&
=\int_0^{+\infty} ds \int_{0}^{s}g_\eps\left(ds_1\right)
\frac{e^{\la_{\iota_2}(\ell) (s-s_1)} \exp\left\{-(\eps \la
  /2-i\xi)s\right\}}{\eps \la
  /2-i\xi-\la_{\iota_1}(k)}\\
&
=\frac{\tilde g_\eps(\eps \la
  /2-i\xi)}{[\eps \la
  /2-i\xi-\la_{\iota_1}(k)][\eps \la
  /2-i\xi-\la_{\iota_2}(\ell)]} .
\end{align*}
Similarly
\begin{align*}
&
\int_0^{+\infty}e^{-(\eps \la/2+i\xi)
  t'}dt' \int_0^{t'}
  g_\eps\left(ds'\right)e^{\la_{\iota_3}(\ell')
  (t'-s')} 
=\frac{\tilde g_\eps(\eps \la/2+i\xi)}{\eps \la/2+i\xi-\la_{\iota_3}(\ell')}.
\end{align*}
Taking the above into account we obtain
\begin{align}
\label{011305-19}
&\Big\|\widetilde{
 I\!I}^{\iota_1,\iota_2,\iota_3}_\eps (\la)\Big\|_{L^1(\bbT)} =\frac{\eps^2}{2\pi}\int_{\bbT}dk\Big|\int_{\bbR}d\xi
  \int_{\bbT^2}
d\ell d\ell' \langle\hat\psi^\star(\ell)\hat\psi(\ell')
  \rangle_{\mu_\eps}|\tilde g_\eps(\eps \la
  /2-i\xi)|^2\\
&
\times\left\{\vphantom{\int_0^1}[\eps (\la
  /2+\ga_0R(k))-i(\xi+\iota_1\om_\eps(k))][\eps (\la
  /2+\ga_0R(k))-i(\xi+\iota_2\om_\eps(\ell))][\eps( \la/2+\ga_0R(k))+i(\xi-\iota_3\om_\eps(\ell'))]\right\}\Big|.\nonumber
\end{align}
Consider only the case $\iota_1=\iota_2=\iota_3=+$, as the other ones
 can be done in a similar fashion. We omit writing the superscripts in
what follows.  
Change variables $\ell:=k'-\eps\eta/2$, $\ell'=k'-\eps\eta/2$ and obtain
\begin{align*}
&\Big\|\widetilde{
 I\!I}_\eps (\la)\Big\|_{L^1(\bbT)} =\frac{\eps^2}{\pi}\int_{\bbT}dk\Big|\int_{\bbR}d\xi
  \int_{T_\eps}
d\eta dk'  \widehat W_\eps(\eta,k')|\tilde g_\eps(\eps \la
  /2-i\xi)|^2 \Big\{[\eps (\la
  /2+\ga_0R(k))-i(\xi+\om_\eps(k))]\\
&
[\eps (\la
  /2+\ga_0R(k))-i(\xi+\om_\eps(k'-\eps\eta/2))][\eps( \la/2+\ga_0R(k))+i(\xi-\om_\eps(k'+\eps\eta/2))]\Big\}^{-1}\Big|.
\end{align*}
Thanks to \eqref{011812aa} and
\eqref{G} we can estimate
\begin{align*}
&\Big\|\widetilde{
 I\!I}_\eps (\la)\Big\|_{L^1(\bbT)} \preceq \eps^2\int_{\bbR}d\xi\int_{\bbR}\varphi(\eta)d\eta
  \int_{\bbT^2}\left|[\eps (\la
  /2+\ga_0R(k))-i(\xi+\om_\eps(k))]\right.
\\
&\left.
\times [\eps (\la
  /2+\ga_0R(k))-i(\xi+\om_\eps(k'-\eps\eta/2))][\eps(
  \la/2+\ga_0R(k))+i(\xi-\om_\eps(k'+\eps\eta/2))]\right|^{-1}dk
  dk',\quad \eps\in(0,1].
\end{align*}
We need to prove that the right hand side vanishes, as
$\eps\to0$. Similarly to what has been done in the proof of Lemma
\ref{lm010304-19}, it suffices only to show that
\begin{equation}
\label{J-eps}
\lim_{\eps\to0}J_\eps=0,
\end{equation}
 where
\begin{align*}
&
J_\eps:=\eps^2\int_{\bbR}d\xi\int_{\bbR}\varphi(\eta)d\eta
  \int_{\bbT^2}\left|[\eps \la
  /2-i(\xi+\om(k))]\right.
\\
&\left.
\times [\eps \la
  /2-i(\xi+\om(k'-\eps\eta/2))][\eps
  \la/2+i(\xi-\om(k'+\eps\eta/2))]\right|^{-1}dk dk'
\end{align*}
Changing variables $\xi+\om(k):=\eps\xi'$ we conclude that
\begin{align*}
&
J_\eps
\approx
\int_{\bbR}d\xi\int_{\bbR}\varphi(\eta)d\eta
  \int_{\bbT^2}
\left\{\vphantom{\int_0^1}(1+|\xi|)(
  1+|\xi+\eps^{-1}(\om(k'-\eps\eta/2)-\om(k))|)\right.\\
&
\left. \vphantom{\int_0^1} (1+|\xi+\eps^{-1}(\om(k)-\om(k'+\eps\eta/2)))\right\}^{-1}dkd
  k'
\le J_\eps^1+J_\eps^2,
\end{align*}
with
\begin{align*}
&
J_\eps^1:=
\frac12\int_{\bbR}d\xi\int_{\bbR}\varphi(\eta)d\eta
  \int_{\bbT^2}
\left\{(1+|\xi|)(
  1+|\xi+\eps^{-1}(\om(k'-\eps\eta/2)-\om(k))|)^2\right\}^{-1}dkd
  k'
\end{align*}
and
\begin{align*}
&
J_\eps^2:=\frac12
\int_{\bbR}d\xi\int_{\bbR}\varphi(\eta)d\eta
  \int_{\bbT^2}
\left\{(1+|\xi|)(1+|\xi+\eps^{-1}(\om(k)-\om(k'+\eps\eta/2)))^2\right\}^{-1}dkd
  k'
\end{align*}
Using an elementary estimate
\begin{equation}
\label{ab}
\int_{\bbR}\frac{1}{1+|x+a|}\times\frac{dx}{1+x^2}\preceq
\frac{1}{1+|a|},\quad a\in\bbR
\end{equation}
we conclude that
\begin{align*}
&
J_\eps^1\preceq
\frac12\int_{\bbR}\varphi(\eta)d\eta
  \int_{\bbT^2}
\left\{
  1+|\eps^{-1}(\om(k'-\eps\eta/2)-\om(k))|\right\}^{-1}dkd
  k'\to0,
\end{align*}
by virtue of the dominated convergence theorem. Estimates for
$J_\eps^2$ are similar.
\qed

\subsubsection{Proof of formula (\ref{010104})}

\label{d-eps2}

From \eqref{021207-19} we have
\begin{eqnarray*}
&&
2{\rm Re}\,\bar{\frak d}_\eps^2(\la,k)
=-\gamma_1  \eps\left\langle\int_0^{+\infty}e^{-\la\eps t}e^{-2\ga_0\eps
   R(k) t} \frac{d}{dt}\left\{\left[\int_0^{t}  e^{\ga_0\eps R(k) ts}
   \cos(\om_\eps(k)s)g_\eps*{\bar p}_0^0(s)ds\right]^2\right.\right.\\
&&
\left.\left.+\left[\int_0^{t}  e^{\ga_0\eps R(k) ts} \sin(\om(k)s)g_\eps*{\bar p}_0^0(s)ds\right]^2
\right\}dt \right\rangle_{\mu_\eps}.
\end{eqnarray*}
Integrating by parts, we obtain
\begin{eqnarray}
&&
2{\rm Re}\,{\frak d}_\eps^2(\la,k)
=C_\eps(\la,k)+S_\eps(\la,k).\label{feb1518}
\end{eqnarray}
The first term in the right side is
\begin{align}
\label{012803}
&
C_\eps(\la,k)
=-\frac{\gamma_1}{4\pi} \eps (\la+2\ga_0\eps R(k))\int_{\bbR}d\xi \int_{\bbT^2}\eps\langle \hat\psi(\ell)\hat\psi^*(\ell')\rangle_{\mu_\eps} \Xi_\eps(\ell,k,\la,\xi)\Xi_\eps^\star(\ell',k,\la,\xi)d\ell d\ell' ,
\end{align}
with
$$
\Xi_\eps(\ell,k,\la,\xi):=\int_0^{+\infty}e^{\ga_0\eps R(k) s}
\cos(\om_\eps(k)s)ds\left\{\int_0^s e^{-i\om_\eps(\ell)
    (s-\tau)}g_\eps(d\tau)\int_{s}^{+\infty}e^{-[(\la/2+\ga_0 R(k))\eps-i\xi] t}dt\right\}.
$$

Integrating out first the $t$ variable, and then the $s$ varable,
we obtain 
\begin{eqnarray*}
&&
\Xi_\eps(\ell,k,\la,\xi)
=\frac{1}{2[(\la/2+\ga_0R(k))\eps-i\xi]}\left\{\frac{\tilde g_\eps\left(\la\eps/2-i[\xi+\om_\eps(k)]\right) }{\la\eps/2-i(\xi+\om_\eps(k)-\om_\eps(\ell))}+\frac{\tilde g_\eps\left(\la\eps/2-i[\xi-\om_\eps(k))]\right)}{\la\eps/2-i[\xi+\om_\eps(k)+\om_\eps(\ell)]}\right\}.
\end{eqnarray*}
Hence, after a change of variables $\xi:=\eps\xi'$, we get
\begin{eqnarray}\label{feb1524}
&&
C_\eps(\la,k)
=-\frac{\gamma_1 (\la+2\ga_0\eps R(k))}{2^4\cdot\pi\eps^2}\int_{\bbR}\frac{d\xi}{(\la/2+\ga_0R(k))^2+\xi^2} \int_{\bbT^2}d\ell d\ell' \eps\langle \hat\psi(\ell)\hat\psi^*(\ell')\rangle_{\mu_\eps}\\
&&
\times \left\{\frac{\tilde g_\eps\left(\la\eps/2-i[\eps\xi+\om_\eps(k)]\right) }{\la/2-i\{\xi+\eps^{-1}[\om_\eps(k)-\om_\eps(\ell)]\}}+\frac{\tilde g_\eps\left(\la\eps/2-i[\eps\xi-\om_\eps(k)]\right)}{\la/2-i\{\xi+\eps^{-1}[\om_\eps(k)+\om_\eps(\ell)]\}}\right\}
\nonumber\\
&&
\times \left\{\frac{\tilde g_\eps\left(\la\eps/2+i[\eps\xi+\om_\eps(k)]\right) }{\la/2+i\{\xi+\eps^{-1}[\om_\eps(k)-\om_\eps(\ell')]\}}+\frac{\tilde g\left(\la\eps/2+i[\eps \xi+\om_\eps(k)])\right)}{\la/2+i\{\xi+\eps^{-1}[\om_\eps(k)+\om_\eps(\ell')]\}}\right\}.\nonumber
\end{eqnarray}
A similar calculation leads to  
\begin{eqnarray}\label{feb1526}
&&
S_\eps(\la,k)
=\frac{\gamma_1 (\la+2\ga_0\eps R(k))}{2^4\pi\eps^2}\int_{\bbR}\frac{d\xi}{(\la/2+\ga_0R(k))^2+\xi^2} \int_{\bbT^2}d\ell d\ell' \eps\langle \hat\psi(\ell)\hat\psi^*(\ell')\rangle_{\mu_\eps}\\
&&
\times \left\{\frac{\tilde g_\eps\left(\la\eps/2-i[\eps\xi+\om_\eps(k)]\right) }{\la/2-i\{\xi+\eps^{-1}[\om_\eps(k)-\om_\eps(\ell)]\}}-\frac{\tilde g_\eps\left(\la\eps/2-i[\eps\xi-\om_\eps(k)]\right)}{\la/2-i\{\xi+\eps^{-1}[\om_\eps(k)+\om_\eps(\ell)]\}}\right\}
\nonumber\\
&&
\times \left\{\frac{\tilde g_\eps\left(\la\eps/2+i(\eps\xi-\om_\eps(k)])\right) }{\la/2+i\{\xi-\eps^{-1}[\om_\eps(k)+\om_\eps(\ell')]\}}-\frac{\tilde g_\eps\left(\la\eps/2+i[\eps\xi+\om_\eps(k)]\right)}{\la/2+i\{\xi+\eps^{-1}[\om_\eps(k)-\om_\eps(\ell')]\}}\right\}.
\nonumber
\end{eqnarray}
Putting (\ref{feb1518}) --  (\ref{feb1526}) together,
gives (\ref{010104}).

\section{Proof of Proposition \ref{cor010304-19}}

\label{appC}

Let (cf \eqref{j})
\begin{eqnarray}
\label{jl}
&&
\tilde{ j}_\eps(\la,k) :=\int_0^{+\infty}e^{-\la t}j_\eps(t,k)dt, \quad {\rm Re}\,\la>0.
\end{eqnarray}
From \eqref{jl} and \eqref{j}  it follows that
\begin{equation}
\label{jl1}
\tilde{ j}_\eps(\la,k)
=\frac{\la\sqrt{1-(\eps\beta)^2}}{2(
   \la+\eps\ga_0R(k) )}\left\{\frac{
   1}{\la+\eps\ga_0R(k)+i\om_\eps(k)}+\frac{
   1}{\la+\eps\ga_0R(k)-i\om_\eps(k)}\right\}.
\end{equation}
Also, since $\Om_0(k)-\Om_\eps(k)=\eps\ga_0R(k){\frak f}\otimes{\frak
  g}$ (see \eqref{Omk}) we have
\begin{eqnarray}
\label{020204-19a}
&\tilde{ j}_0(\la,k)-\tilde{
  j}_\eps(\la,k)=\frac12\left[(\la-\Om_0(k))^{-1}
  -(\la-\Om_\eps(k))^{-1}\right]{\frak f}\cdot{\frak f}\\
&
=\frac12(\la-\Om_0(k))^{-1}(\Om_0(k)-\Om_\eps(k))
  (\la-\Om_\eps(k))^{-1}{\frak f}\cdot{\frak f}
=\frac12 \eps\ga_0R(k)\tilde{ j}_\eps(\la,k) \tilde{ j}_0(\la,k).\nonumber
\end{eqnarray}
The following result holds.
\begin{prop}
\label{prop010204-19}
We have
\begin{equation}
\label{010204-19}
\tilde J(\la)=\tilde J_\eps(\la)+\eps\ga_0\tilde{\cal R}_\eps(\la),\quad\eps\in(0,1],\,
\la\in\mathbb C_+,
\end{equation}
where
\begin{equation}
\label{010204-19a}
\tilde{\cal  R}_\eps(\la)=\int_{\bbT}R(k)\tilde j_\eps(\la,k)\tilde j_0(\la,k)dk,\quad
\la\in\mathbb C_+.
\end{equation}
In addition, for any $p\in(1,2)$ and a uniformly continuous and bounded function $\xi:\bbR\to\bbR_+$ satisfying
\begin{equation}
\label{lim-is}
\xi(\eta)\ge
\xi_0,\quad\,\eta\in\bbR
\end{equation}
 we have
\begin{equation}
\label{010204-19b}
\lim_{\eps\to0}\eps^p\int_{\bbR}|\tilde{\cal R}_\eps(\eps\xi(\eta)+i\eta)|^pd\eta=0.
\end{equation}
\end{prop}
\proof
Identity \eqref{010204-19} follows directly from \eqref{020204-19a}.
Letting $\la:=\eps\xi(\eta)+i\eta$, using \eqref{jl1} together with \eqref{010204-19a}  and the change of variables $k\mapsto
v=\om(k)$ we can
write
\begin{align}
\label{030204-19a}
&
\tilde{ J}(\eps\xi(\eta)+i\eta)-\tilde{ J}_\eps(\eps\xi(\eta)+i\eta)
=\frac12 \sum_{\iota=\pm}\left\{ \int_{\bbR}
  \frac{ \chi_{\iota,\eps}(v)dv }{\eps\xi(\eta)+i(\eta+v)}+\int_{\bbR}
  \frac{ \chi_{\iota,\eps}(v)dv }{\eps\xi(\eta)+i(\eta-v)}\right\},
\end{align}
with
$$
\chi_{\iota,\eps}(v)=\iota \eps\ga_0R(\om_+(v))\tilde{ j}_\eps(\eps\xi(\eta)+i\eta, \om_+(v))\frac{1_{[\om_{\rm min}, \om_{\rm max}]}(v)}{\om'(\om_+(v))}.
$$
Here, (see Section \ref{sec2.2.1}) $\om_{\pm}$ are the two branches of
inverses of the unimodal dispersion relation
$\om$, with  $\om_+: [\om_{\rm min}, \om_{\rm max}]\to [0,1/2]$ and $\om_-:=-\om_+$.
In addition to \eqref{020707-19}, in the optical case, we also have
\begin{equation}
\label{020707-19a}
|\om'(\om_\pm(u))|\approx (u-\om_{\rm min})^{1/2},
\hbox{ for $u-\om_{\rm min}\ll1$.}
\end{equation}

Hence,
\begin{align}
\label{040204-19}
 &
\eps\ga_0R(k)|\tilde{ j}_\eps(\eps\xi(\eta)+i\eta,k)|\\
&\le \frac12\left\{\left|\frac{
   \eps\ga_0R(k)}{\eps(\xi(\eta)+\ga_0R(k))+i(\eta+\om_\eps(k))}\right|+\left|\frac{
   \eps\ga_0R(k)}{\eps(\xi(\eta)+\ga_0R(k))+i(\eta-\om_\eps(k))}\right|\right\}.\nonumber
\end{align}
The above allows us to conclude that
\begin{align*}
 &
\eps\ga_0R(k)|\tilde{ j}_\eps(\eps\xi(\eta)+i\eta,k)|\le C
\end{align*}
for some $C>0$ independent of $\eta,\eps,k$ and, as a result,
$$
|\chi_{\iota,\eps}(v)|\le \frac{C1_{[\om_{\rm min}, \om_{\rm max}]}(v)}{\om'(\om_+(v))}.
$$
Thus 
\begin{equation}
\label{030404-19}
\lim_{\eps\to0+}\chi_{\iota,\eps}(v)=0
\end{equation}
both a.s. and in the $L^p$ sense for any $p\in[1,2)$, see
\eqref{020707-19} and \eqref{020707-19a}.
Let $\xi_0$ be as in \eqref{lim-is}.
Note that for any $\eta\in\bbR$
\begin{align*}
&
\Big|\int_{\bbR}
  \frac{ \chi_{\iota,\eps}(v-\eta)dv }{\eps\xi(\eta)+iv}\Big|\le
 \Big|\int_{[|v|\ge \eps\xi_0]}
  \frac{ \chi_{\iota,\eps}(v-\eta)dv }{iv}\Big|
\\
&
+ \Big|\int_{[|v|\ge \eps\xi_0]}
  \frac{ \chi_{\iota,\eps}(v-\eta)dv }{\eps\xi(\eta)+iv}-\int_{[|v|\ge \eps\xi_0]}
  \frac{ \chi_{\iota,\eps}(v-\eta)dv }{iv}\Big|+\Big|\int_{[|v|<\eps\xi_0]}
  \frac{ \chi_{\iota,\eps}(v-\eta)dv }{\eps\xi(\eta)+iv}\Big|
\end{align*}
Denote the expressions in the utmost right hand side by
${\cal J}_{1,\eps}(\eta)$, ${\cal J}_{2,\eps}(\eta)$ and ${\cal
  J}_{3,\eps}(\eta)$, respectively.

 Using Theorem 3.2, p. 35 of \cite{stein} we conclude that for any
 $p\in(1,+\infty)$ there
 exists a constant $C>0$, independent of $\eps>0$ such that
\begin{equation}
\label{010404-19}
\|{\cal J}_{1,\eps}\|_{L^p(\bbR)}\le C \|\chi_{\iota,\eps}\|_{L^p(\bbR)}.
\end{equation}
Considering ${\cal J}_{3,\eps}$, there exists $C>0$ such that
$$
  \frac{ 1_{[|v|<\eps\xi_0]} }{|\eps\xi(\eps)+iv|}\le \frac{C 1_{[|v|<\eps\xi_0]}}{\eps}
$$
thus, again  (using Young's inequality)
for any $p\in(1,+\infty)$ there exists $C>0$ such that
\begin{equation}
\label{020404-19}
\|{\cal J}_{3,\eps}\|_{L^p(\bbR)}\le C  \|\chi_{\iota,\eps}\|_{L^p(\bbR)} ,\quad\eps\in(0,1].
\end{equation}

For $\rho\in(0,1)$ we can write
\begin{align*}
 &
|{\cal J}_{2,\eps}(\eta)|\le \int_{[|v|\ge \eps\xi_0]}
  \frac{\eps\xi(\eta) |\chi_{\iota,\eps}(v-\eta)|dv
   }{[(\eps\xi(\eta))^2+v^2]^{1/2}|v|}\le
\|\xi\|_\infty\int_{[|v|\ge \eps\xi_0]}
  \frac{\eps^{\rho}|\chi_{\iota,\eps}(v-\eta)|dv }{|v|^{1+\rho}}.
\end{align*}
By an application of Young's inequality for convolutions we have
\begin{equation}
\label{020404-19}
\|{\cal J}_{2,\eps}\|_{L^p(\bbR)}\le
2\eps^{\rho}\|\xi\|_\infty \|\chi_{\iota,\eps}\|_{L^p(\bbR)}\int_{\eps\xi_0}^{+\infty}\frac{dv}{v^{1+\rho}}
\le C \|\chi_{\iota,\eps}\|_{L^p(\bbR)},\quad\eps\in(0,1],
\end{equation}
with constant $C>0$ independent of $\eps$.

Taking into account all the above we conclude that
 $$
\lim_{\eps\to0+}\int_{\bbR}|\tilde{ J}(\eps\xi(\eta)+i\eta)-\tilde{ J}_\eps(\eps\xi(\eta)+i\eta)|^pd\eta=0
$$
for any $p\in(1,2)$. 
\qed


Proposition \ref{cor010304-19} follows from the following result.
\begin{prop}
\label{cor010304-19a}
Suppose that $\xi:\bbR\to\bbR_+$ is a uniformly continuous and bounded function  satisfying
\begin{equation}
\label{lim-is1}
\inf_{\eta\in\bbR}\xi(\eta)>0
\end{equation}
and $\tilde r_\eps(\la)$ is given by \eqref{010304-19}.
Then, for any $p\in(1,+\infty)$, we have
$$
\lim_{\eps\to0+}\eps^p\int_{\bbR}|\tilde r_\eps(\eps\xi(\eta)+i\eta)|^pd\eta=0,
$$
\end{prop}
Indeed, assume the above result and let $K$ be as in the statement
of Proposition \ref{cor010304-19}.  
Define $\xi_\pm(\eta):=K(\om_\pm(\eta))$, $\eta\in\bar \bbT_\pm$ and
$\xi_{\pm}(\eta)=1$, elsewhere. We conclude that in particular
$\eps|\tilde r_\eps(\eps\xi_\pm(\eta)+i\eta)$ converges to $0$ in the
Lebesgue measure on $[\om_{\rm min},\om_{\rm max}]$. This obviously implies that
$\eps|\tilde r_\eps(\eps K(k)-i\om(k))$ convergence in the Lebesgue
measure on $\bbT$
to $0$. The $L^p$ convergence 
follows from the fact that the functions are bounded.

\subsection*{Proof of Proposition \ref{cor010304-19a}}

We have (cf \eqref{010204-19a})
$$
\tilde g_\eps(\la)-\tilde g(\la)
=\eps\ga_0\ga_1\left(1+\ga_1\tilde
  {J}_\eps(\la)\right)^{-1}\left(1+\ga_1\tilde
  {J}(\la)\right)^{-1}\tilde{\cal R}_\eps(\la).
$$
From the above identity we conclude that
\begin{align*}
&
|\tilde g_\eps(\eps\xi(\eta)+i\eta)-\tilde g(\eps\xi(\eta)+i\eta)|\le \eps\ga_0\ga_1|\tilde{\cal R}_\eps(\eps\xi(\eta)+i\eta)|.
\end{align*}
From \eqref{010204-19b} we conclude that
\begin{align*}
&
\lim_{\eps\to0+}\int_{\bbR}\Big|\tilde g_\eps(\eps\xi(\eta)+i\eta)-\tilde g(\eps\xi(\eta)+i\eta)\Big|^pd\eta=0
\end{align*}
for any $p\in(1,2)$. On the other hand, thanks to \eqref{G}, for $p\ge
2$ we get
\begin{align*}
&
\int_{\bbR}|\tilde g_\eps(\eps\xi(\eta)+i\eta)-\tilde
  g(\eps\xi(\eta)+i\eta)|^pd\eta\le 2^{p-3/2}\int_{\bbR}|\tilde g_\eps(\eps\xi(\eta)+i\eta)-\tilde
  g(\eps\xi(\eta)+i\eta)|^{3/2}d\eta\to0
\end{align*}
as $\eps\to0+$.
\qed

\end{document}